\documentclass[floatfix,reprint,superscriptaddress,showpacs,preprintnumbers,nofootinbib,amsmath,amssymb,epsfig,pre,floats]{revtex4-1}

\usepackage{graphicx}
\usepackage{epstopdf}
\usepackage{hyperref}
\usepackage{dcolumn}
\usepackage{bm}

\PassOptionsToPackage{caption=false}{subfig}
\usepackage{subfig}

\usepackage{rotating}
\usepackage{color} 

\begin{document}

\title{Simulated evolution of protein-protein interaction networks with realistic topology}

\author{Jack Peterson}
\email{jack@tinybike.net}
\affiliation{Biophysics Graduate Group, University of California, San Francisco, CA 94158}
\author{Steve Press\'e}
\affiliation{Department of Pharmaceutical Chemistry, University of California, San Francisco, CA 94158}
\author{Kristin S.~Peterson}
\affiliation{Department of Forest Ecosystems and Society, Oregon State University, Corvallis, OR 97331}
\author{Ken A.~Dill}
\affiliation{Laufer Center for Physical and Quantitative Biology, Stony Brook University, NY 11794}

\begin{abstract}
We model the evolution of eukaryotic protein-protein interaction (PPI) networks.  In our model, PPI networks evolve by two known biological mechanisms: (1) Gene duplication, which is followed by rapid diversification of duplicate interactions.  (2) Neofunctionalization, in which a mutation leads to a new interaction with some other protein.  Since many interactions are due to simple surface compatibility, we hypothesize there is an increased likelihood of interacting with other proteins in the target protein's neighborhood.  We find good agreement of the model on 10 different network properties compared to high-confidence experimental PPI networks in yeast, fruit flies, and humans.  Key findings are: (1) PPI networks evolve modular structures, with no need to invoke particular selection pressures. (2) Proteins in cells have on average about 6 degrees of separation, similar to some social networks, such as human-communication and actor networks.  (3) Unlike social networks, which have a shrinking diameter (degree of maximum separation) over time, PPI networks are predicted to grow in diameter.  (4) The model indicates that evolutionarily old proteins should have higher connectivities and be more centrally embedded in their networks.  This suggests a way in which present-day proteomics data could provide insights into biological evolution.
\end{abstract}

\maketitle

We are interested in the evolution of protein-protein interaction (PPI) networks.  PPI network evolution accompanies cellular evolution, and may be important for processes such as the emergence of antibiotic resistance in bacteria~\cite{Hughes_2003, Cirz_2005}, the growth of cancer cells~\cite{Taylor_2009}, and biological speciation~\cite{Lynch_2001_Genetics,Ting_2004,Dutkowski_2009}.  In recent years, increasingly large volumes of experimental PPI data have become available \cite{Ito_2000,Uetz_2000,Krogan_2006,Yu_2008}, and a variety of computational techniques have been created to process and analyze these data~\cite{Marcotte_1999,Pellegrini_1999,Valencia_2002,Gomez_2003,Jothi_2005,Liu_2005,Shoemaker_2007,Burger_2008}.  Although these techniques are diverse, and the experimental data are noisy~\cite{Deane_2002}, a general picture emerging from these studies is that the evolutionary pressures shaping protein networks are deeply interlinked with the networks' topology~\cite{Yamada_2009}.  Our aim here is to construct a minimal model of PPI network evolution which accurately captures a broad panel of topological properties.

In this work, we describe an evolutionary model for eukaryotic PPI networks.  In our model, protein networks evolve by two known biological mechanisms: (1) a gene can duplicate, putting one copy under new selective pressures that allow it to establish new relationships to other proteins in the cell, and (2) a protein undergoes a mutation that causes it to develop new binding or new functional relationships with existing proteins.  In addition, we allow for the possibility that once a mutated protein develops a new relationship with another protein (called the target), the mutant protein can also more readily establish relationships with other proteins in the target's neighborhood.  One goal is to see if random changes based on these mechanisms could generate networks with the properties of present-day PPI networks.  Another goal is then to draw inferences about the evolutionary histories of PPI networks.

\begin{figure}[!ht]
\begin{center}
\includegraphics[width=0.45\textwidth]{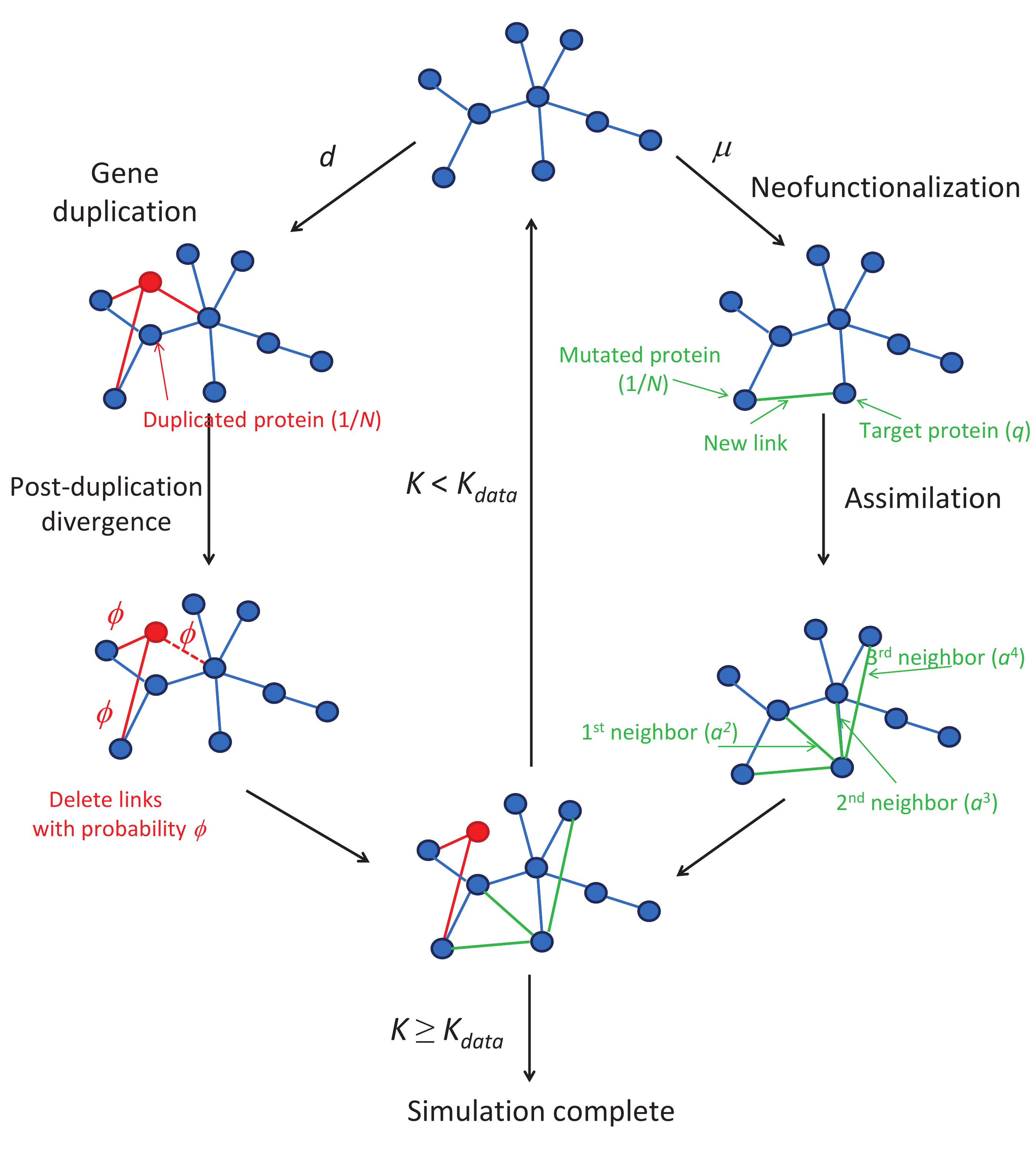}
\end{center}
\caption{{\bf DUNE model flowchart.}  At each time step, the simulated network undergoes a duplication or neofunctionalization event.  Red nodes/links indicate nodes/links that have been created by duplication during the current time step.  Green links indicate links that have been created by neofunctionalization during the current time step.  A dashed line indicates a duplicated link that has been deleted during the post-duplication divergence.  Only 3 neighbors are shown for the assimilation mechanism; however, the actual simulations included up to 20th neighbors.  The simulated network evolves until its number of links ($K$) meets or exceeds the number of links in the data ($K_\text{data}$).}
\label{fig:flowchart}
\end{figure}

\section*{The Model}

We represent a PPI network as a graph.  Each node on the graph represents one protein.  A link (edge) between two nodes represents a physical interaction between the two corresponding proteins.  The links are undirected and unweighted.  To model the evolution of the PPI graph, we simulate a series of steps in time.  At time $t$, one protein in the network is subjected to either a gene duplication or a neofunctionalizing mutation, leading to an altered network by time $t + \Delta t$.  We refer to this model as the DUNE (DUplication \& NEofunctionalization) model.

\subsection*{Gene duplication}  One mechanism by which PPI networks change is gene duplication (DU) \cite{Ohno_1970,Zhang_2003,Xiao_2008}.  In DU, an existing gene is copied, creating a new, identical gene.  In our model, duplications occur at a rate $d$, which is assumed to be constant for each organism.  All genes are accessible to duplication, with equal likelihood.  For simplicity, we assume that one gene codes for one protein.  One of the copies continues to perform the same biological function and remains under the same selective pressures as before.  The other copy is superfluous, since it is no longer essential for the functioning of the cell~\cite{Wagner_2003}.

The superfluous copy of a protein/gene is under less selective pressure; it is free to lose its previous function and to develop some other function within the cell.  Due to this reduced selective pressure, further mutations to the superfluous protein are more readily accepted, including those that would otherwise have been harmful to the organism~\cite{Koch_1972,Taylor_2004}.  Hence, a superfluous protein diverges rapidly after its DU event~\cite{Lynch_2000,Maslov_2004}.  This well-known process is referred to as the \emph{post-duplication divergence}.  Following \cite{Vazquez_2003}, we assume that the link of each such superfluous protein/gene to its former neighbors is deleted with probability $\phi$.  The post-duplication divergence tends to be fast; for simplicity, we assume the divergence occurs within the same time step as the DU.  The divergence is asymmetric~\cite{Kellis_2004,Gu_2005}: one of the proteins diversifies rapidly, while the other protein retains its prior activity.  We delete links from the original or the duplicate with equal probability because the proteins are identical.  As discussed in the supporting information (SI), this is closely related to the idea of \emph{subfunctionalization}, where divergence freely occurs until redundancy is eliminated.  In our model, $\phi$ is an adjustable parameter.

In many cases, the post-duplication divergence results in a protein which has lost all its links.  These `orphan' proteins correspond to silenced or deleted genes in our model.  As discussed below, our model predicts that the gene loss rate should be slightly higher than the duplication rate in yeast, and slightly lower in flies and humans.

We simulate a gene duplication event at time $t$ as follows:

1a) Duplicate a randomly-chosen gene with probability $d \Delta t$.

2a) Choose either the original (50\%) or duplicate (50\%), and delete each of its links with probability $\phi$.

3a) Move on to the next time interval, time $t + \Delta t$.

\subsection*{Neofunctionalization}  Our model also takes into account that DNA can be changed by random mutations.  Most such mutations do not lead to changes in the PPI network structure.  However, some protein mutations lead to new interactions with some other protein (which we call the \emph{target protein}).  The formation of a novel interaction is called a \emph{neofunctionalization} (NE) event.  NE refers to the creation of new interactions, not to the disappearance of old ones.  Functional deletions tend to be deleterious to organisms~\cite{Lynch_1998}.  We do not account for loss-of-function mutations (link deletions) except during post-duplication divergence because damaged alleles will, in general, be eliminated by purifying selection.  In our model, NE mutations occur at a rate $\mu$, which is assumed to be constant.  All proteins are equally likely to be mutated.

How does the mutated protein choose a target protein to which it links?  We define a probability $q$ that any protein in the network is selected for receiving the new link from the mutant protein.  To account for the possibility of homodimerization, the mutated protein may also link to itself \cite{Wagner_2003,Ispolatov_2005}.  Random choice dictates that $q = 1/N$ (see SI).

Many PPI's are driven by a simple geometric compatibility between the surfaces of the proteins \cite{Jones_1996}.  The simplest example is the case of PPI's between flat, hydrophobic surfaces \cite{Tovchigrechko_2001}, a type of interaction which is very common \cite{Wu_2007}.  These PPI's have a simple planar interface, and the binding sites on the individual proteins are geometrically quite similar to one another.  One consequence of these similar-surface interactions is that if protein A can bind to proteins B and C, then there is a greater-than-random chance that B and C will interact with each other.  We refer to this property as \emph{transitivity}: if A binds B, and A binds C, then B binds C.  The number of triangles in the PPI network should correlate roughly with transitivity.  As discussed below, the number of triangles (as quantified by the global clustering coefficient) is about 45 times higher in real PPI networks than in an equally-dense random graph.  This suggests that transitivity is quite common in PPI networks.  Another source of transitivity is gene duplication.  If A binds B, then A is copied to create a duplicate protein A', then A' will (initially) also bind B.  If A interacts with A', then a triangle exists.  However, duplication is unlikely to be the primary source of transitivity; recent evidence shows that, due to the post-duplication divergence, duplicates tend to participate in fewer triangles than other proteins \cite{Vinogradov_2009}.

A concrete example of transitivity is provided by the evolution of the retinoic acid receptor (RAR), an example of neofunctionalization which has been characterized in detail \cite{Escriva_2006}.  Three paralogs of RAR exist in vertebrates (RAR$\alpha$, $\beta$, and $\gamma$), as a result of an ancient duplication.  The interaction profiles of these proteins are quite different.  Previous work indicates that RAR$\beta$ retained the role of the ancestral RAR \cite{Escriva_2006}, while RAR$\alpha$ and $\gamma$ evolved new functionality.  RAR$\alpha$ has several interactions not found in RAR$\beta$.  RAR$\alpha$ has novel interactions with a histone deacetylase (HDAC3) as well as seven of HDAC3's nearest-neighbors (HDAC4, MBD1, Q15959, NRIP1, Q59FP9, NR2E3, GATA2).  None of these interactions are found in RAR$\beta$.  The probability that all of these novel interactions were created independently is very low.  RAR$\alpha$ has 65 known PPI's and HDAC3 has 83, and the present-day size of the human PPI network is a little over 3000 proteins.  Therefore, the chance of RAR$\alpha$ randomly evolving novel interactions with 7 of HDAC3's neighbors is less than 1 in a billion.  This strongly suggests that when a protein evolves an interaction to a target, it has a greater-than-random chance of also linking to other, neighboring proteins.

How do similar-surface interactions affect the evolution of PPI networks?  First, consider how an interaction triangle would form.  Suppose proteins A and B bind due to physically similar binding sites.  Protein X mutates and evolves the capacity to bind A.  There is a reasonable chance that X has a surface which is similar to both A and B.  If so, protein X is likely to also bind to B, forming a triangle.  Denote the probability that two proteins interact due to a simple binding site similarity by $a$.  The probability that A binds B (and X binds A) in this manner is $a$.  Assuming these probabilities are identical and independent, the probability that X binds B is $a^2$.

So far, we have discussed transitivity as it affects the PPI's in which protein A is directly involved (A's first-neighbors).  We now introduce a third protein to the above example, resulting in a chain of interactions: protein A binds B, B binds C, but C does not bind A.  Protein X mutates and gains an interaction with A (with probability $a^2$).  What is the probability that X will also bind C?  The probability that B binds C due to surface similarity is $a$.  Thus, X will bind C (A's second-neighbor) with probability $a^3$.  In general, the probability that X will bind one of A's $j^\text{th}$ neighbors is $a^{j+1}$.  We refer to this process as \emph{assimilation}, and the `assimilation parameter' $a$ is a constant which varies between species.  As discussed in SI, it is primarily mutliple-partner proteins which bind to their partners at different times and/or locations which are affected by this process; consequently, at most one link is created by assimilation at the first-neighbor level, second-neighbor level, etc.  Assimilation is assumed to act on a much shorter time scale than duplication and neofunctionalization; in our model, it is instantaneous.

Our hypothesized assimilation mechanism makes several predictions that could be tested experimentally: (1) the probability of a protein assimilating into a new pathway should be $a^2$ (at the first-neighbor level), $a^3$ (at the second-neighbor level), and so on, where $a$ is a constant which varies between species; (2) weak, nonspecific binding and planar interfaces should be overrepresented in interaction triangles (and longer cycles) between non-duplicate proteins; (3) competitive inhibitors should be overrepresented in interaction triangles; and (4) domain shuffling should be associated with assimilation.  (See SI for discussion of (3) and (4).)

We simulate a neofunctionalization event at time $t$ as follows:

1b) Mutate a randomly-chosen gene with probability $\mu \Delta t$.

2b) Link to a randomly-chosen target protein.

3b) Add a second link to one of the target's first-neighbor proteins, chosen randomly, with probability $a^2$.

4b) Add a link to one of the target's second-neighbor proteins, with probability $a^3$, etc.

5b) Move on to the next time interval, time $t + \Delta t$.

\begin{table}[b]
\caption{{\bf Network sizes and model parameters}}
\begin{tabular}{| c | c c | c c c c |}
\hline
 & $N_\text{data}$ & $K_\text{data}$ & ${d}$ & $\mu$ & $\phi$ & $a$ \\
\hline
Yeast & 2170 & 3819 & 0.01 & $7.86 \times 10^{-4}$ & 0.555 & 0.690 \\
Fly & 878 & 1140 & 0.0014 & $5.89 \times 10^{-4}$ & 0.866 & 0.546 \\
Human & 3165 & 5547 & 0.0037 & $7.62 \times 10^{-4}$ & 0.652 & 0.727 \\
\hline
\end{tabular}
\begin{flushleft}$N$ and $K$ are the numbers of proteins and links, respectively.  ($K_\text{data}$ is used to stop the simulation.  $N_\text{data}$ is not used as a constraint.)  $d$ and $\mu$ have units of per gene per million years (Myr).  $\phi$ and $a$ are probabilities (unitless).  $K_\text{data}$ and $d$ are constraints from the data, while $\mu$, $\phi$, and $a$ are adjustable parameters.  We used Monte Carlo simulations to optimize the parameter values, by minimizing the total symmetric mean absolute percentage error values of the simulated versus the experimental data (see SI).  Our values of $\mu$ are substantially lower than $d$ because $\mu$ is the rate of mutations leading to the creation of a new PPI (rather than being a simple mutation rate, which would be much higher).
\end{flushleft}
\label{tab:params}
\end{table}

\subsection*{Model simulation and parameters}  A flowchart of how PPI networks evolve in our model is shown in Figure~\ref{fig:flowchart}.  To simulate the network's evolution, one of the two mechanisms above is used at each time step, using~\cite{Gillespie_1977}.  We call each possible time series a \emph{trajectory}.  We begin each trajectory starting from two proteins sharing a link (the simplest configuration that is still technically a network).  Each simulated trajectory ends when the model network has grown to have the same total number of links, $K$, as found in the experimental data, $K_\text{data}$.  Here, we perform sets of simulations for three different organisms: yeast (\emph{Saccharomyces cerevisiae}), fruit flies (\emph{Drosophila melanogaster}), and humans (\emph{Homo sapiens}).  Because evolution is stochastic, there are different possible trajectories, even for identical starting conditions and parameters.  We simulated 50 trajectories for each organism.  Our figures show the median values of each feature as a heavy line, and individual trajectories as light lines.

For a given data set, the number of links ($K_\text{data}$) is known.  We estimate the duplication rate $d$ from literature values.  There have been several empirical estimates of duplication rates, mostly falling within an order of magnitude of each other \cite{Lynch_2000,Gu_2002,Gao_2004,Lynch_2003,Osada_2008,Lynch_2003,Cotton_2005,Pan_2007}.  We averaged together the literature values to estimate $d$ for each species (Table~\ref{tab:params}).

The quantity $\mu$ is not as well known.  Its value relative to $d$ has been the topic of considerable debate \cite{Wagner_2003,Beltrao_2007,Lynch_2008,Gibson_2009}.  Although, in principle, $\mu$ is a measurable quantity, it has proven difficult to obtain an accurate value, in part because the fixation rate of neofunctionalized alleles varies with population size \cite{Kimura_1957,Walsh_1995}.  In the absence of a consensus order-of-magnitude estimate, in our model, we treat $\mu$ as a fitting parameter.  Consistent with the findings of \cite{He_2005} and \cite{Beltrao_2007}, our best-fit values of $\mu$ are within an order of magnitude of each other for yeast, fruit fly, and human networks.  Best-fit parameter values are given in Table~\ref{tab:params}.

\begin{figure}[t]
\begin{center}
\includegraphics[width=0.45\textwidth]{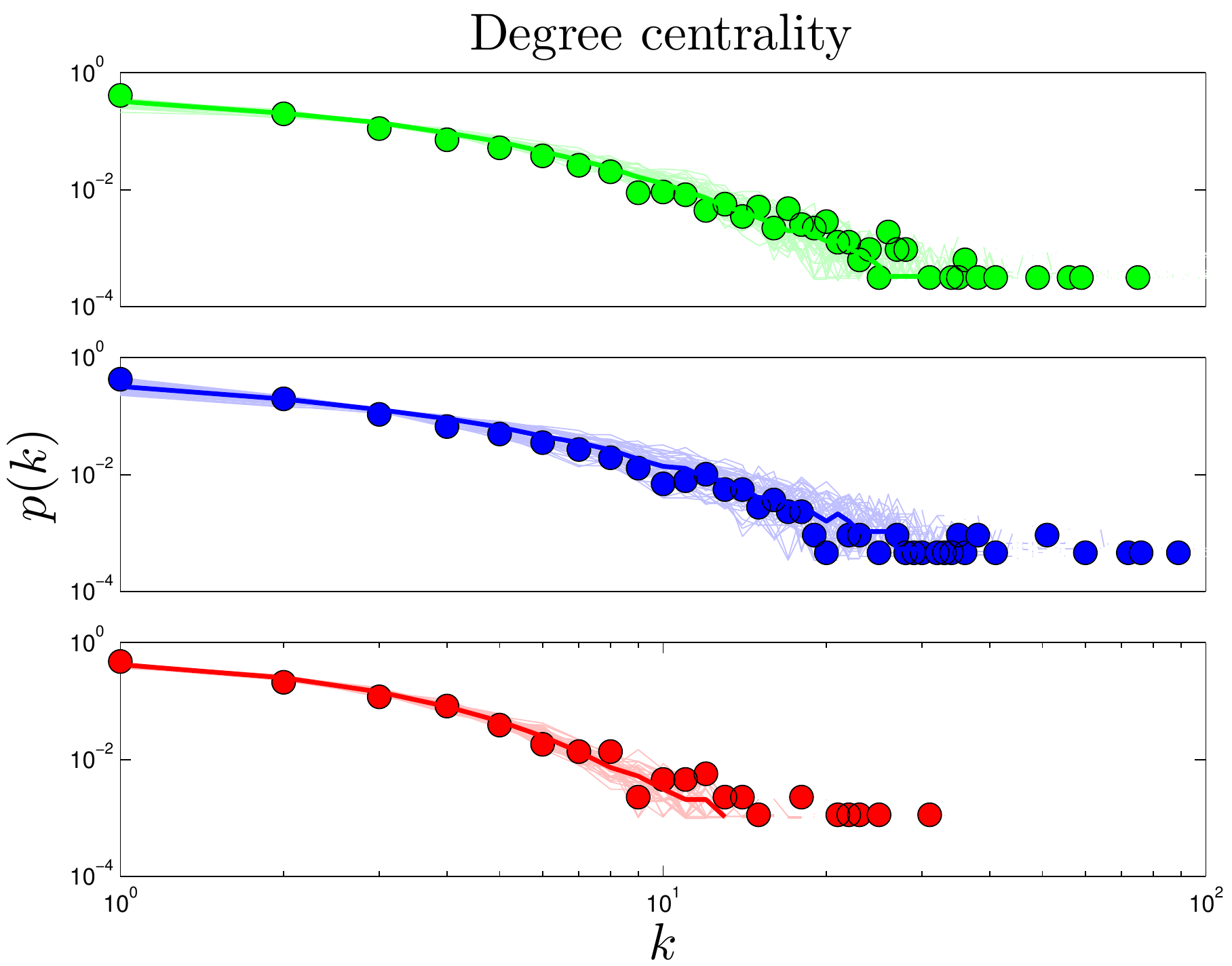}
\end{center}
\caption{{\bf Degree centrality.}  Degree ($k$) distributions in human (green), yeast (blue), and fly (red).  Heavy lines are the median values from 50 simulations, and light lines are results of individual simulations.  Points represent high-confidence empirical data for each organism (see Methods).  Unless otherwise noted, color coding in the same in all plots.  Quantitative comparisons between simulation and experiment (for DUNE and several other models) are detailed in SI.}
\label{fig:degree}
\end{figure}

\begin{table}[b]
\begin{tabular}{| c | c c c c c |}
\hline
 & $Q$ & $D$ & $f_1$ & $\langle C \rangle$ & $\langle k \rangle$ \\
\hline
\textbf{Yeast data} & \textbf{0.75} & \textbf{15} & \textbf{0.89} & \textbf{0.09} & \textbf{3.65} \\
DUNE & ${0.74(7)}$ & ${17(6)}$ & $0.8(1)$ & ${0.041(9)}$ & ${4.0(8)}$ \\
V\'azquez & $0.80(4)$ & $21(5)$ & $0.2(1)$ & $0.045(5)$ & $2.6(4)$ \\
Berg & $0.518(4)$ & $12.0(7)$ & $0.990(3)$ & $0.0027(9)$ & $4.10(3)$ \\
RG & $0.910(3)$ & $36(3)$ & $0.987(6)$ & $0.475(8)$ & $5.31(8)$ \\
MpK & $0.58(6)$ & $24(5)$ & $1.000(2)$ & $0.08(3)$ & $4.4(6)$ \\
ER & $0.588(8)$ & $13.0(9)$ & $0.995(2)$ & $0.002(1)$ & $3.5(6)$ \\
\hline
\textbf{Fly data} & \textbf{0.86} & \textbf{23} & \textbf{0.73} & \textbf{0.10} & \textbf{2.93} \\
DUNE & $0.82(2)$ & $20(2)$ & $0.81(3)$ & $0.09(1)$ & $2.36(9)$ \\
\hline
\textbf{Human data} & \textbf{0.75} & \textbf{15} & \textbf{0.88} & \textbf{0.08} & \textbf{3.69} \\
DUNE & $0.74(6)$ & $17(2)$ & $0.88(4)$ & $0.09(1)$ & $3.7(4)$ \\
\hline
\end{tabular}
\caption{{\bf Comparison of network features.} Modularity $Q$, diameter $D$, fraction of nodes in the largest component $f_1$, global clustering coefficient $\langle C \rangle$, and $\langle k \rangle$ is the average degree of proteins the largest component.  `Data' is the empirical data, `DUNE' is the model described here, `V\'azquez' is the duplication-only model of \cite{Vazquez_2003}, `Berg' is the link dynamics model \cite{Berg_2004}, `RG' is random geometric \cite{Przulj_2004}, `MpK' is the physical desolvation model presented in \cite{Deeds_2006}, and `ER' is an Erd\H{o}s-R\'{e}nyi random graph \cite{Erdos_1960}.  Simulated values are the median ($\pm$ standard deviation) over 50 simulations.  (See SI for details of each model's setup and optimization.)}
\label{tab:SVF}
\end{table}

\section*{Results}

\subsection*{Present-day network topology}  One test of an evolutionary model is its predictions for present-day PPI network topologies.  Current large-scale PPI data sets have a high level of noise, resulting in significant problems with false positives and negatives \cite{Deane_2002,Deeds_2006}.  To mitigate this, we compare only to `high-confidence' experimental PPI network data gathered in small-scale experiments (see Methods).  We computed 10 topological features, quantifying various static and dynamic aspects of the networks' global and local structures: degree, closeness, eigenvalues, betweenness, modularity, diameter, error tolerance, largest component size, clustering coefficients, and assortativity.  8 of these properties are described below (see SI for others).

The \emph{degree} $k$ of a node is the number of links connected to it.  For protein networks, a protein's degree is the number of proteins with which it has direct interactions.  Some proteins interact with few other proteins, while other proteins (called `hubs') interact with many other proteins.  Previous work indicates that hubs have structural and functional characteristics that distinguish them from non-hubs, such as increased proportion of disordered surface residues and repetitive domain structures \cite{Patil_2010}.  The high degree of a protein hub could indicate that protein has unusual biological significance \cite{Jeong_2001}.  The network's overall link density is described by its mean degree, $\langle k \rangle$ (Table~\ref{tab:SVF}).  The \emph{degree distribution} $p(k)$ is the probability that a protein will have $k$ links.  PPI networks have a few hub proteins and many relatively isolated proteins.  The heavy tail of the degree distribution shows that PPI networks have significantly more hubs than random networks have.  Simulated and experimental degree distributions are compared in Figure~\ref{fig:degree}.  (For quantitative comparisons, see SI.)

\begin{figure*}
\begin{center}
\includegraphics[width=0.45\textwidth]{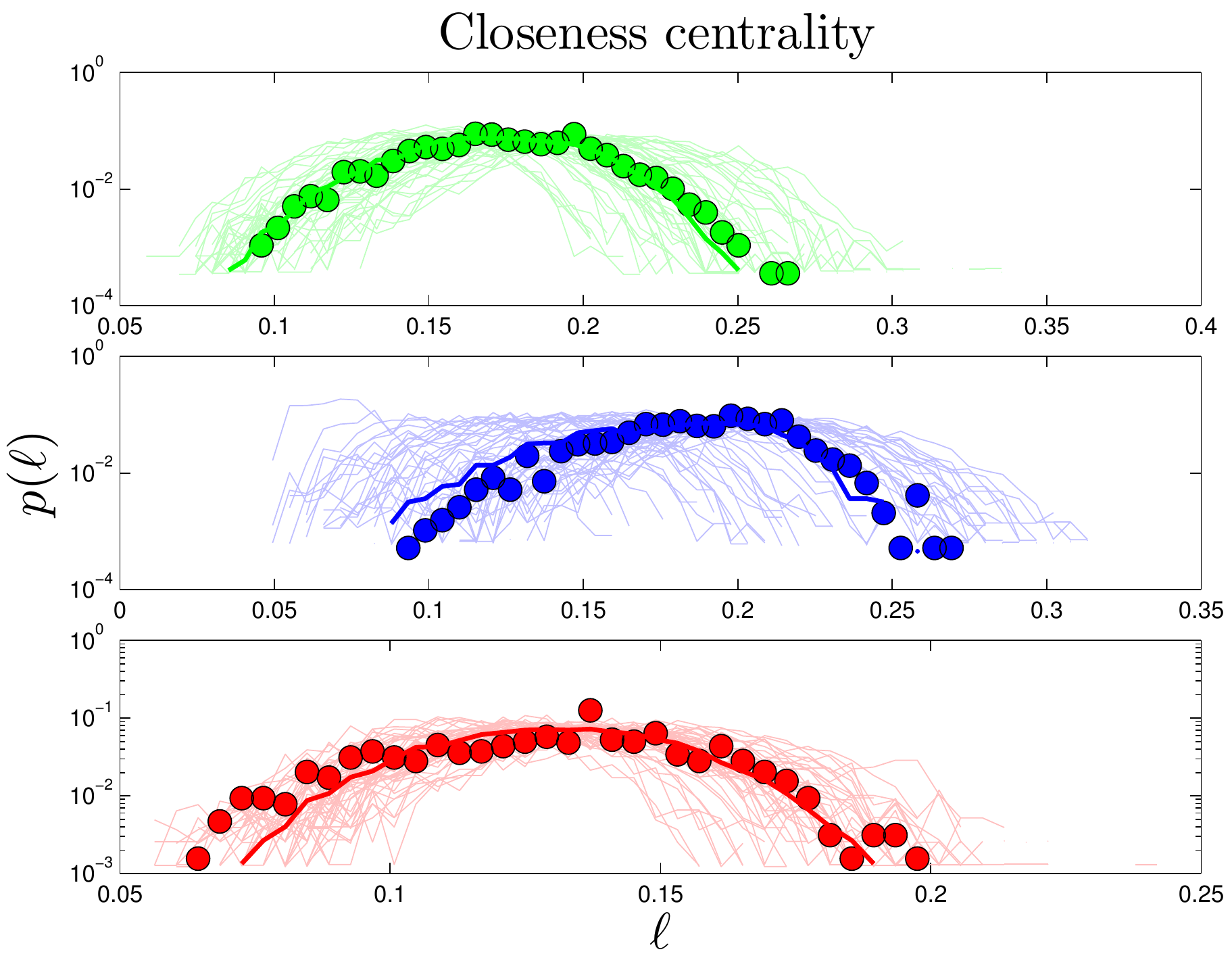}
\includegraphics[width=0.23\textwidth]{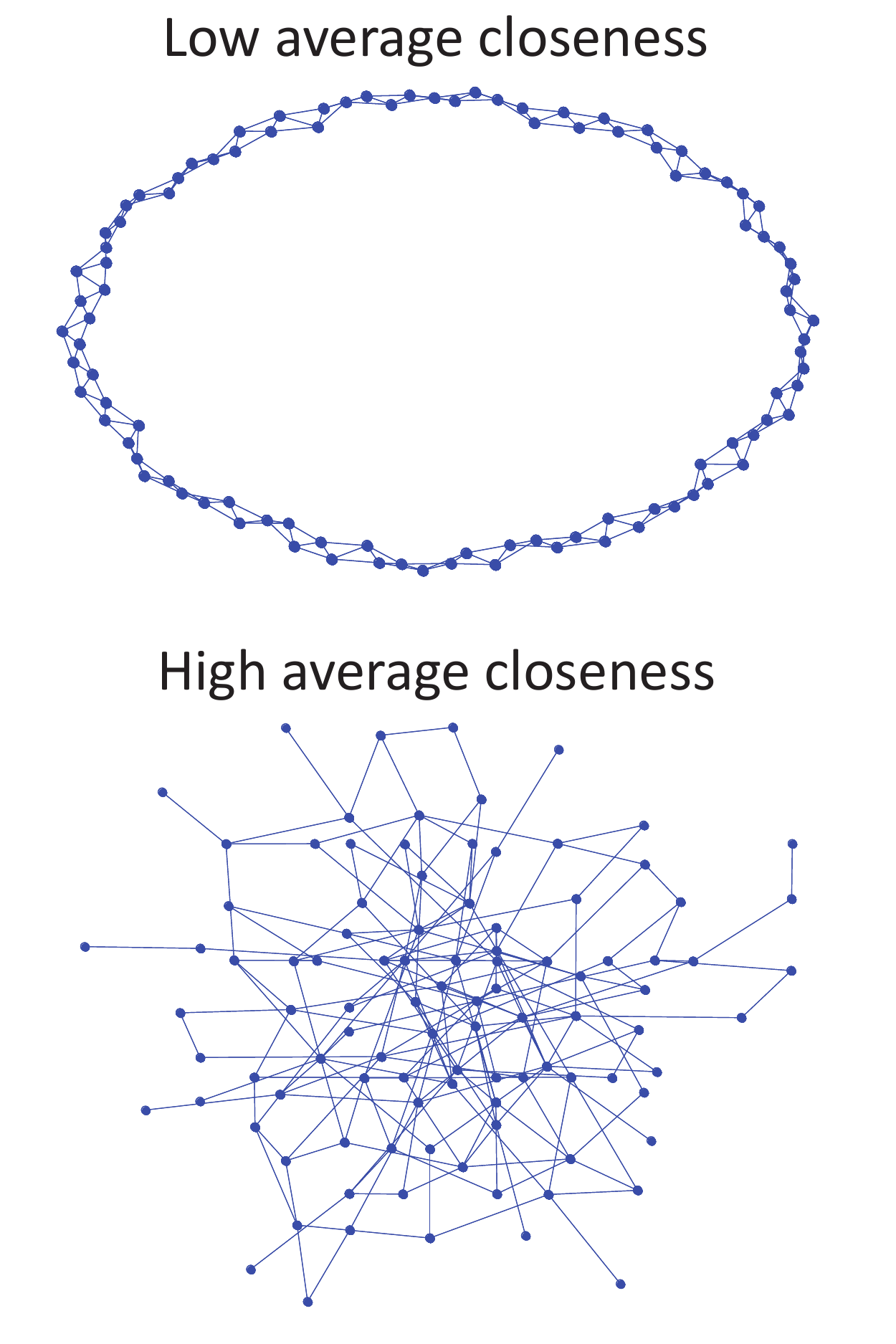}
\end{center}
\caption{{\bf Closeness centrality.}  (A) Closeness ($\ell$) distributions in human (green), yeast (blue), and fly (red).  Heavy lines are the median values from 50 simulations, and light lines are results of individual simulations. (B) Examples of networks with low average closeness $\langle \ell \rangle = 0.06$ (top; each node is generally far away from most other nodes because there are no `short cuts') and high average closeness $\langle \ell \rangle = 0.28$ (bottom; the random connections allow each node to be only a short distance from the other nodes).  Note that both networks pictured here have the same number of nodes ($N = 100$) and roughly the same average degree (top: $\langle k \rangle = 4$, bottom: $\langle k \rangle = 3.7$).}
\label{fig:closeness}
\end{figure*}

\begin{figure}[b]
\begin{center}
\includegraphics[width=0.45\textwidth]{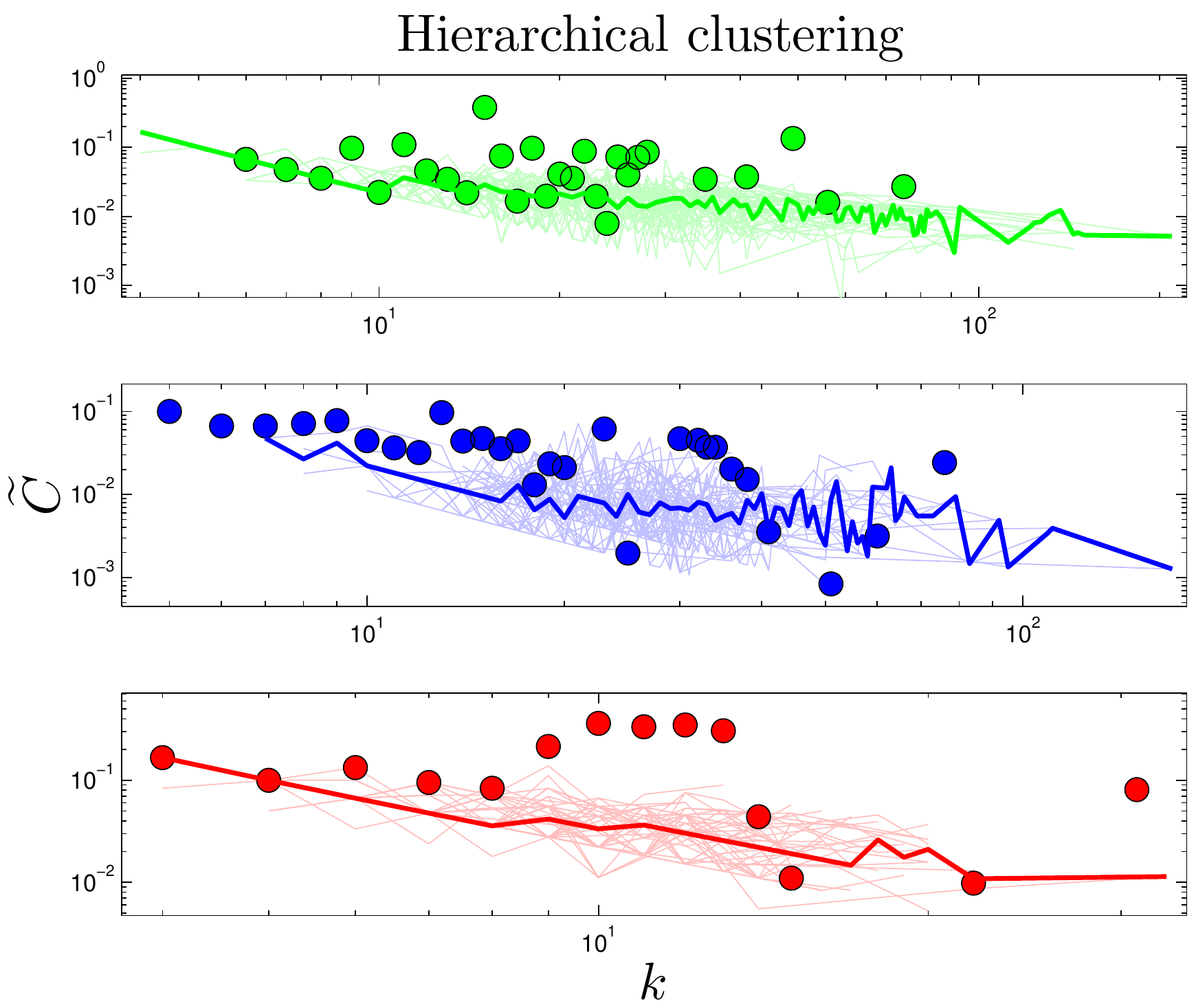}
\end{center}
\caption{{\bf Hierarchical clustering.}  Median clustering coefficient vs.~degree in human (green), yeast (blue), and fly (red).  Heavy lines are the median values from 50 simulations, and light lines are results of individual simulations.}
\label{fig:clustering}
\end{figure}

\begin{figure}[b]
\begin{center}
\includegraphics[width=0.45\textwidth]{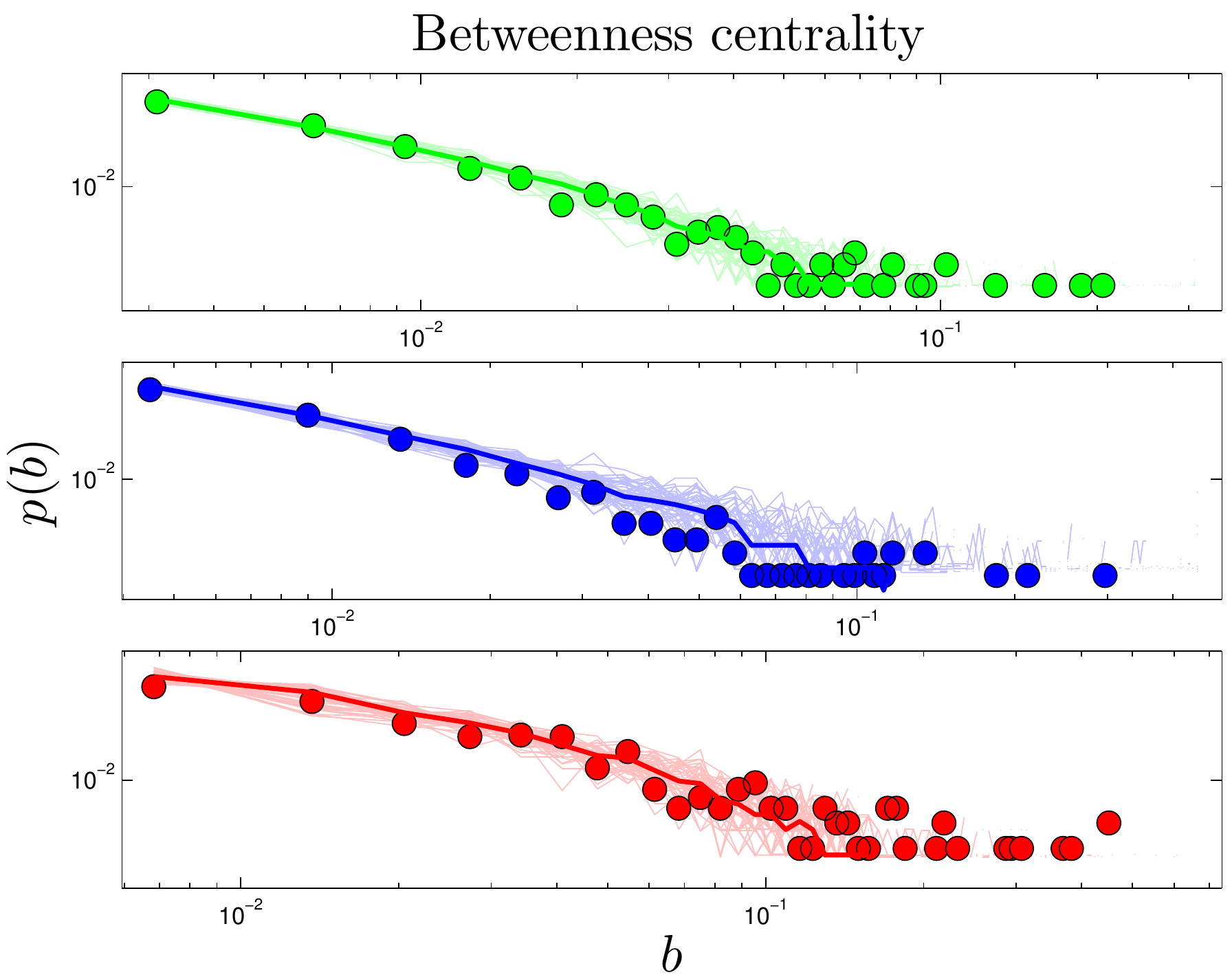}
\end{center}
\caption{{\bf Betweenness centrality.}  Betweenness ($b$) distributions in human (green), yeast (blue), and fly (red).  Heavy lines are the median values from 50 simulations, and light lines are results of individual simulations.}
\label{fig:betweenness}
\end{figure}

\emph{Component} refers to a set of reachable proteins.  If any protein is reachable from any other protein (by hopping from neighbor to neighbor), then the network only has one component.  If there is no path leading from protein A to B, then A and B are in different components.  The fraction of nodes in the largest component ($f_1$) is a measure of network fragmentation (Table~\ref{tab:SVF} and Figure~\ref{fig:dynamics}).  Note that, although silent genes (proteins with no links) exist in real systems, these genes do not appear in data sets consisting only of PPI's.  Therefore, calculations of $f_1$ for all models exclude orphan proteins (proteins with $k=0$).

Gene loss, the silencing or deletion of genes, is known to play an important role in evolution.  The loss of a functioning gene will damage an organism, making the gene loss unlikely to be passed on.  The exception is if the gene is redundant.  Consistent with this reasoning, evidence suggests that many gene loss events are losses of one copy of a duplicated gene \cite{Ku_2000,Kellis_2004}.  Although empirical estimates of the gene loss rate varied considerably, a consistent finding across several studies is that the rates of gene duplication and loss are of the same order-of-magnitude \cite{Lynch_2000,Cotton_2005,Gao_2004}.  This broad picture is in good agreement with our model.  In our model, a gene is considered lost when it has degree zero.  Our model predicts that the ratio of orphan to non-orphan proteins is $1.6\pm0.4$ in yeast, $0.58\pm0.06$ in flies, and $0.67\pm0.09$ in humans.  The gene loss rate has been previously estimated to be about half the duplication rate in both flies and humans \cite{Lynch_2000,Cotton_2005}, consistent with our model's prediction.

The \emph{distance} between nodes $i$ and $j$ is defined as the number of node-to-node steps that it takes along the shortest path to get from node $i$ to $j$.  The \emph{closeness centrality} of a node $i$, $\ell_i$, is the inverse of the average distance from node $i$ to all other nodes in the same component.  The \emph{diameter}, $D$, of a network is the longest distance in the network.  Simulated closeness distributions are compared to experiments in Figure~\ref{fig:closeness}.  Interestingly, proteins have about `six degrees of separation', similar to social networks \cite{Travers_1969,Leskovec_2008}. The closeness distributions $p(\ell)$ have peaks around $1/\ell \approx 5 - 7$.

\begin{figure}[b]
\begin{center}
\includegraphics[width=0.45\textwidth]{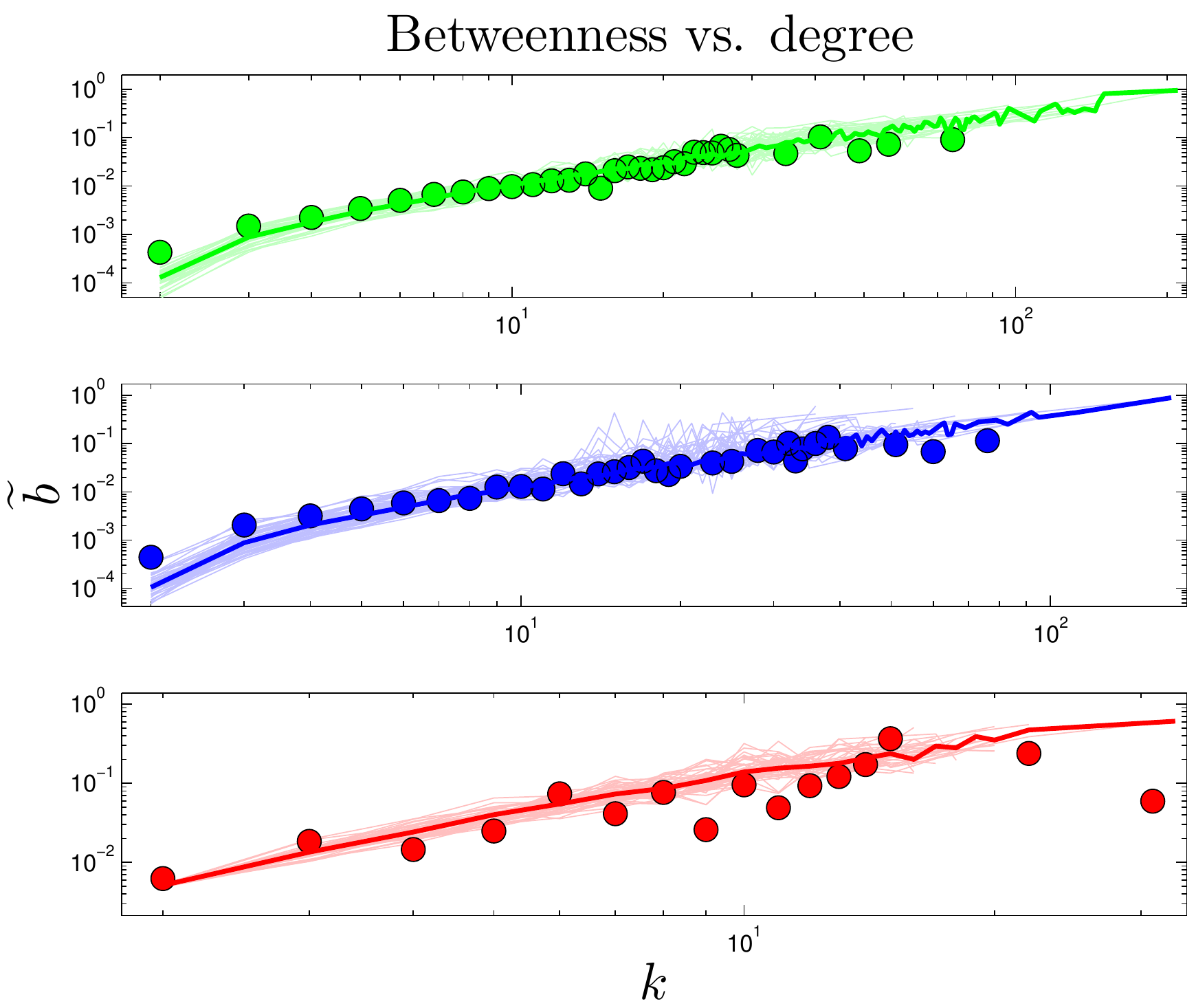}
\end{center}
\caption{{\bf Betweenness vs.~degree.} Shown are median betweenness vs.~degree values in human (green), yeast (blue), and fly (red).  Heavy lines are the median values from 50 simulations, and light lines are results of individual simulations.}
\label{fig:bvk}
\end{figure}

Another property of a network is its \emph{modularity} \cite{Yook_2004}.  Networks are modular if they have high densities of links (defining regions called modules), connected by lower densities of links (between modules).  One way to quantify the extent of modular organization in a network is to compute the modularity index, $Q$ \cite{Newman_2004,Newman_2006}:
\begin{equation}
Q \equiv \frac{1}{K} \sum_{i,j}^N \left( A_{ij} - \frac{k_ik_j}{K} \right) \delta (u_i,u_j),
\end{equation}
where $k_i$ and $k_j$ are the degrees of nodes $i$ and $j$, $u_i$ and $u_j$ denote the modules to which nodes $i$ and $j$ belong, $\delta(u_i,u_j) = 1$ if $u_i = u_j$ and $\delta(u_i,u_j) = 0$ otherwise, and $A_{ij} = 1$ if nodes $i$ and $j$ share a link, and $A_{ij} = 0$ otherwise.  $Q$ quantifies the difference between the actual within-module link density to the expected link density in a randomly connected network.  $Q$ ranges between $-1$ and $1$; positive values of $Q$ indicate that the number of links within modules is greater than random.  The numerical value of $Q$ required for a network to be considered `modular' depends on the number of nodes and links and method of computation.  To calibrate baseline $Q$ values given our particular network data, we used the null model described in \cite{Maslov_2002}.  Our non-modular baseline values are $Q = 0.603$ for the human PPI net, $Q = 0.590$ for yeast, and $Q = 0.722$ for flies (see SI).  As shown in Table~\ref{tab:SVF}, PPI networks are highly modular, and our simulated $Q$ values are in good agreement with those of experimental data.

\begin{figure}[t]
\begin{center}
\includegraphics[width=0.45\textwidth]{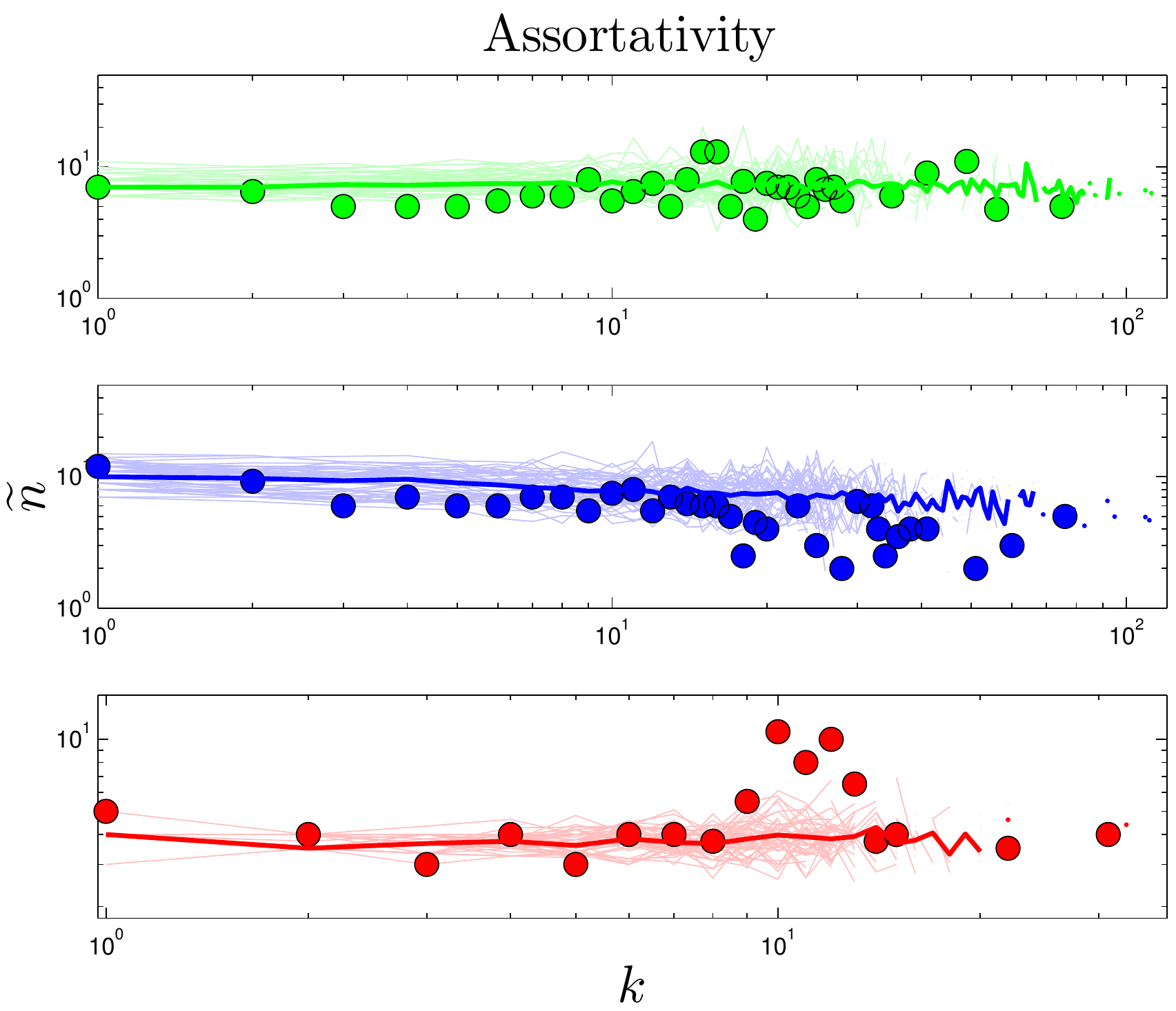}
\end{center}
\caption{{\bf Assortativity.}  Median nearest-neighbor degree vs.~degree  in human (green), yeast (blue), and fly (red).  Heavy lines are the median values from 50 simulations, and light lines are results of individual simulations.}
\label{fig:assortativity}
\end{figure}

\begin{figure*}[t]
\begin{center}
\includegraphics[width=0.45\textwidth]{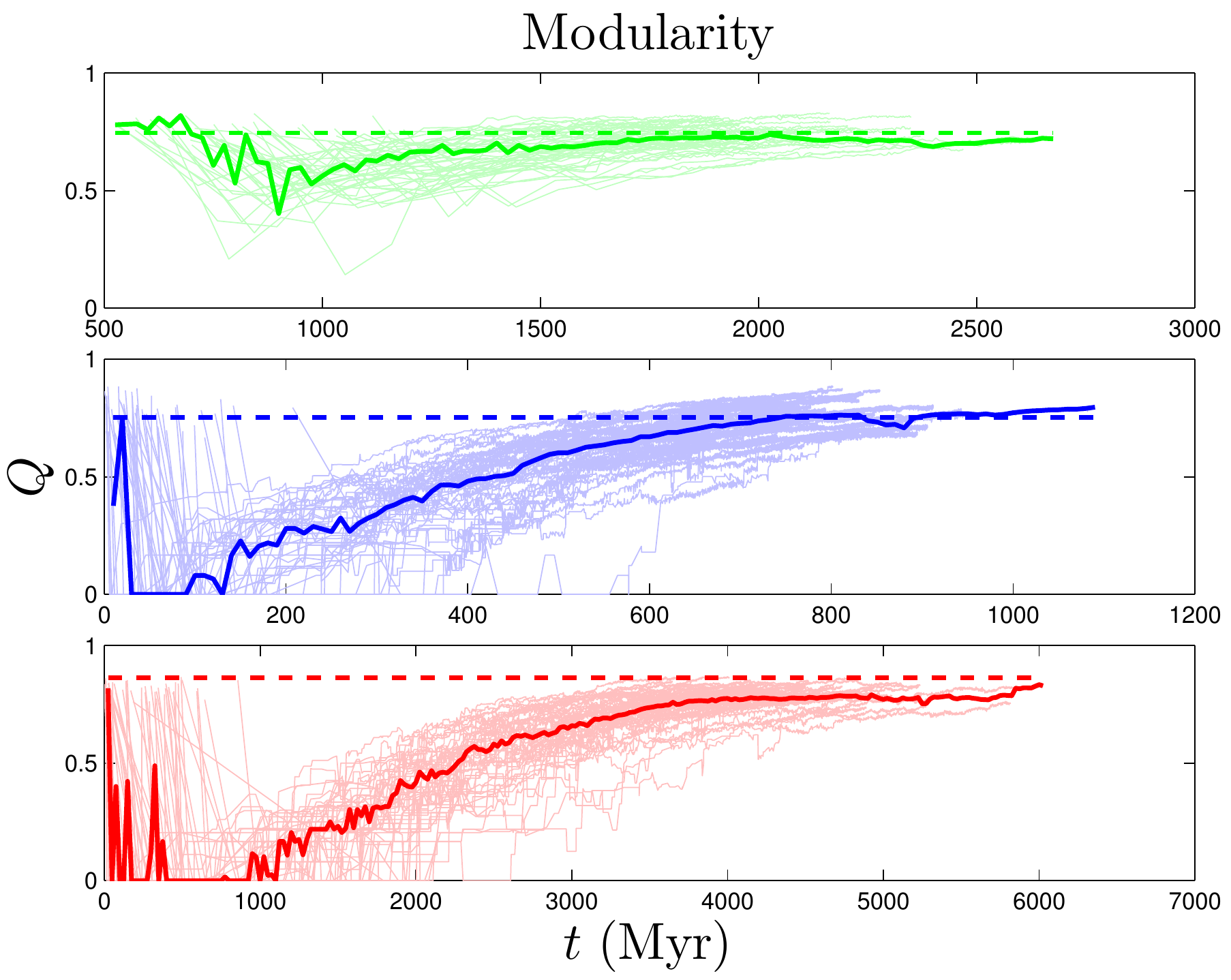}
\includegraphics[width=0.45\textwidth]{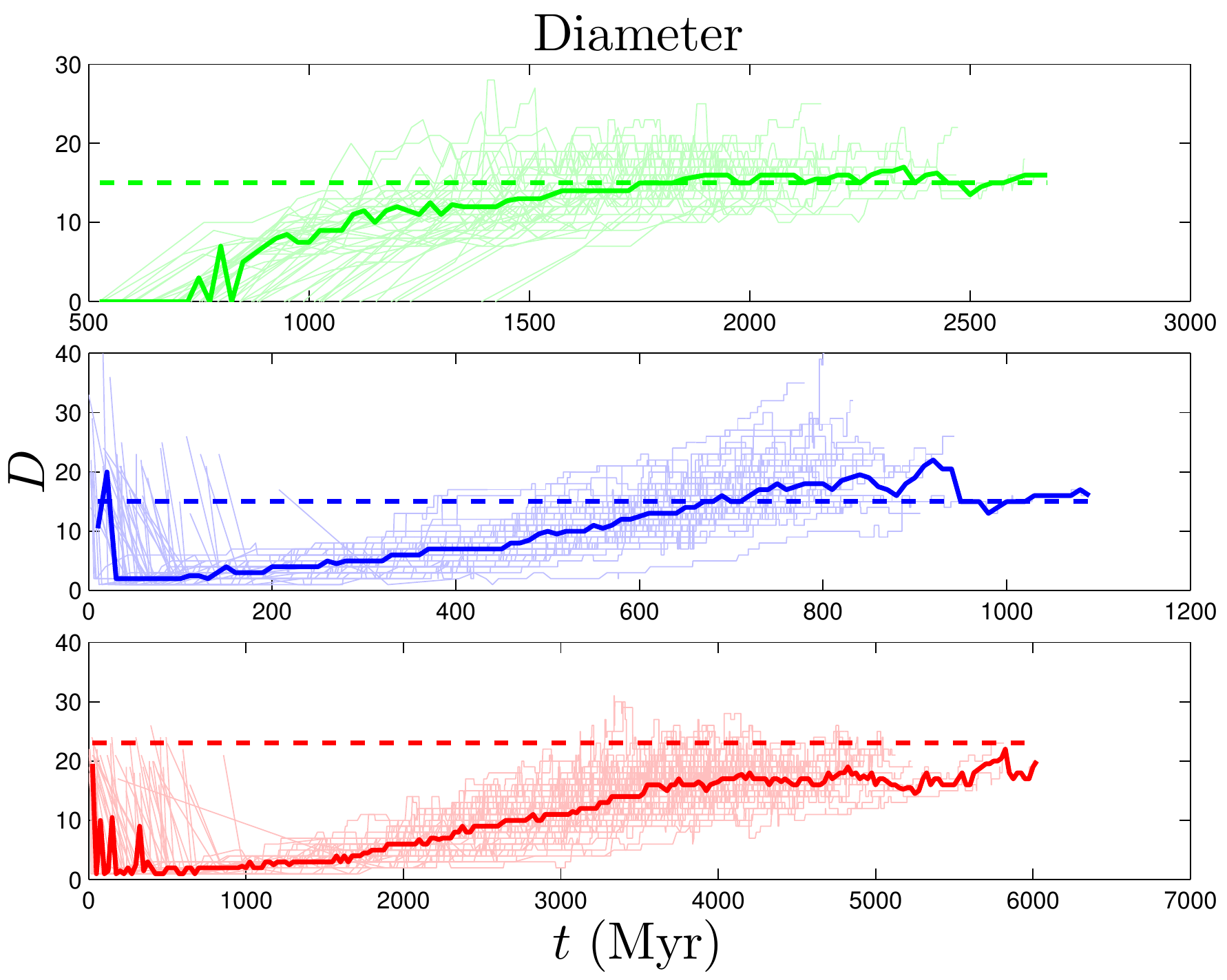}
\end{center}
\caption{{\bf Modularity and diameter.}  (A) Modularity $Q$ and (B) diameter $D$ are both predicted to grow with time in human (green), yeast (blue), and fly (red).  Light lines indicate the evolutionary trajectories of 50 individual simulations, and the heavy line is the median value.  The modularity and diameter of the empirical data are shown as dashed horizontal lines.  Time traces occasionally do not start at $t=0$ because these simulations spend the first few time steps in a completely disconnected state, so the dynamical quantities are undefined.  (See Figure~\ref{fig:dynamics} for other dynamical plots.)}
\label{fig:modularity}
\end{figure*}

The \emph{clustering coefficient}, $C_i$, for a protein $i$, is a measure of mutual connectivity of the neighbors of protein $i$.  $C_i$ is defined as the ratio of the actual number of links between neighbors of protein $i$ to the maximum possible number of links between them,
\begin{equation}
C_i = \frac{\text{\# edges between neighbors of node } i}{ k_i (k_i - 1)}.
\end{equation}
In a PPI network, clustering is thought to reflect the high likelihood that proteins of similar function are mutually connected \cite{vonMering_2002}.  The average (or global) clustering coefficient, $\langle C \rangle$, quantifies the extent of clustering in the network as a whole.  As shown in Table~\ref{tab:SVF}, PPI networks have large global clustering coefficient values; the yeast PPI network, for example, has a value of $\langle C \rangle$ which is 45 times higher than that of a random graph of equivalent link density.  In flies and humans, our simulated networks have $\langle C \rangle$ values in excellent agreement with the data; in yeast, our predicted value is slightly low.

A network is said to be `hierarchically clustered' if the clustering coefficient and degree obey a power-law relation, $C \sim k^{-\xi}$ \cite{Ravasz_2002} (Figure~\ref{fig:clustering}), indicating that nodes are organized into small-scale modules, and the small-scale modules are in turn organized into larger-scale modules following the same pattern \cite{Barabasi_2004}.  By plotting each node's clustering coefficient against its degree, we observed a trend consistent with hierarchical clustering, although data in the tail is very limited.

The \emph{betweenness} of a node measures the extent to which it `bridges' between different modules.  \emph{Betweenness centrality}, $b$, is defined as:
\begin{equation}
b_i \equiv \frac{\text{\# shortest paths passing through node }i}{\text{\# total shortest paths}}.
\end{equation}
Betweenness has been proposed as a uniquely functionally-relevant metric for PPI networks because it relates local and global topology.  It has been argued that knocking out a protein that has high betweenness may be more lethal to an organism than knocking out a protein of high degree \cite{Joy_2005}.  Betweenness distributions are shown in Figure~\ref{fig:betweenness}.

If a network's well-connected nodes are mostly attached to poorly-connected nodes, the network is called \emph{disassortative}.  A simple way to quantify disassortativity is by determining the median degree of a protein's neighbors ($n$) as a function of its degree ($k$).  Previous work has found that yeast networks are disassortative \cite{Maslov_2002}.  It has been argued that disassortativity is an essential feature of PPI network evolution, and recent modeling efforts have heavily emphasized this feature \cite{Zhao_2007,Wan_2010}.  However, it was noted by \cite{Aloy_2002} that disassortativity may simply be an artifact of the yeast two-hybrid technique, and \cite{Hakes_2008} pointed out that this trend is quite different among different yeast datasets, and in some cases is completely reversed, resulting in \emph{assortative} mixing, where high degree proteins prefer to link to other high-degree proteins.  As shown in Figure~\ref{fig:assortativity} and Table~\ref{tab:exponents}, the empirical data shows no evidence of disassortativity in flies or humans, and even the trend in yeast is quite weak.  This conclusion is based solely on analysis of the empirical data, and casts further doubt on the role of disassortative mixing in PPI network evolution.

Some higher-order features of the network are simply a result of its degree sequence, and other features might be important in their own right.  As discussed in \cite{Maslov_2002}, it is possible to isolate the effects of the degree sequence by `rewiring' (detaching then reattaching links) the network at random, subject to the restriction that the degree sequence must be preserved.  If a property contains extra information about the network's structure, then it should be different in the rewired network.  On the other hand, if the network is rewired many times, and the property is always the same, then it is likely to just be a result of the degree sequence.  We randomly rewired the empirical network $4K_\text{data}$ times.\footnote{Network rewiring was carried out using a script downloaded from \texttt{http://www.cmth.bnl.gov/$\sim$maslov/matlab.htm}.}  As expected, modularity is decreased by random rewiring.  Upon rewiring, we find $Q = 0.603 \pm 0.002$ in humans, $Q = 0.590 \pm 0.003$ in yeast, and $Q = 0.722 \pm 0.007$ in flies (median $\pm$ standard deviation from 50 repeats of the rewiring algorithm).  Rewiring also shrinks the diameters of PPI networks to $D = 13 \pm 0.9 $ in humans, $D = 12 \pm 1.0$ in yeast, and $D = 15 \pm 1.1$ in flies.  These results suggest that these features contain important structural information about the network, and are not merely consequences of the degree sequence.

One reason we are interested in calculating $Q$ and $D$ is simply to check that the values are comparable between the simulated and experimental networks.  However, on a more qualitative level, we would also like to have some idea of what the threshold is for a network to be considered `modular' or `small-diameter'.  The rewired $Q$ and $D$ values are useful because these features are dependent on the size of the network (number of nodes $N$ and links $K$); given an identical network construction method, $Q$ and $D$ will generally be different in sparse versus dense networks.  We use these $Q$ and $D$ values as baseline values with which the experimental and simulated networks can be compared; we considered $Q$ and $D$ values differing from the rewired values by more than a standard deviation to be significantly different.

Comparisons of simulated and experimental eigenvalue spectra and error tolerance curves are shown in SI (Figures \ref{fig:spectrum} and \ref{fig:errortolerance}).  As discussed in SI, the various per-node network properties we have analyzed are largely uncorrelated (Figure \ref{fig:pca}).

\subsection*{Evolutionary trajectories}  We now consider the question of how PPI networks evolve in time.  The present-day networks show a rich-get-richer structure: PPI networks tend to have both more well-connected nodes and more poorly connected nodes than random networks have.  In our model, the rich-get-richer property has two bases: duplication and assimilation.  The equal duplication chance per protein means the probability for a protein with $k$ links to acquire a new link via duplication of one of its interaction partners is proportional to $k$.  Likewise, the probability of a protein to receive a link from the first-neighbor assimilation probability $a$ is proportional to its degree $k$.  `Rich' proteins get richer because the probability of acquiring new links rises with the number of existing links.

First, we discuss two dynamical quantities for which experimental evidence exists: the rate of gene loss, and the relation between a protein's age and its centrality.  Gene losses in our model correspond to `orphan' proteins which have no interactions with other proteins.  As shown in Figure~\ref{fig:dynamics}, the fraction of orphan proteins grows quickly at first, then levels off.  This is consistent with the findings of \cite{Cotton_2005}: in humans, while the overall duplication rate is higher than the loss rate, when only data from the past 200 Myr are considered, the loss rate is slightly higher than the duplication rate.  In our model, after the initial rapid expansion, the rate of gene loss stabilizes relative to the duplication rate.

\begin{figure}
\begin{center}
\includegraphics[width=0.49\textwidth]{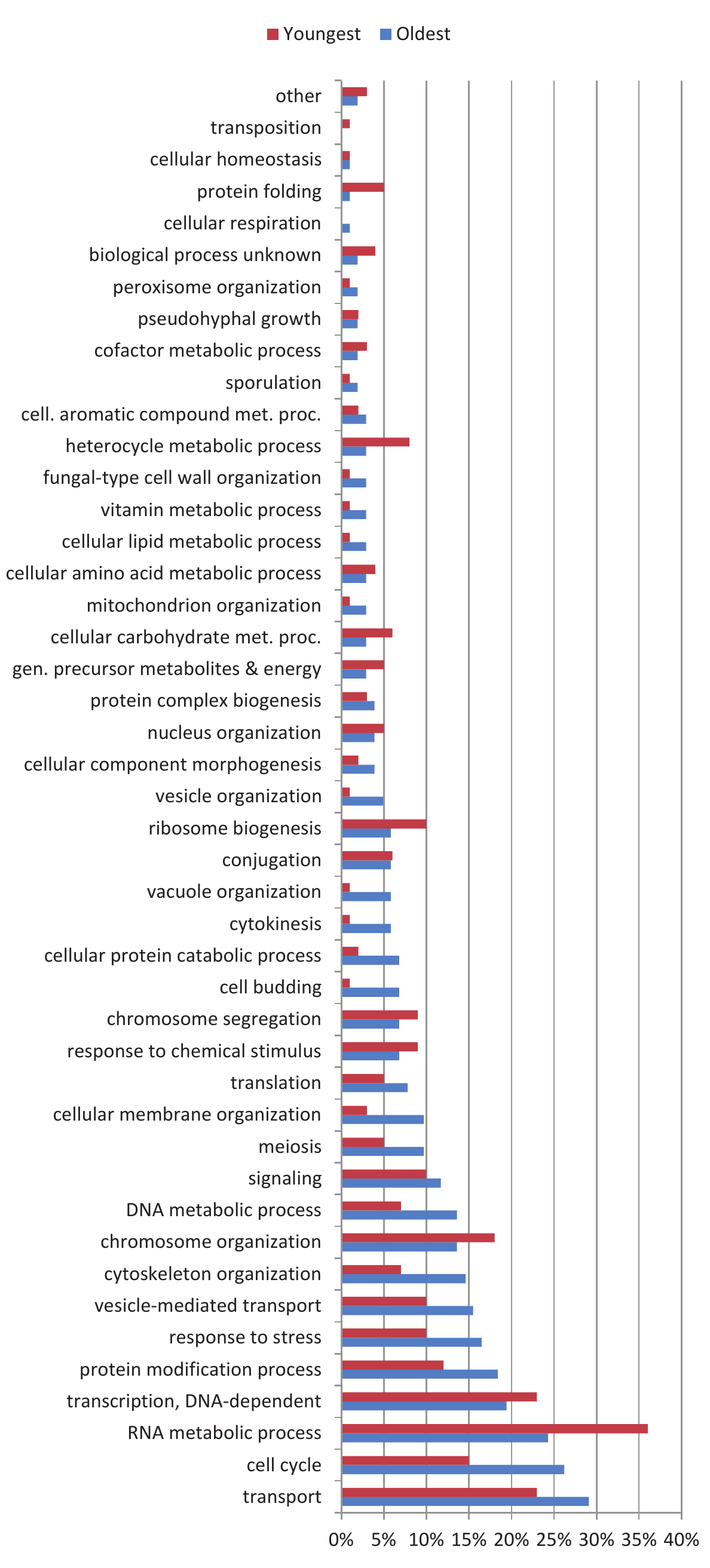}
\end{center}
\caption{{\bf Gene ontology.}  Shown are GO-slim profiles for the 100 oldest and 100 youngest proteins, as measured by $S$-value, in the yeast PPI network.}
\label{fig:goslim}
\end{figure}

\begin{figure*}
\begin{center}
\includegraphics[width=0.32\textwidth]{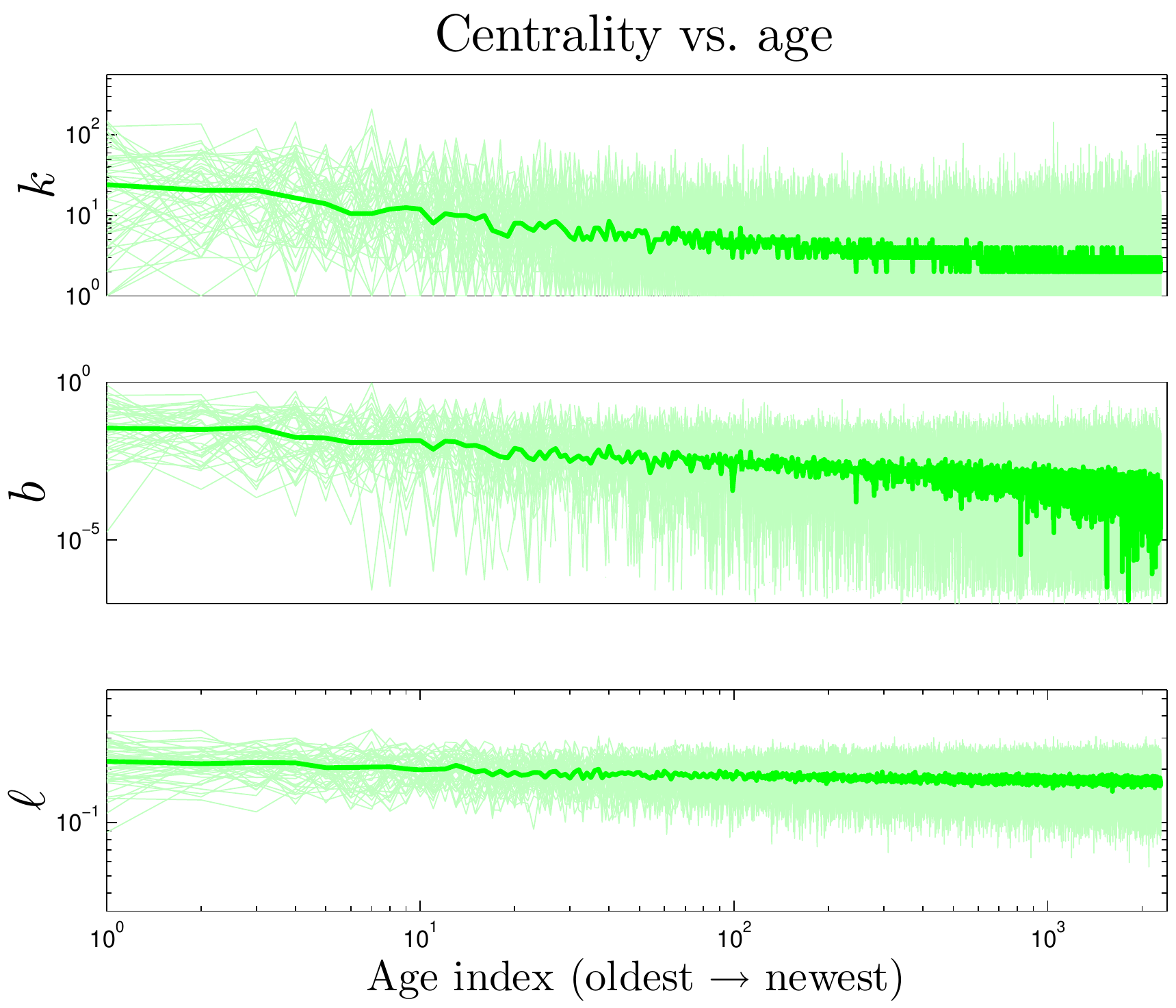}
\includegraphics[width=0.32\textwidth]{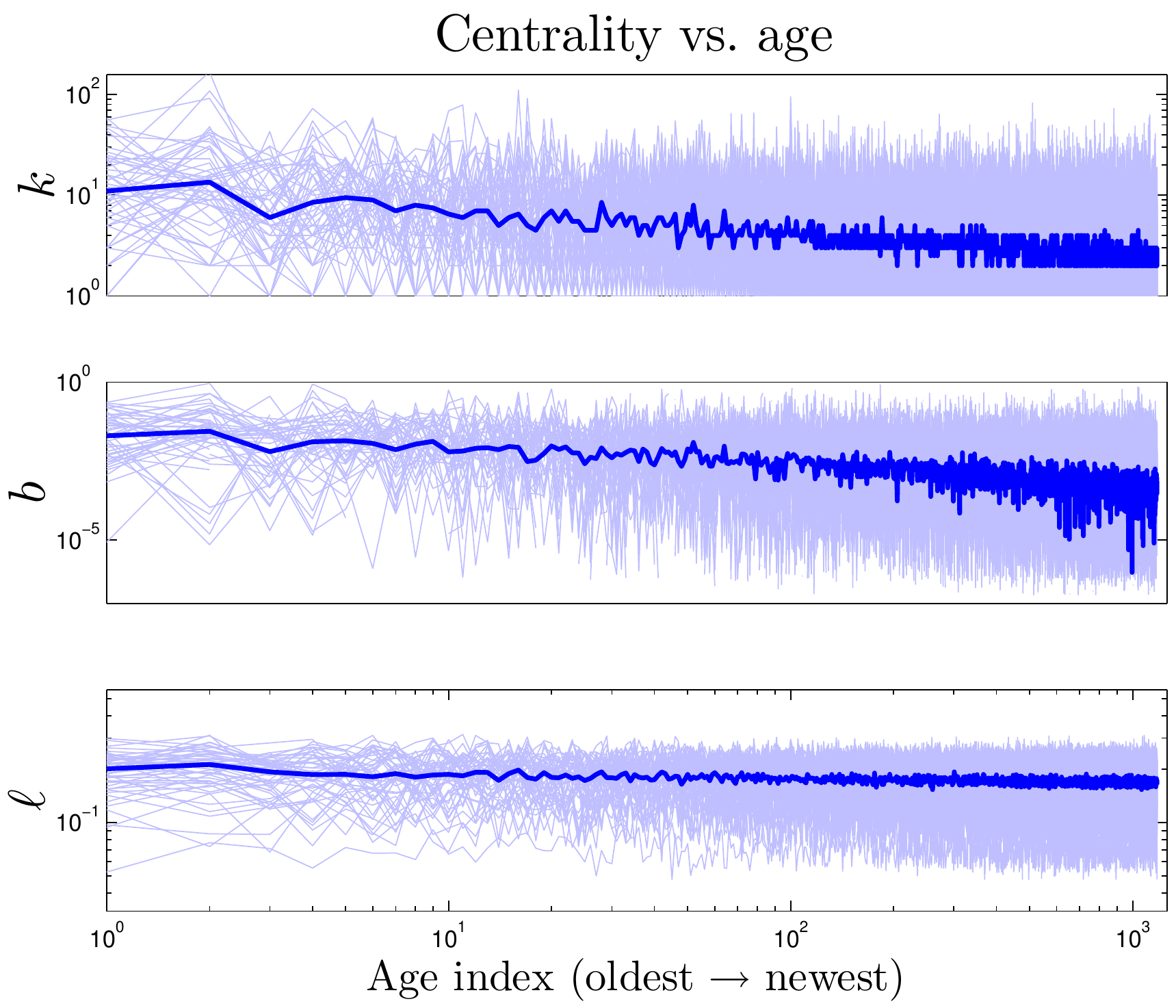}
\includegraphics[width=0.32\textwidth]{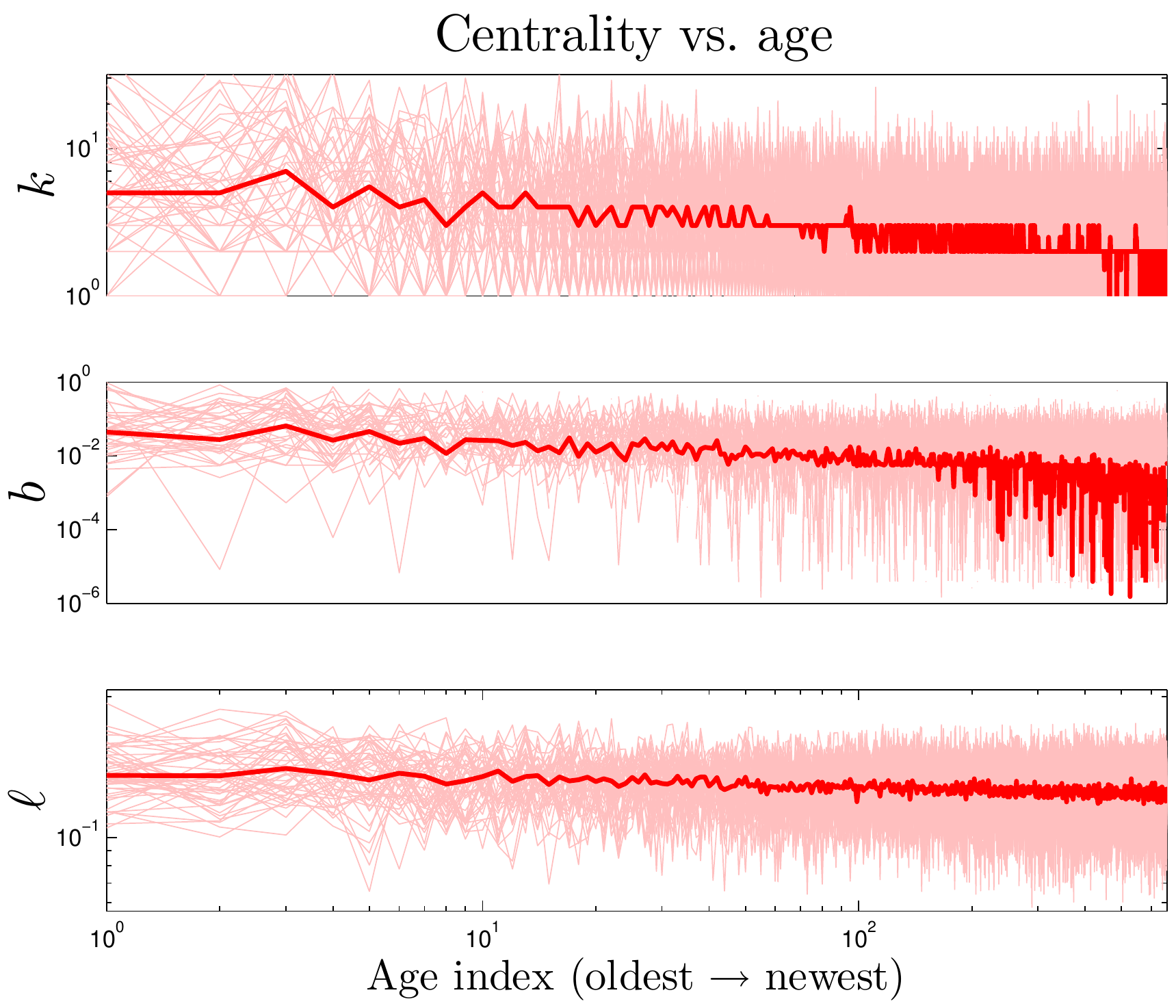}
\end{center}
\caption{{\bf Older proteins are more central.}  Simulations of a protein's age index (time since introduction into the network) vs. degree ($k$), betweenness ($b$), and closeness ($\ell$) centrality, for human (green), yeast (blue), and fly (red).  The oldest proteins are on the \emph{left} in this figure, and the proteins get younger moving to the right.  There is an approximately monotonic increase in centrality with age.}
\label{fig:agecentrality}
\end{figure*}

We define the `age' of a protein in our simulation according to the order in which proteins were added to the network.  Our model shows that a protein's age correlates with certain network properties.  Consistent with earlier work \cite{Woese_1987,Krapivsky_2001,Qin_2003,Zhu_2011}, we find that older proteins tend to be more highly connected.  We plotted the `age index' of a protein (the time step at which the protein was introduced) versus its centrality scores.  As shown in Figure~\ref{fig:agecentrality}, the age index negatively correlates with degree, betweenness, and closeness centralities: older proteins tend to be more central than younger proteins.  Figure~\ref{fig:agecentrality} shows our model's prediction that a protein's age correlates with degree, betweenness, and closeness centrality.  We confirmed this prediction by following the evolutionary trajectories of individual proteins (Figure~\ref{fig:tracking}).  These results are consistent with the eigenvalue-based aging method described in \cite{Zhu_2011} (Figure~\ref{fig:lap_age}).  Phylogenetic protein age estimates indicate that older proteins tend to have a higher degree \cite{Woese_1987,Zhu_2011}, which our model correctly predicts.  Interestingly, the eigenvalue-based scores are only modestly correlated with other centrality scores (0.36 degree, 0.47 betweenness, and 0.10 closeness correlations).  Using the eigenvalue method in tandem with our centrality-based method could provide stronger age-discriminating power for PPI networks than either method alone.

The correlation between centrality and age suggests that static properties of present-day networks may be used to estimate relative protein ages.  Suppose each normalized centrality score ($k' \equiv k/\max (k)$, $\ell' \equiv\ell/\max (\ell)$, $b'\equiv b/\max (b)$) represents a coordinate in a 3-D `centrality space'.  We can then define a composite centrality score ($S$) as $S^2 \equiv (k')^2 + (\ell')^2 + (b')^2$.

Do older proteins typically have different functions than newer proteins?  We classified \emph{S.~cerevisiae} proteins using the GO-slim gene ontology system in the Saccharomyces Genome Database.  As shown in Figure~\ref{fig:goslim}, GO-slim enrichment profiles were somewhat different between the oldest and youngest proteins (as measured by their $S$ values).  Several categories which were more enriched for the oldest proteins were the cell cycle, stress response, cytoskeletal and cell membrane organization, whereas younger proteins were overrepresented in several metabolic processes.  Overall, the differences were not dramatic, suggesting that cellular processes generally require both central and non-central proteins to function.  Consistent with this, ancient proteins tend to be centrally located with modules, as their betweenness values gradually decline over time (Figure~\ref{fig:tracking}).  The roughly linear relation between degree and betweenness also suggests that ancient proteins do not occupy structurally `special' positions within the network, such as stitching together separate modules (Table \ref{tab:exponents} and Figure \ref{fig:bvk}).  This may indicate that modules tend to accumulate around the most ancient proteins, which act as a sort of nucleus.  Thus, ancient proteins are involved in all kinds of pathways, because they have each nucleated their own pathway.

In contrast to the two dynamical quantities discussed so far, most structural properties of PPI networks have only been measured for the present-day network.  Although our model accurately reproduces the present-day values of these quantities, there is no direct evidence that the simulated trajectories are correct; rather, these are predictions of our model.  Figure~\ref{fig:modularity} shows that both modularity $Q$ and diameter $D$ increase with time.  These are not predictions that can be tested yet for biological systems, since there is no time-resolved data yet available for PPI evolution.  Time-resolved data is only currently available for various social networks (links to websites, co-authorship networks, etc.).  Interestingly, the diameters of social networks are found to shrink over time \cite{Leskovec_2005_2}.  Our model predicts that PPI networks differ from these social networks in that their diameters grow over time.  In addition to $Q$ and $D$, we tracked the evolutionary trajectories of several other quantities: the evolution of the global clustering coefficient, the rate of signal propagation, the size of the largest connected component (Figure~\ref{fig:dynamics}), as well as betweenness and degree values for individual nodes (Figure~\ref{fig:tracking}).  See SI for details.

\begin{figure*}[t]
\begin{center}
\includegraphics[width=0.24\textwidth]{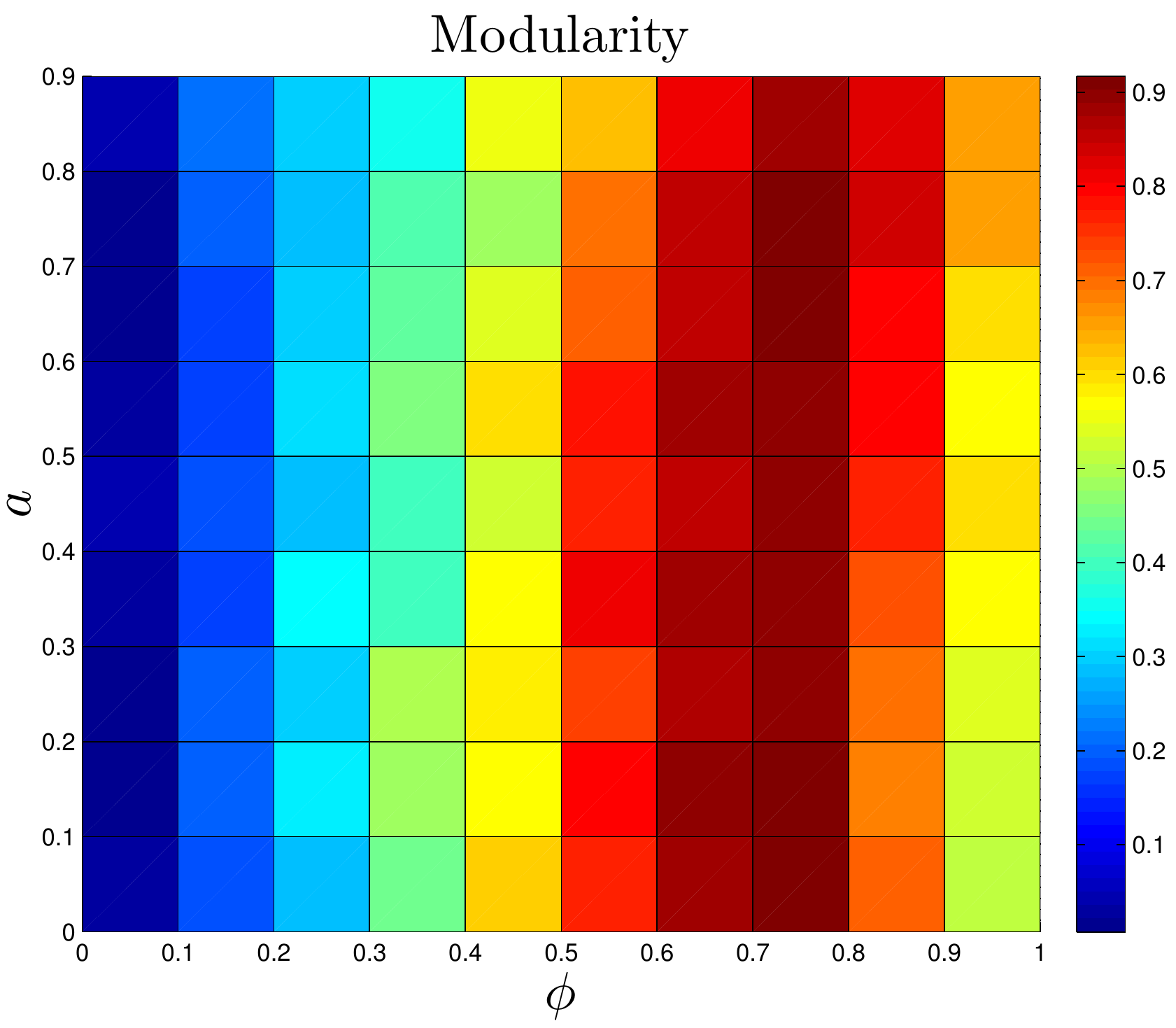}
\includegraphics[width=0.24\textwidth]{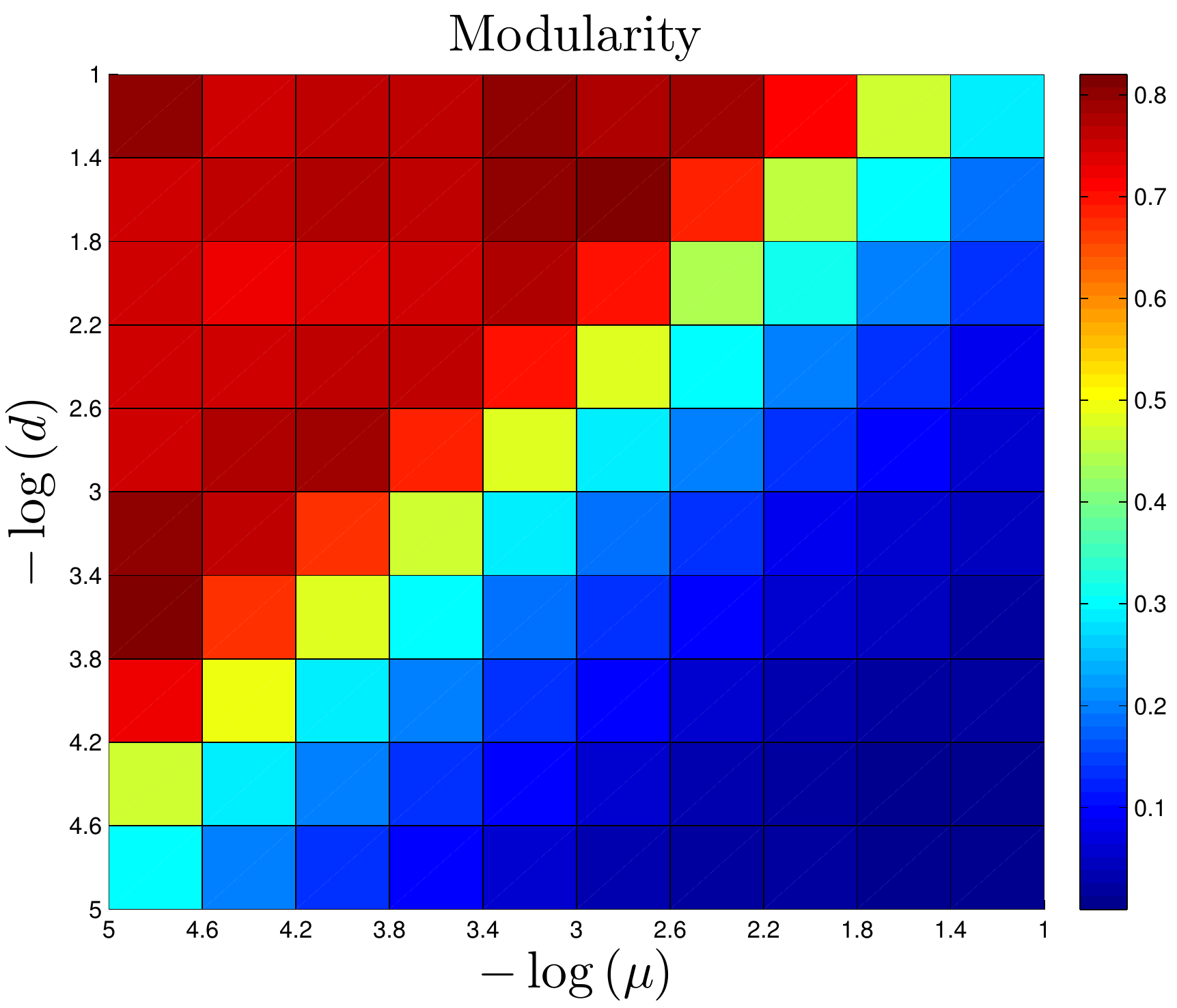}
\includegraphics[width=0.24\textwidth]{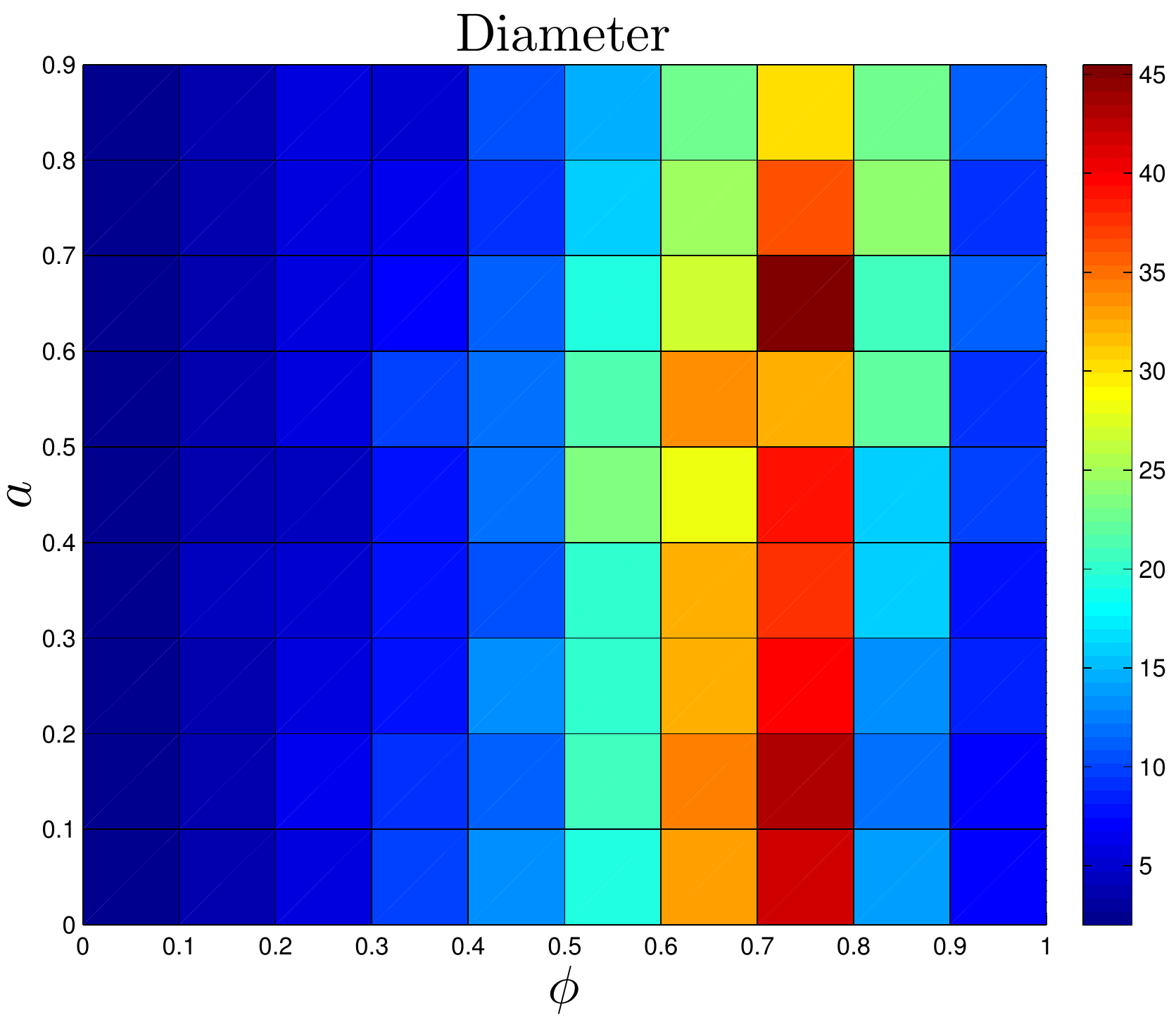}
\includegraphics[width=0.24\textwidth]{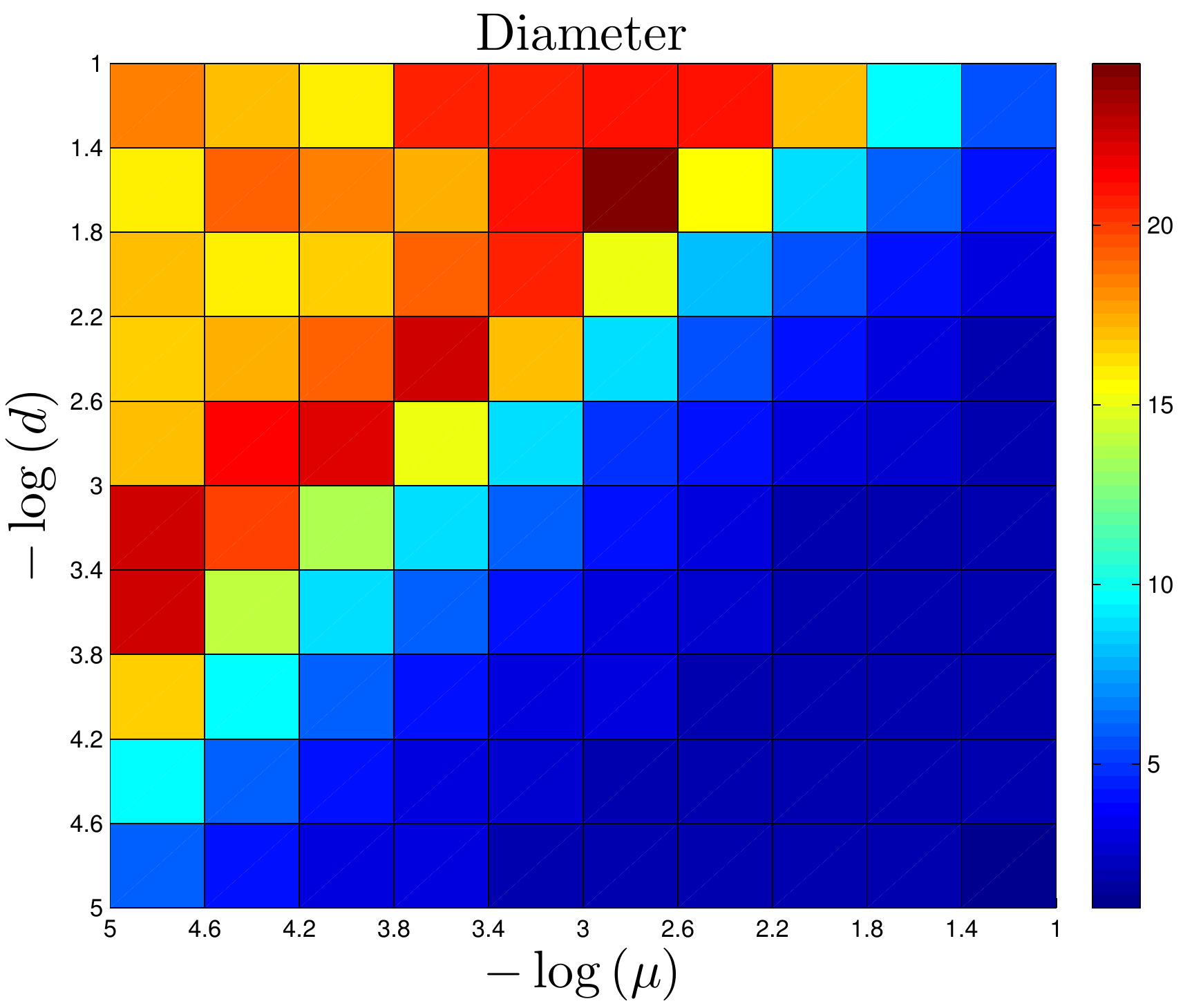}
\includegraphics[width=0.24\textwidth]{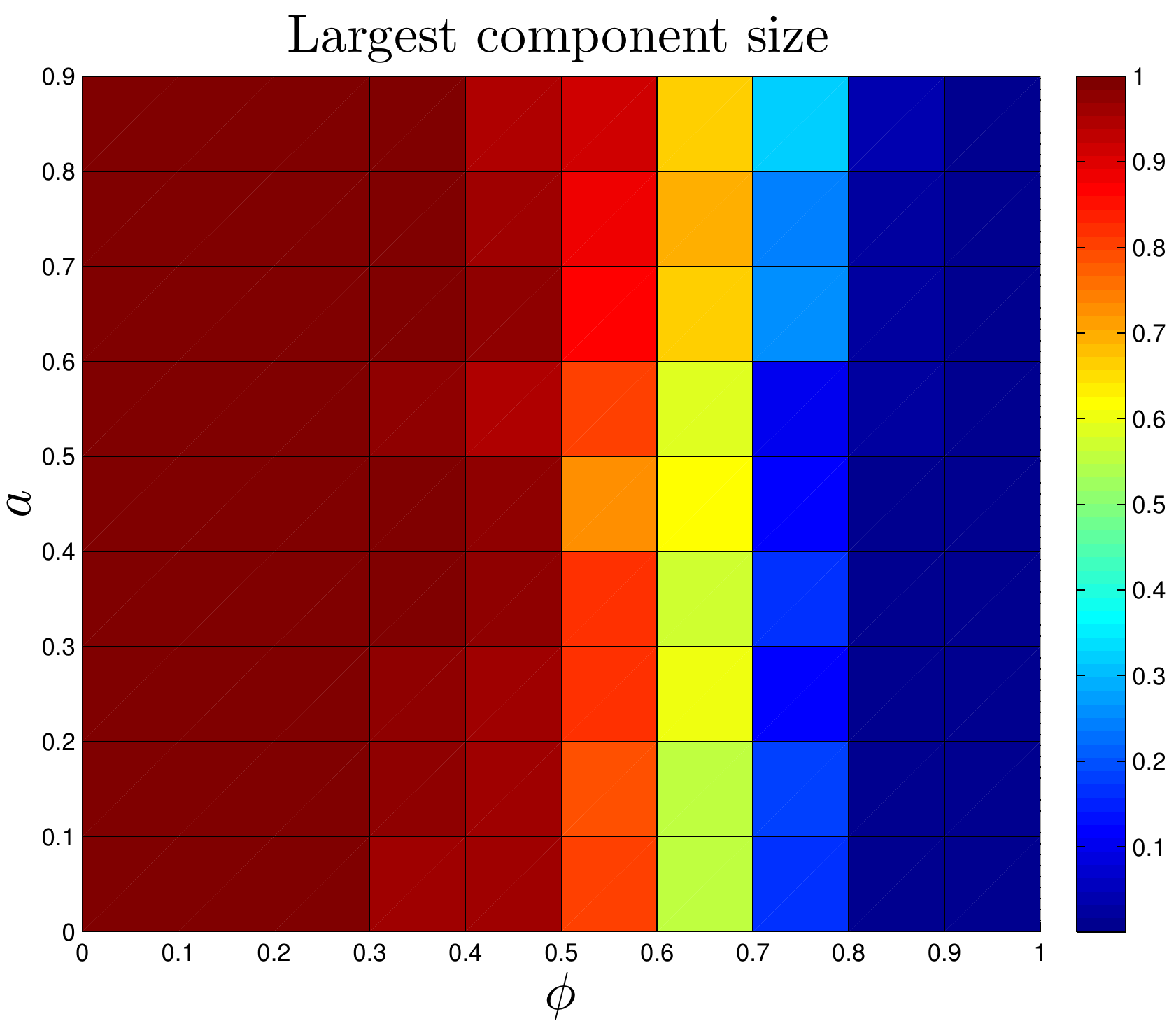}
\includegraphics[width=0.24\textwidth]{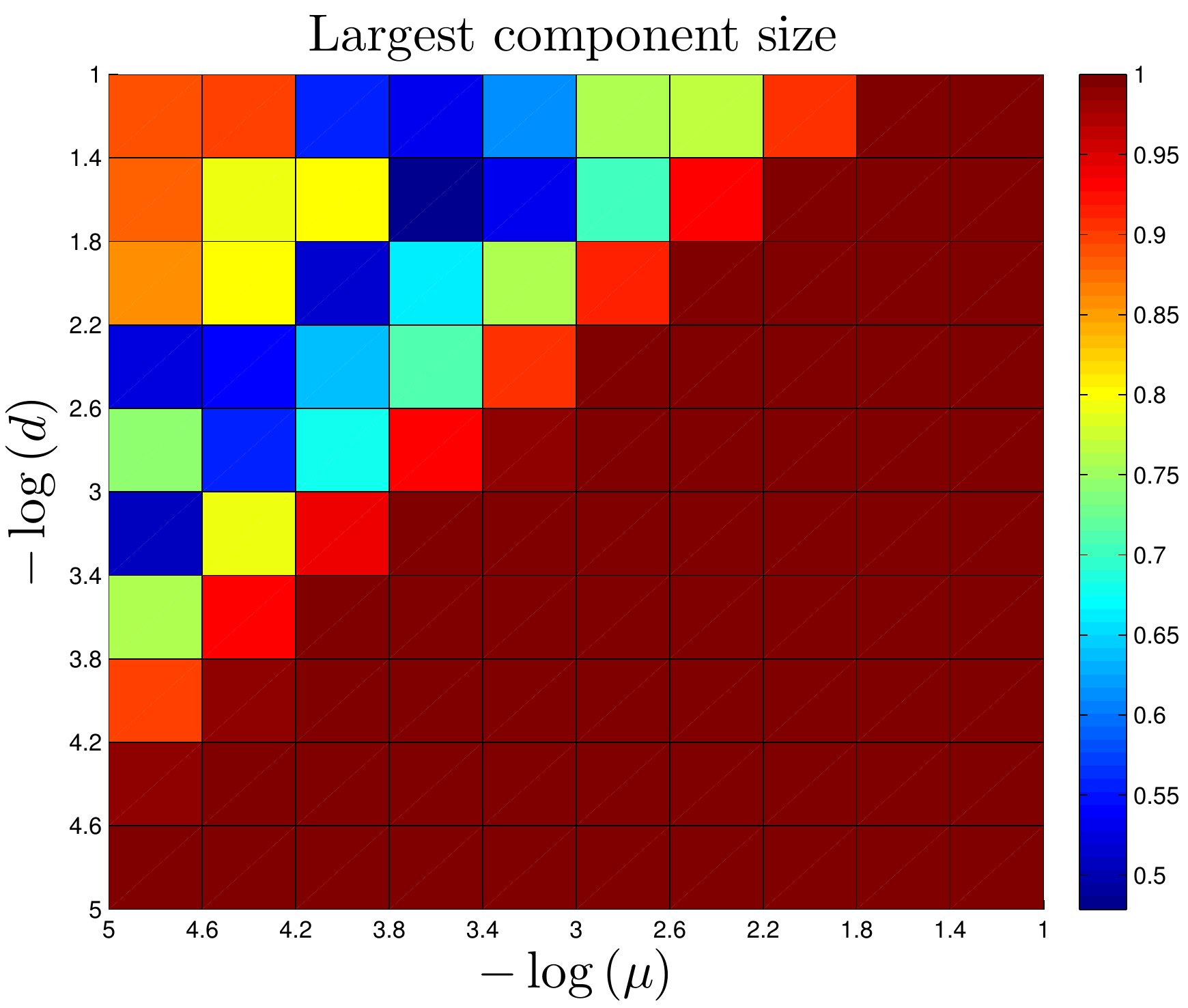}
\includegraphics[width=0.24\textwidth]{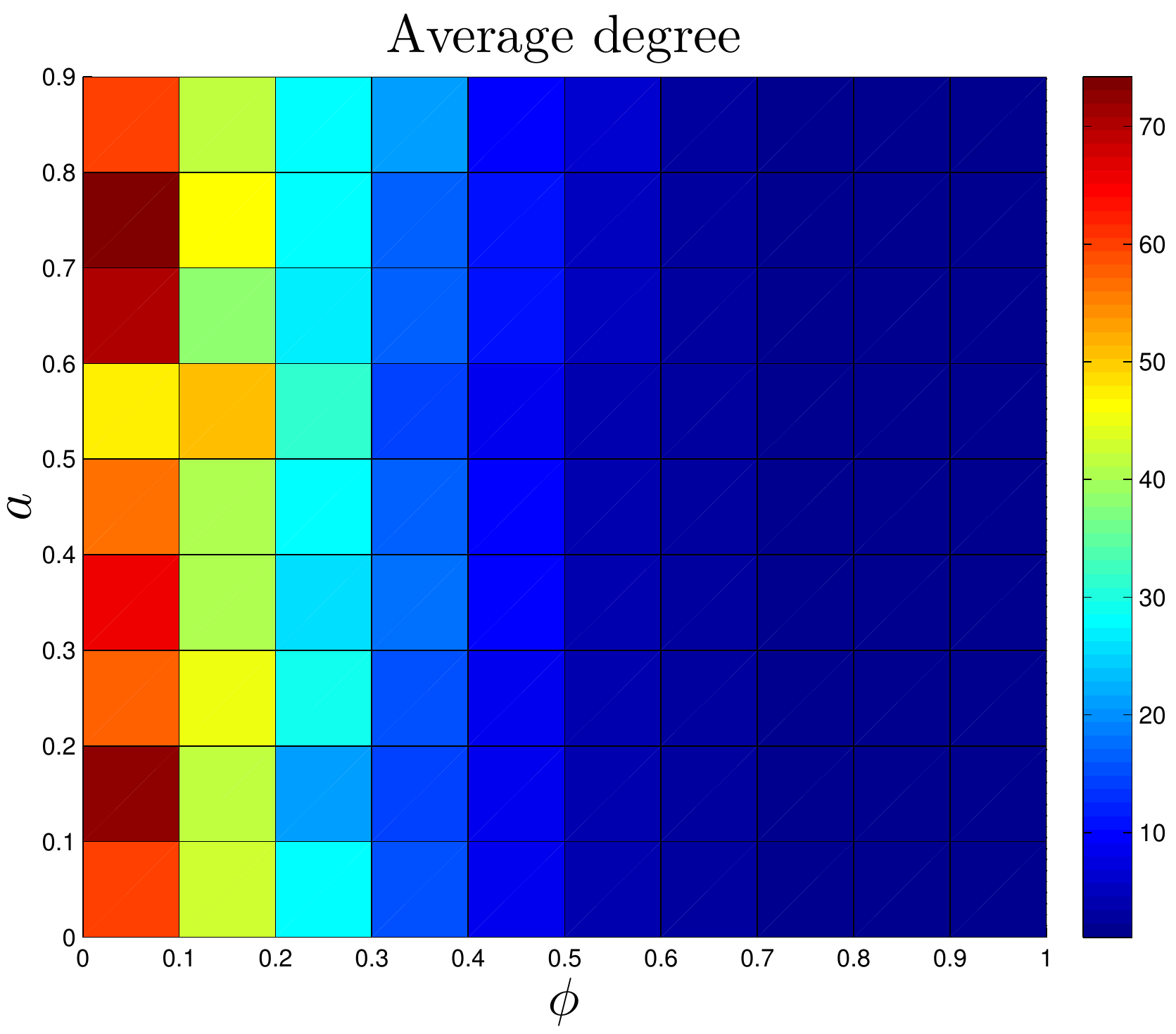}
\includegraphics[width=0.24\textwidth]{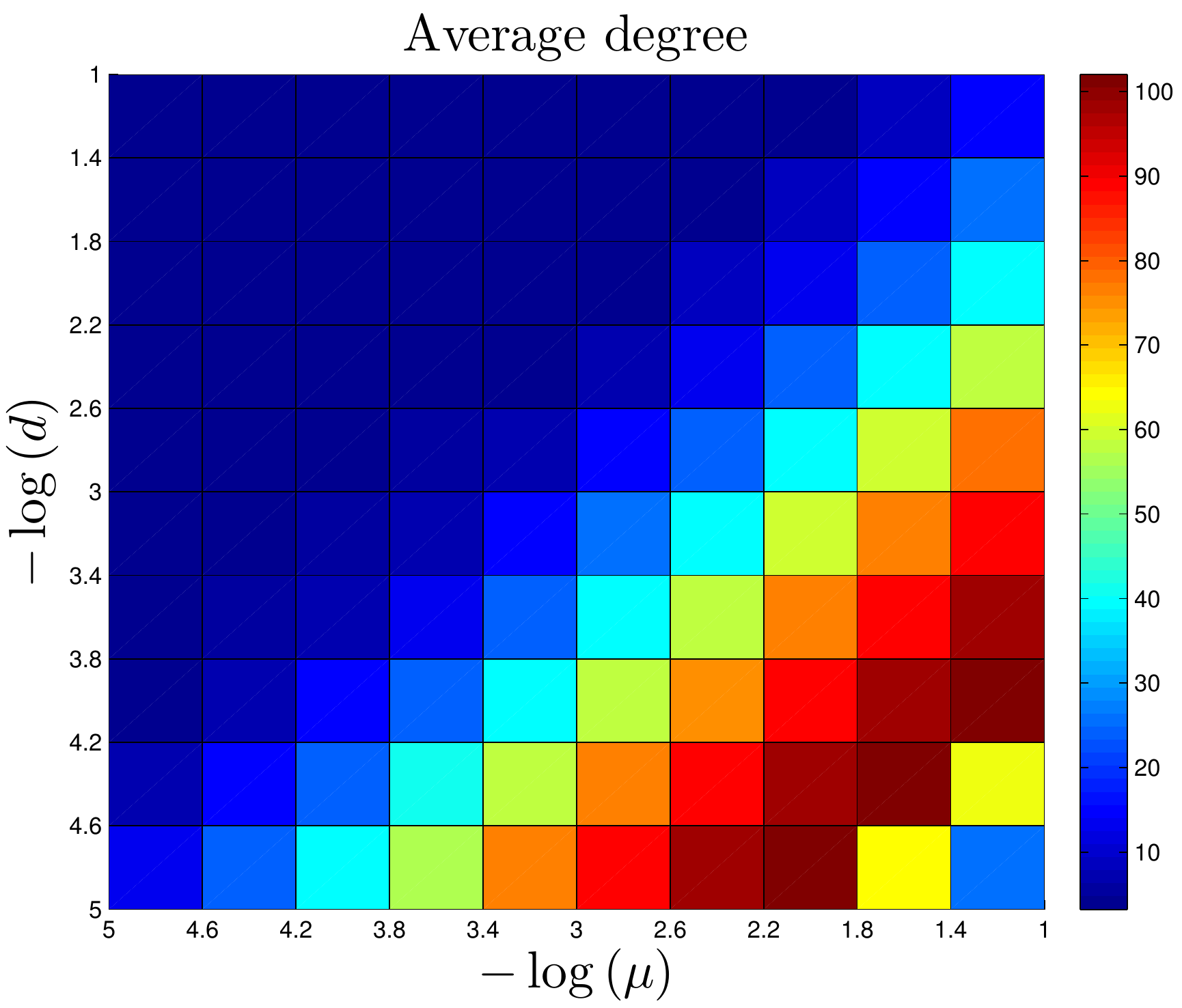}
\end{center}
\caption{{\bf Sensitivity analysis.}  Heat maps represent median values for 10 simulations per parameter combination of the yeast network.  Left: $\phi$ and $a$ are varied, $d$ and $\mu$ values are kept fixed.  Right: $d$ and $\mu$ varied, $\phi$ and $a$ kept fixed.}
\label{fig:sens2d}
\end{figure*}

The number of `extra' links created by each assimilation event should independent of $N$.  Proteins with multiple interaction partners may interact with their partners simultaneously (so-called `party hubs') or at different times/locations (`date hubs') \cite{Han_2004}.  Since they bind to several partners simultaneously, party hubs typically have multiple binding sites, each specific to one binding partner \cite{Kim_2006}.  This specificity suggests that a protein which evolved the capability to bind to a party hub would be unlikely to undergo assimilation.  By contrast, the binding sites of date hubs are often disordered regions which are able to form transient interactions with multiple partners \cite{Ekman_2006,Singh_2007}.  If a protein evolves the capability to bind to a date hub, it is likely to share the physical characteristics of the hub's neighbors, leading to assimilation.  However, to avoid competition for the same binding site, the interaction partners of date hubs tend not to be coexpressed \cite{Kim_2006}.  One consequence of this is that assimilating proteins will likely only bind one of the target protein's neighbors -- whichever neighbor happens to be present at that time and place.  Although the capability to bind to the hub protein's other neighbors may initially be present, these will presumably remain unused in the cell.  Our expectation is that the assimilating protein will therefore be unlikely to retain this capability, as it evolves.  Similarly, only a single extra link should be generated at the second-neighbor level, third-neighbor level, etc.  Consistent with the evidence discussed above, the number of links created by assimilation is approximately independent of the total network size.  Party hubs typically are centrally-located within modules, while date hubs often function to stitch together large-scale modules in the cell.  It may be that duplication-only models are unrealistically fragmented (Table~\ref{tab:SVF}) because their modules are not properly attached with date hubs; instead, the modules are disconnected components.

\subsection*{Sensitivity analysis}

The DUNE model has four parameters.  One parameter, the DU rate $d$, is estimated from empirical data.  The other three are adjustable parameters: the NE rate $\mu$, the divergence probability $\phi$, and the assimilation probability $a$.  To gain a better understanding of how these parameters affect the final network structure, starting with each organism's set of parameters, we systematically adjusted all 4 parameters.  Results are shown in Figure~\ref{fig:sens2d}.

As expected, the divergence parameter $\phi$ was positively correlated with both the modularity $Q$ and the diameter.  When $\phi \approx 0$, the network quickly reaches a fully-connected state, where all proteins are linked to all other proteins.  Consequently, the network is not organized into modules, and all distances in the network are equal to 1. The `thinning out' of duplicate links is therefore essential to generate non-trivial network features.  The opposite occurs when the NE rate $\mu$ is too low: $f_1 \approx 0$, and the network evolves towards a completely disconnected state.

Why does gene duplication lead to modularity?  Consider an initially uniform network.  When a node is duplicated at random, this causes the original node, the copy, and their immediate neighbors to share more links internally than they do with the rest of the network.  Subsequent duplications amplify this effect: if a node that has 10 links within a module but only 2 external links is duplicated, there are now 20 internal and 4 external links (prior to post-duplication divergence).

Interestingly, $Q$ has a weak negative correlation with the assimilation parameter $a$.  When $a$ is large, the probability to link to distant neighbors of the target protein is relatively high, and mutated proteins have a non-negligible chance to generate links to proteins outside of their target's pathway.  This causes modules to blur at the edges; their member proteins will share a higher number of links to other modules than for a low $a$ network.  Although the modularity is reduced for a high $a$ network, it does not disappear entirely.  Similarly, there is a sharp decrease in $Q$ as the NE rate surpasses the DU rate, indicating the important role of the DU mechanism in modular organization.

Diameter is also negatively correlated with $a$.  When a single NE event has a significant chance to generate links to the target protein's neighbors, this tends to reduce the overall separation of proteins in the network.

\section*{Discussion}

The relevance of selection to PPI network evolution has been a topic of considerable debate \cite{Lynch_2007}, particularly in the context of higher-order network features, such as modularity.  A number of authors have argued that specific selection programs are required to generate modular networks, such as oscillation between different evolutionary goals \cite{Lipson_2002,Kashtan_2005,Komurov_2007,Wagner_2007,Espinosa-Soto_2010,Soyer_2010}.  However, previous work has shown that gene duplication by itself, in the absence of both natural selection and neofunctionalization, can generate modular networks \cite{Hallinan_2004,Sole_2008}.  Consistent with the findings of \cite{Hallinan_2004,Sole_2008}, modularity in our model is primarily generated by gene duplications (Figure~\ref{fig:sens2d}; see SI for sensitivity analysis).  Unfortunately, duplication-only models err in their predictions of other network properties (Tables~\ref{tab:SVF} and \ref{tab:SMAPE}; Figure \ref{fig:othermodels}).  A well-known problem with duplication models is that they generate excessively fragmented networks, with only about 20\% of the proteins in the largest component.  This is in sharp contrast to real PPI networks, which have 73\% to 89\% of their proteins in the largest component.  Neofunctionalization-only models have most of their proteins in the largest component, but are significantly less modular than real networks.  As shown in Table~\ref{tab:SVF}, by modeling duplication and neofunctionalization simultaneously, the DUNE model generates networks which have the modularity found in duplication-only models, while retaining most proteins in the largest component.  This lends support to the idea that gene duplication contributes to the modularity found in real biological networks, and that protein modules can arise under neutral evolution, without requiring complicated assumptions about selective pressures.  This is consistent with recent experimental work characterizing a real-world fitness landscape, showing that it is primarily shaped by neutral evolution \cite{Hietpas_2011}.

One model for the fate of duplicate genes is \emph{subfunctionalization} (SF).  In SF, the original and the duplicate genes are both free to lose their redundant functions, so they can evolve freely until they exactly reproduce the ancestral function \cite{Lynch_2000_Genetics}.  The post-duplication divergence in our model is similar in spirit to SF, but it differs in two significant ways: (1) in our model, the link loss is completely asymmetric, and (2) a fraction ($1-\phi$) of the redundant links are retained, so, unlike SF, not all of the redundancy is eliminated.  For the first point, empirical evidence suggests that the divergence is asymmetric \cite{Gu_2005}, although the assumption of \emph{complete} asymmetry would likely need to be revisited to build a finer-grained model.  Second, genetic regulatory networks have been shown to be robust to random link deletions, indicating that these networks retain some degree of redundancy \cite{Siegal_2002,Bergman_2003,Wagner_2007_BioSystems}.  \emph{In silico} evidence suggests that a more accurate picture may be of a transient period of functional divergence, followed by prolonged neofunctionalization, resulting in only a partial loss of redundancy \cite{MacCarthy_2007}.  This is consistent with our model.

One example of a known biological mechanism which should lead to assimilation is domain shuffling, the copy-and-pasting of part of one protein into another \cite{Moran_1999,Patthy_1999}.  The neofunctionalization mechanism described here is quite general, and includes domain shuffling, among other methods of PPI creation.  A PPI formed via domain shuffling will often be the result of a binding site duplication.  Assuming the binding is due to simple surface similarity, the initial link will be to the protein which had its domain copied.  The likelihood of binding to neighbors of the original protein should depend only on the probability that each interaction is due to surface similarity because the copied binding site will be identical to the original.

The role of domain shuffling in assimilation raises the question of whether domains should be modeled explicitly, rather than representing proteins as integral units.  Previous work indicates that overall PPI network topology is robust to the details of domain shuffling \cite{Evlampiev_2007}.  Moreover, while proteins which have experienced domain shuffling have a higher average degree than other proteins, high- and low-degree proteins are equally likely to acquire new interactions this way \cite{Cancherini_2010}.  Because the creation of new links by domain shuffling should be topologically very similar to the creation of new links by other neofunctionalization events, we believe our model is a reasonable implementation of this mechanism, as it applies to the evolution of network topology.

Previous estimates of NE rates in eukaryotes have varied widely, generally falling in the range of 100 to 1000 changes/genome/Myr \cite{Wagner_2003,Berg_2004,Beltrao_2007}, or on the order of 0.1 change/gene/Myr.  However, more recent empirical work has identified several problems with the methods used to obtain these estimates, suggesting that \emph{de novo} link creation is much less common than previously thought \cite{Gibson_2009}.  This is consistent with our model.  The best-fit values of our NE rate $\mu$ are in the range of $10^{-5}$ to $10^{-4}$/gene/Myr (Table~\ref{tab:params}), which in all three organisms are considerably slower than the duplication rates $d$.

Biologically, many of the interactions created by our neofunctionalization mechanism are expected to initially be weak, non-functional interactions.  The results of \cite{Heo_2011} suggest that strong functional interactions are correlated with hydrophobicity, which in turn is correlated with promiscuity.  We posit that initially weak, non-functional interactions are an essential feature of PPI evolution, as they provide the `raw material' for the subsequent evolution of functional interactions.  If this reasoning is correct, one consequence should be that hub proteins are, on average, more important to the cell than non-hub proteins.  This has been found to be true: both degree \cite{Jeong_2001} and betweenness centrality \cite{Joy_2005} have positive correlations with essentiality, indicating that hub proteins are often critical to the cell's survival.

We have described here a model for how eukaryotic protein networks evolve.  The model, called DUNE, implements two biological mechanisms: (1) gene duplications, leading to a superfluous copy of a protein that can change rapidly under new selective pressures, giving new relationships with other proteins and (2) a protein can undergo random mutations, leading to neofunctionalization, the \emph{de novo} creation of new relationships with other proteins.  Neofunctionalization can lead to assimilation, the formation of extra novel interactions with the other proteins in the target's neighborhood.  Biological evidence suggests that this type of mechanism exists.  Our specific implementation is based on a simple geometric surface-compatibility argument for the observed transitivity in PPI networks.  This is, of course, a heavily simplified model of PPI network evolution, and there are many biological factors which have not been included.  However, our relatively simple model shows good agreement with 10 topological properties in yeast, fruit flies, and humans.  One finding is that PPI networks can evolve modular structures, just from these random forces, in the absence of specific selection pressures.  We also find that the most central proteins also tend to be the oldest.  This suggests that looking at the structures of present-day protein networks can give insight into their evolutionary history.

\subsection*{Methods}

Genome-wide PPI screens have a high level of noise \cite{Deane_2002}, and specific interactions correlate poorly between data sets \cite{Deeds_2006}.  We found that several large-scale features differed substantially between types of high-throughput experiments (see SI).  Due to concerns about the accuracy and precision of data obtained through high-throughput screens, we chose to work with `high-confidence' data sets consisting only of pairwise interactions confirmed in small-scale experiments, which we downloaded from the public HitPredict database \cite{Patil_2011}.  We found sufficient high-confidence data in yeast (\emph{S. cerevisiae}), fruit flies (\emph{D. melanogaster}), and humans (\emph{H. sapiens}).

All simulations and network feature calculations were carried out in Matlab.  Our scripts are freely available for download at \texttt{http://interacto.me}.  We computed betweenness centralities, clustering coefficients, shortest paths, and component sizes using the MatlabBGL package.  Modularity values were calculated with the algorithm of \cite{Blondel_2008}.  All comparisons (except the degree distribution) are between the largest connected components of the simulated and experimental data.

Due to the human network's somewhat larger size, most dynamical features were calculated once per 50 time steps for the human network, but were updated at every time step in the yeast and fly networks.  For dynamical plots, the $y$ coordinates of the trend line are medians-of-medians.  The amount of time elapsed per time step (the $x$ coordinate) varies between simulations.  We binned the time coordinates to the nearest 10 million years for yeast and fly, and 25 million years for human.  When multiple values from the same simulation fell within the same bin, we used the median value.  We then calculated the median value between simulations.  Scatter plot trend lines are calculated in a similar way.  The trend line represents the median response variable ($C$, $b$, or $\ell$) value over all nodes within a single simulation with degree $k$.  The $y$ coordinate of the trend line is therefore the median (across 50 simulations) of these median response variables.  This median-of-medians includes all simulations that have nodes of a given degree.

\subsection*{Empirical data}

We downloaded large-scale data sets from BioGRID \cite{Stark_2006}, and used the Wilcoxon rank-sum test to compare aggregate statistical features across various experimental types in yeast (\emph{S. cerevisiae}) and humans (\emph{H. sapiens}) \cite{Han_2004}.  As expected, we found that data obtained by affinity capture was significantly different than pairwise experimental data (primarily yeast two-hybrid and \emph{in vitro} complexation), as the affinity capture interactions represent entire complexes, which is somewhat different information than the pairwise interactions we are attempting to capture using our model.  However, more surprisingly, the only feature to show significant agreement between pair-wise techniques was the eigenvalue distribution of the walk matrix ($P > 0.05$).  Further sub-dividing the individual techniques into smaller data sets containing only results obtained in single experiments, we discovered that, again, only the spectra agreed between different screens.

Note that, due to the small size of the fly network, there may be too many missing links to obtain an accurate description the network's large-scale topology.  Although, by appropriate parameter tuning, our model is able to accurately reproduce the fly network, it is possible that different parameters will be required to match the fly network once it becomes more fully characterized experimentally.  The data sets considered here do not include interactions which are enabled through post-translational modifications.  Although these data sets are far from complete, and may be susceptible to false-positive detections, these appear to be the most accurate data available at the present time.

\begin{acknowledgments}
We would like to thank Tom MacCarthy and Anna Panchenko for valuable discussions.
\end{acknowledgments}

\bibliographystyle{unsrt}
\bibliography{PPI}

\setcounter{figure}{0}
\setcounter{table}{0}
\makeatletter
\renewcommand{\thefigure}{S\@arabic\c@figure}
\makeatletter
\renewcommand{\thetable}{S\@arabic\c@table}

\begin{appendix}

\section*{Supporting Information}

\subsection*{Randomness conjecture}

We describe here a test of our randomness conjecture, $q=1/N$.  The implication is that the average number of NE links per protein should be independent of $N$.  Another possibility is that $q$ is constant, implying that the number of NE links per protein is proportional to $N$.  To test this, we compared NE rates in the human and yeast PPI networks.  The total number of proteins in humans is estimated to be 22740 \cite{Scherer_2008} and in yeast, 5616 \cite{Byrne_2005}.  The number of mutations to coding DNA is approximately 0.004/genome/replication in humans and 0.0027/genome/replication in yeast \cite{Drake_1998}.  If the number of new links is proportional to $N$, then, based on the number of proteins and the mutation rate, there should be roughly 600\% more links created by NE in humans than in yeast.  However, by counting the number of nonredundant interactions in duplicate gene pairs, it has been shown empirically that the average number of links created by NE per protein is only about 8\% higher in humans than in yeast \cite{He_2005}.  These results support the conjecture that the probability for a protein to receive a new link via point mutation is approximately independent of $N$, as previously noted in \cite{Beltrao_2007}. Due to the finite copy number of proteins, as well as the compartmentalization of eukaryotic cells, we regard it as unlikely that proteins will simultaneously acquire multiple links to targets in different locations in the cell, or which are involved in divergent biological processes.

\begin{figure}[b]
\begin{center}
\includegraphics[width=0.45\textwidth]{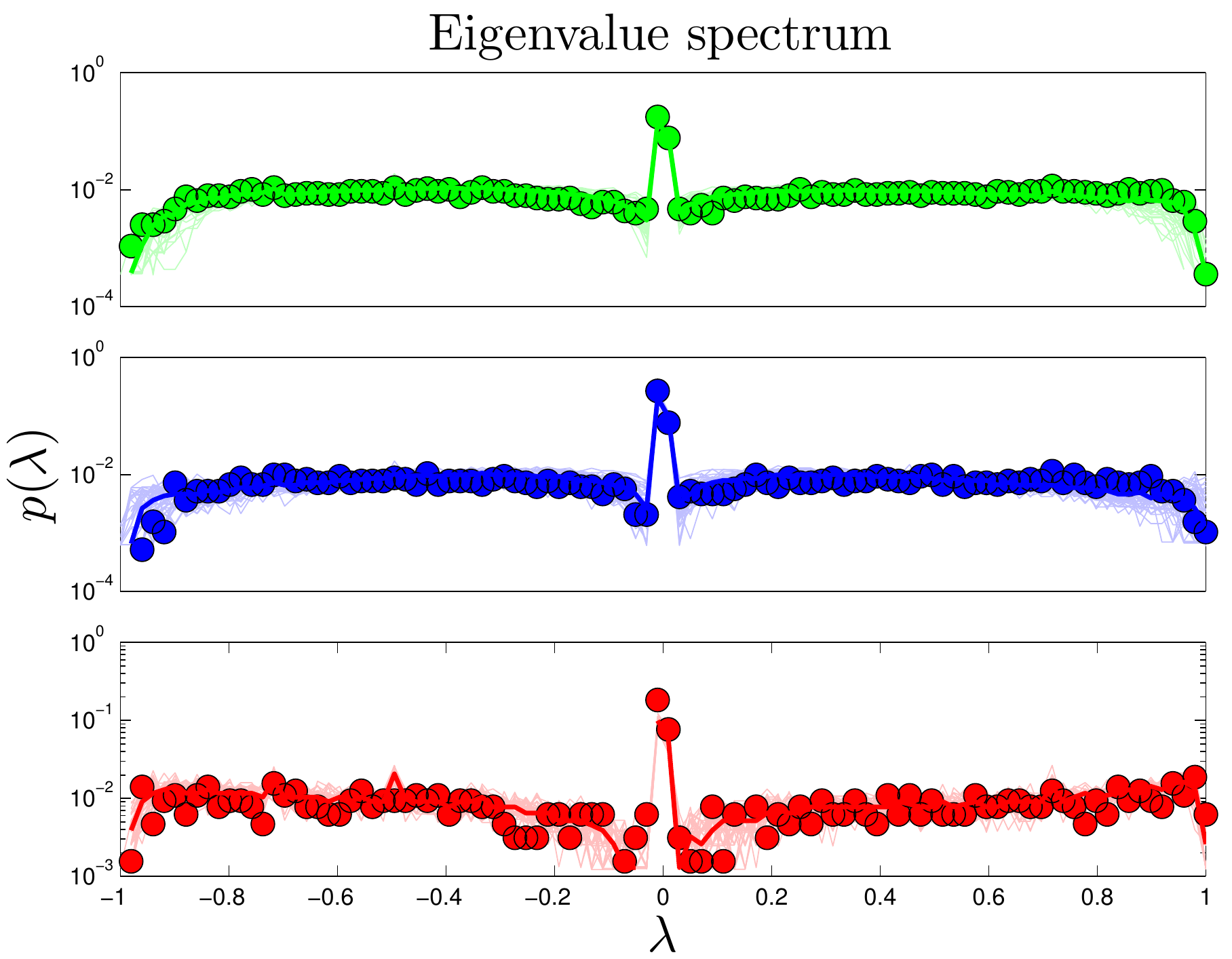}
\end{center}
\caption{{\bf Walk matrix eigenvalues.}  Shown are eigenvalue ($\lambda$) distributions in human (green), yeast (blue), and fly (red).  Heavy lines are the median values from 50 simulations, and light lines are results of individual simulations.}
\label{fig:spectrum}
\end{figure}

\begin{figure*}
\begin{center}
\includegraphics[width=0.45\textwidth]{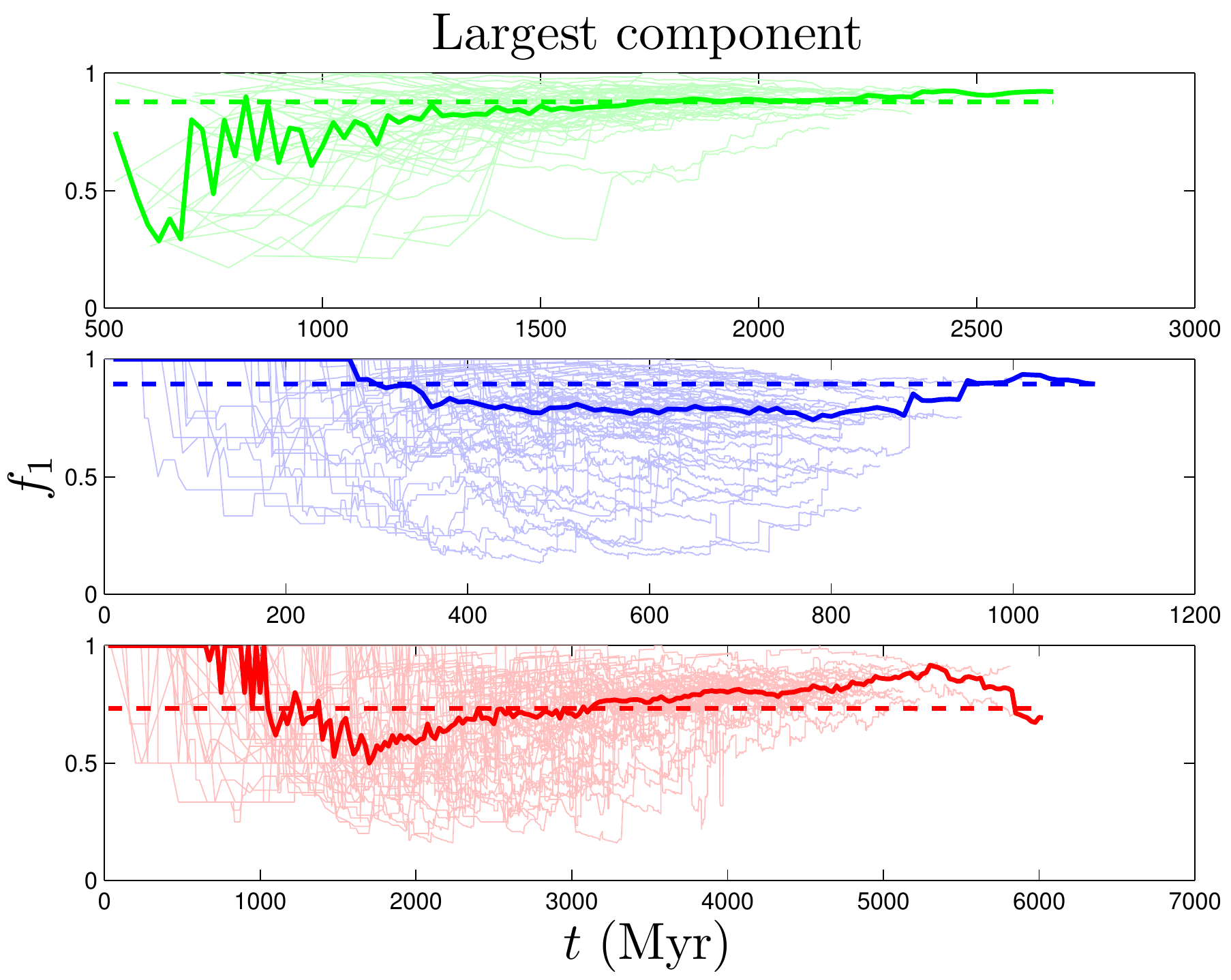}
\includegraphics[width=0.45\textwidth]{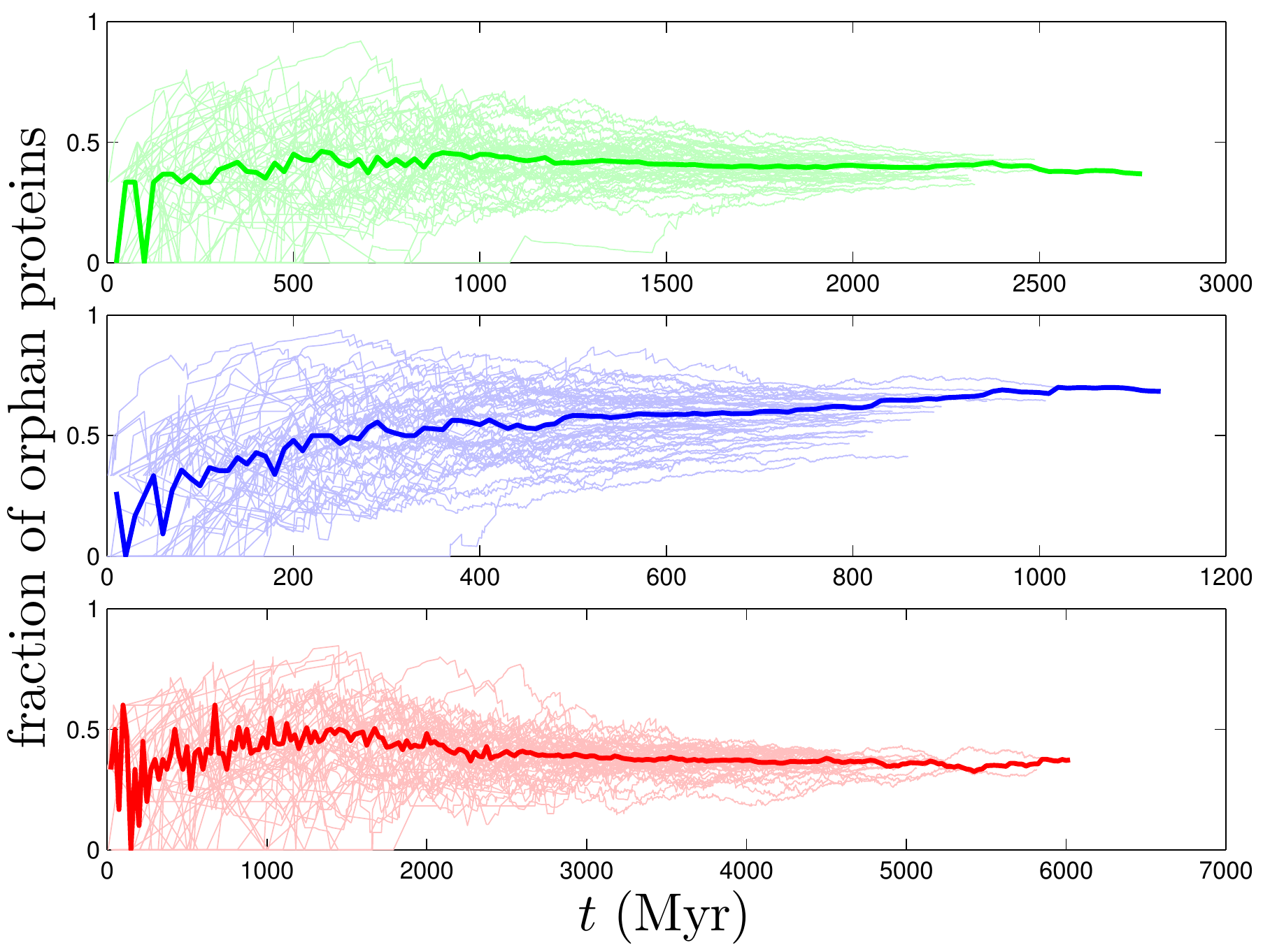}
\includegraphics[width=0.45\textwidth]{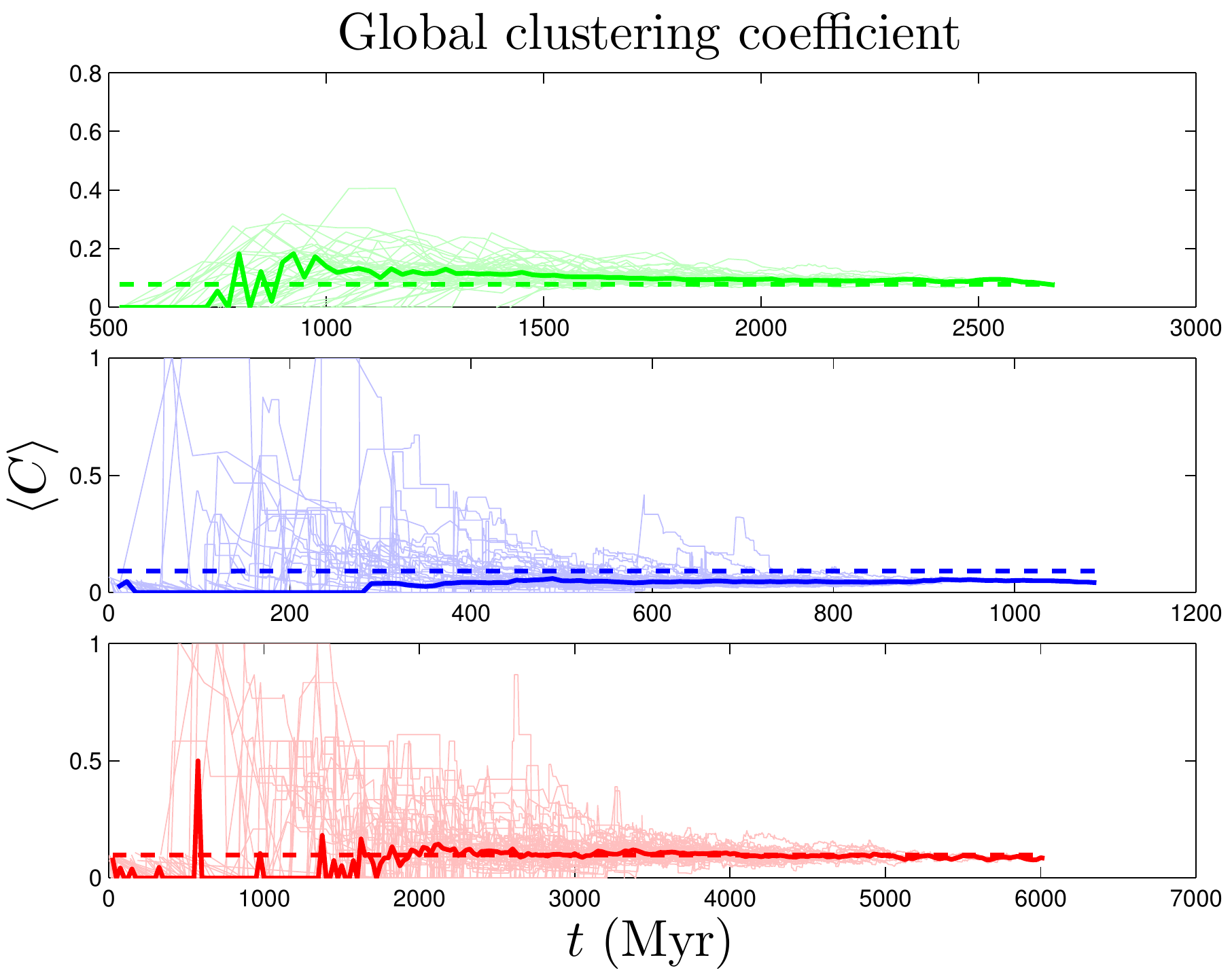}
\includegraphics[width=0.45\textwidth]{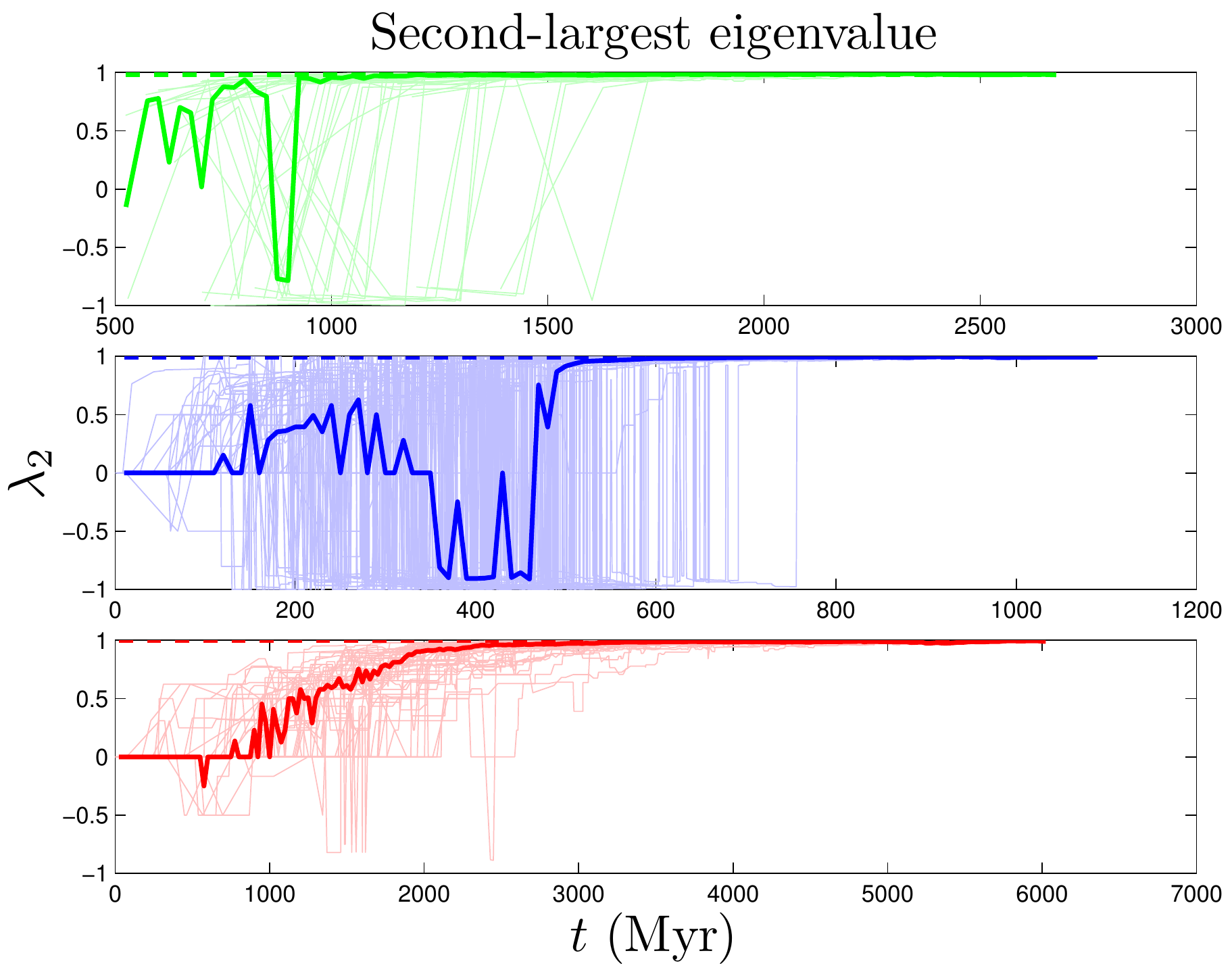}
\end{center}
\caption{{\bf Dynamical features.}  Shown are the evolution of (A) the largest component size, (B) the fraction of orphan proteins, (C) the global clustering coefficient, and (D) the second-largest eigenvalue of the walk matrix, in human (green), yeast (blue), and fly (red).  Light lines indicate the evolutionary trajectories of 50 individual simulations, and the heavy line is the median value.  Empirical data values are shown as a dashed line, where available.}
\label{fig:dynamics}
\end{figure*}

\begin{figure*}
\begin{center}
\includegraphics[width=0.45\textwidth]{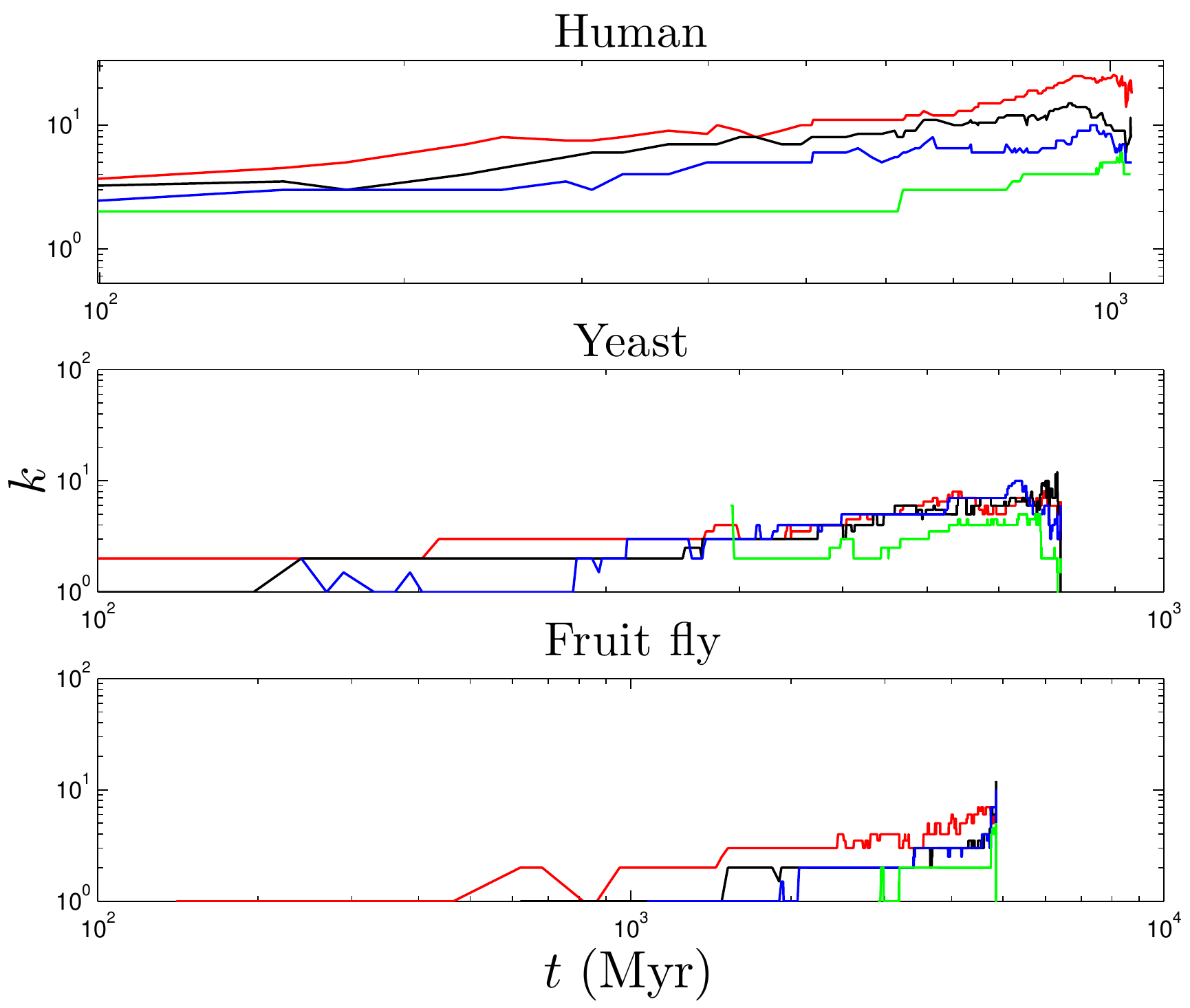}
\includegraphics[width=0.45\textwidth]{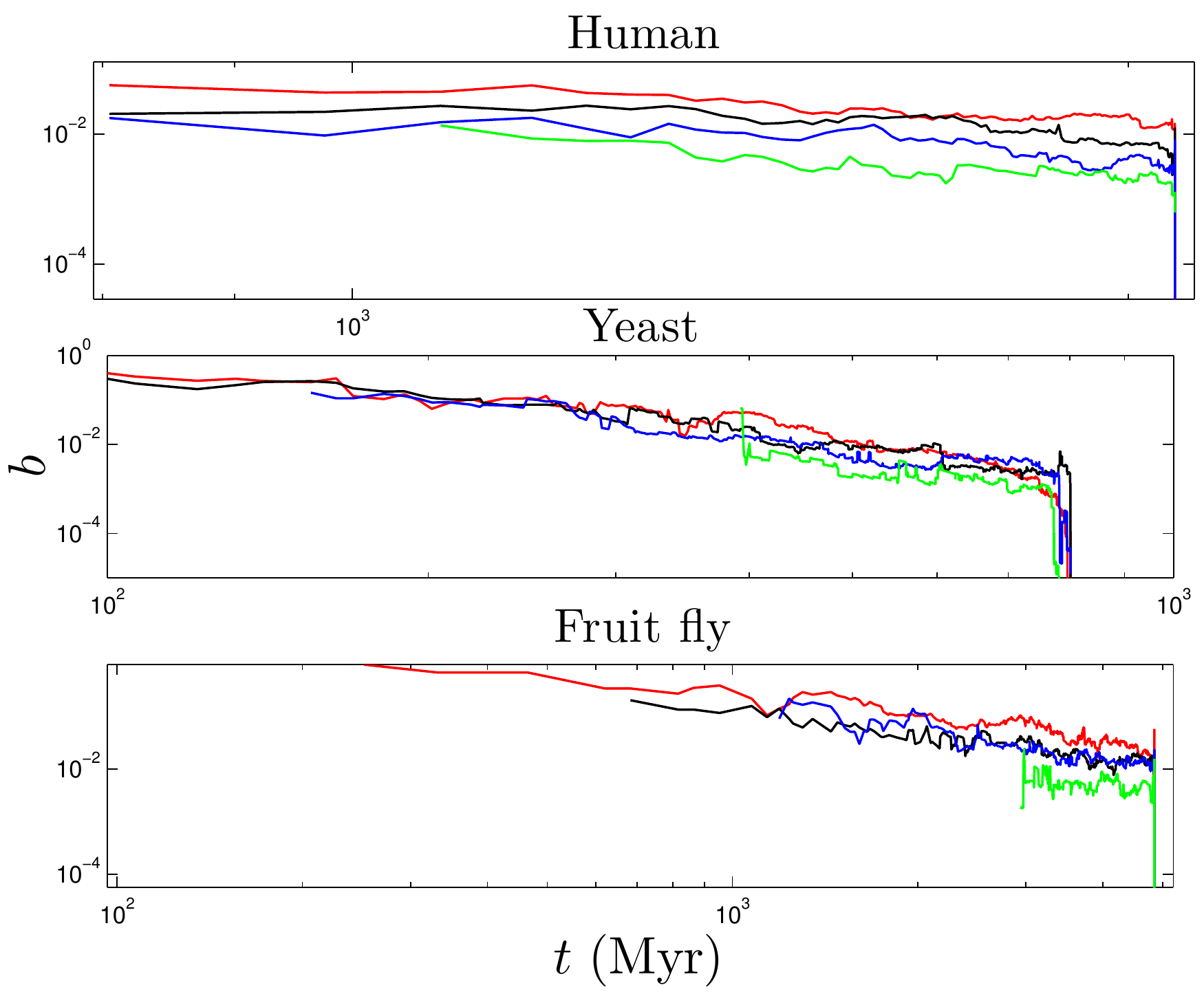}
\end{center}
\caption{{\bf Individual protein centrality scores.}  Evolution of degree (A) and betweenness (B) for proteins introduced to the network at different times in humans (top), yeast (middle), and flies (bottom).  The 1st protein (one of the two initial proteins) is shown in red, the 6th protein in black, the 11th protein in blue, and the 101st protein in green.  Curves are median values from 50 simulations.}
\label{fig:tracking}
\end{figure*}

\subsection*{Large-scale duplications}

Our model does not explicitly consider simultaneous duplication of multiple genes (chromosomal duplications, whole genome duplications, etc.).  However, as shown in Table~\ref{tab:params}, duplication rates in our model are considerably higher than neofunctionalization rates, so that, on average, there are multiple duplications per neofunctionalization event.  Sequential duplications of this type may be thought of as an (imperfect) representation of multi-gene duplications.  The advantage of this approach is that we do not require separate rates for each duplication scale (one could imagine an extremely detailed model which included separate rates for gene duplication, gene-pair duplication, gene-triplet duplication, etc.).  The downside is that our implementation of larger-scale duplications will generally include some genes which have been duplicated multiple times, and others which have not been duplicated at all.  A potential mitigating factor is that the rate of gene loss (and evolution in general) following genome duplication is very high~\cite{Ku_2000,Kellis_2004}, so even a completely faithful large-scale duplication would likely be altered within short order.

\subsection*{Eigenvalues}
The connectivity of a network can be expressed by its \emph{adjacency matrix}, an $N\times N$ matrix $\mathbf{A}$, in which the entries $A_{ij}$ equal 1 if a link exists between proteins $i$ and $j$, and 0 otherwise.  If $\mathbf{A}$ is normalized by column, then the entries describe the rates of a transition from $i$ to $j$ in one time step.  The distribution of eigenvalues $p(\lambda)$ is called the network's \emph{spectrum} (Figure~\ref{fig:spectrum}).  This matrix and its eigenvalues can be interpreted in terms of a process in which a random walker starts on one node $i$, and, over a series of time steps, reaches another node $j$.  Intuitively, this can be thought of as a signal propagation rate: if one protein is affected by an external signal, how long does it take that signal to diffuse through the network?  The eigenvalues $\lambda$ of this `walk matrix' describe the rate at which a random walk on the network reaches steady-state.  The second-largest eigenvalue ($\lambda_2$) determines the rate of convergence of the random walk (Figure~\ref{fig:dynamics}).  A larger value of $\lambda_2$ indicates a slower signal propagation rate.

\begin{figure}[b]
\begin{center}
\includegraphics[width=0.45\textwidth]{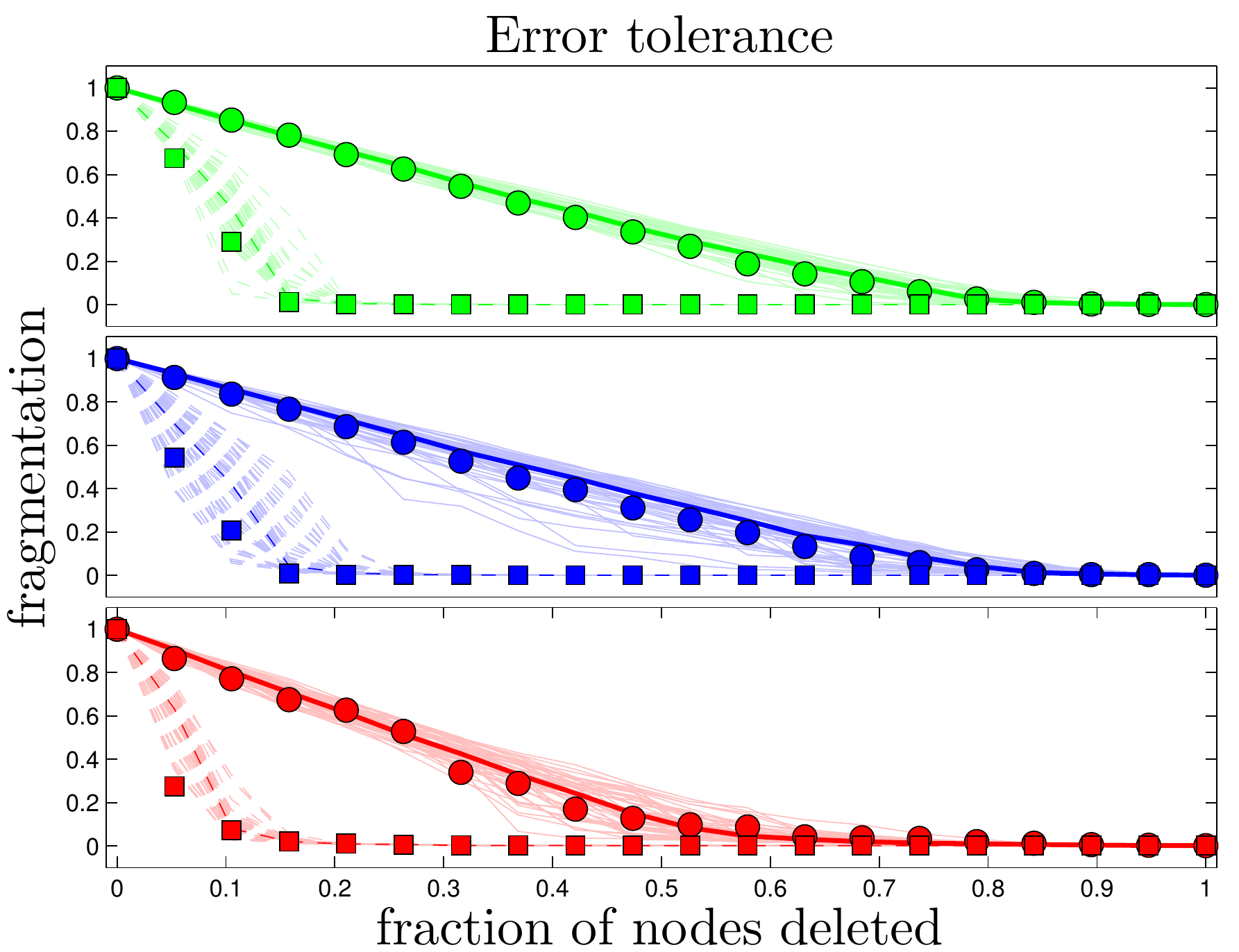}
\end{center}
\caption{{\bf Error tolerance.}  Shown are error tolerance curves in human (green), yeast (blue), and fly (red).  Circles indicate proteins deleted randomly, and squares indicate proteins deleted starting with the most well-connected protein and removing proteins in descending order.}
\label{fig:errortolerance}
\end{figure}

\subsection*{Error tolerance}

We measured the `error tolerance' as described in \cite{Vazquez_2003}: we examined the decrease of $f_1$ when nodes (and their accompanying edges) were deleted from the network either (1) at random, or (2) according to their degree, starting with the most well-connected node.  Results for the simulated and experimental networks were very similar (Figure~\ref{fig:errortolerance}).

\subsection*{Simulation length: early versus late evolution}

The total time elapsed during our simulations varies considerably, with yeast and human simulations running about 1 to 2 billion years, and the fly simulations about 5 billion years.  This is compared to the rough estimate of 3.5 billion years since the origin of life on Earth \cite{Campbell_1999}.  The exceptionally long duration of the fly simulations are due to the very low gene duplication rate ($d = 0.001$/gene/Myr).  The aim of our model is to describe the evolution of PPI networks with all their present-day machinery.  Gene duplication, in the form in which it exists today, certainly would not have existed at the origin of life!  The initial state in our model consists of two interacting proteins.  Biologically, these are two polypeptides (or, more likely, RNA molecules) in a pre-biotic soup, that happen to interact in a way that is mutually beneficial.  Each of these molecules has the ability to replicate.  This autonomous replication of individual proteins corresponds to `gene duplication' in the very early stages of evolution.  However, this is a very different conceptual underpinning for the duplication mechanism, and it seems unlikely to share the present-day values of the duplication rate.  Because, in the early stages of evolution, each time step represents a very long duration in real time, it is likely that this accounts for the discrepancy in total time elapsed.

\begin{table}[b]
\begin{tabular}{| c | c c c c c |}
\hline
 & $\gamma$ & $\beta$ & $\xi$ & $\alpha$ & $\delta$\\
\hline
Yeast & $2.8(2)$ & $2.4(1)$ & $1.8(2)$ & $1.2(2)$ & $0.32(6)$ \\
Fly & $3.1(2)$ & $2.2(4)$ & $0.8(5)$ & $0.8(7)$ & $0.0(3)$ \\
Human & $2.8(1)$ & $2.3(1)$ & $1.5(3)$ & $1.3(2)$ & $0.0(1)$ \\
\hline
\end{tabular}
\caption{{\bf Scaling exponents.}  Distributional exponents ($p(k) \sim k^{-\gamma}$, $p(b) \sim b^{-\beta}$) were estimated using the maximum likelihood method of \cite{Clauset_2009}.  Other exponents (${C} \sim k^{-\xi}$, ${b} \sim k^{\alpha}$, ${n} \sim k^{-\delta}$) were estimated using nonlinear regression.  Due to the relatively small sizes of the data sets, there is considerable uncertainty in these estimates.}
\label{tab:exponents}
\end{table}

\subsection*{Fitting functions}

The degree distribution obeys a power law in its tail, $p(k) \sim k^{-\gamma}$ \cite{Clauset_2009}, with $\gamma \approx 3$ (Table~\ref{tab:exponents}), implying that hub proteins are more common than would be expected for a randomly connected network, which would have an exponentially decaying $p(k)$.  The closeness distribution $p(\ell)$ is approximately Gaussian, with mean 0.17 and standard deviation 0.03 in humans, mean 0.19 and standard deviation 0.03 in yeast, and mean 0.13 and standard deviation 0.03 in flies. Closeness is a measure of distance, indicating that the distances within the network are essentially a random walk in `node space'.  The betweenness distribution also follows a power law in its tail.  This is an indication of modular structure, due to the overrepresentation of `bridge' proteins, relative to a randomly connected network.

All species examined show a power law decay in clustering coefficient as a function of degree, ${C} \sim k^{-\xi}$.  Poorly-connected proteins therefore tend to have \emph{higher} clustering coefficients, meaning that a greater fraction of their neighbors are mutually connected.

Disassortative mixing was quantified for the yeast PPI network in \cite{Maslov_2002} as a power law \emph{decrease} in median neighbor degree, ${n} \sim k^{-\delta}$.  This is consistent with our data, although the very small estimated value of $\delta = 0.32$ indicates only a slight negative relation (Table~\ref{tab:exponents}).  Interestingly, $\delta = 0$ in both human and fly networks, indicating that disassortativity may be a trait unique to the yeast network.

\subsection*{Principal component analysis}

We examined six features which calculate a value for each node in the network: degree centrality, clustering coefficients, closeness centrality, eigenvalue spectrum, betweenness centrality, and mean nearest-neighbor degree.  To quantify the independence of these features, we used principal component analysis (PCA) \cite{Shlens_2009}.  Each feature assigns a value to each node in the network, giving a $6 \times N$ data matrix, where each row represents a feature (signal), and each column is a node (sample).  We subtract the mean and divide by the standard deviation of each row.  This results in a standardized data matrix, denoted by $\mathbf{Y}$.  The $6 \times 6$ correlation matrix for each species is defined as $\mathbf{C} \equiv \frac{1}{N-1} \mathbf{Y}\mathbf{Y}^\mathrm{T}$:
\begin{equation}
\mathbf{C}_\text{h} =
\begin{bmatrix}
      1  &    0.10   &   0.87   &  0.56  &   0.02   &  -0.12 \\
   0.10     &       1  &   -0.04   & 0.09    & 0.01  &   0.03 \\
     0.87  &   -0.04   &         1 &    0.44  & -0.01   &  -0.08 \\
0.56  &   0.09    &    0.44     &        1  &  0.00 &    0.34 \\
  0.02   &  0.01 &  -0.01  & 0.00   &         1 &   -0.02 \\
  -0.12   &  0.03   &  -0.08   &  0.34 &   -0.02   &         1
\end{bmatrix},
\end{equation}
\begin{equation}
\mathbf{C}_\text{y} =
\begin{bmatrix}
       1  &   0.03    &  0.91   &  0.43 &  -0.02  &   -0.21  \\
     0.03   &         1  &  -0.06  &  0.04  &  0.01   &   0.03  \\
      0.91  &  -0.06   &         1  &   0.39  &  -0.01   &  -0.14  \\
      0.43 &  0.04    &    0.39      &    1  &   -0.03 &      0.34 \\
    -0.02  &  0.01  &  -0.01  &   -0.03    &        1 &   -0.04  \\
     -0.21  &   0.03   &  -0.14   &   0.34  &  -0.04    &        1 
\end{bmatrix},
\end{equation}
\begin{equation}
\mathbf{C}_\text{f} =
\begin{bmatrix}
     1     & 0.15  &    0.62    &  0.36   &  -0.11 &    -0.15 \\
     0.15 &            1 &   -0.08   & 0.06 &   -0.10 &   -0.02 \\
      0.62  &  -0.08 &           1   &  0.40   &  0.02   &  -0.16 \\
     0.36    &   0.06  &    0.40     &        1    & 0.00 &   0.30 \\
     -0.11    &-0.10   &  0.02    &    0.00    &        1  &  -0.05 \\
     -0.15    &-0.02    & -0.16  &   0.30   & -0.05 &           1 
\end{bmatrix}.
\end{equation}
The entries of each $\mathbf{C}$ are (from left-to-right, and top-to-bottom):  degree centrality, clustering coefficients, betweenness centrality, closeness centrality, eigenvalue spectrum, and mean nearest-neighbor degree.  Many of the off-diagonal elements of the $\mathbf{C}$ matrices are close to zero, suggesting that the features are to a large extent independent of one another.

\begin{figure}[t]
\begin{center}
\includegraphics[width=0.45\textwidth]{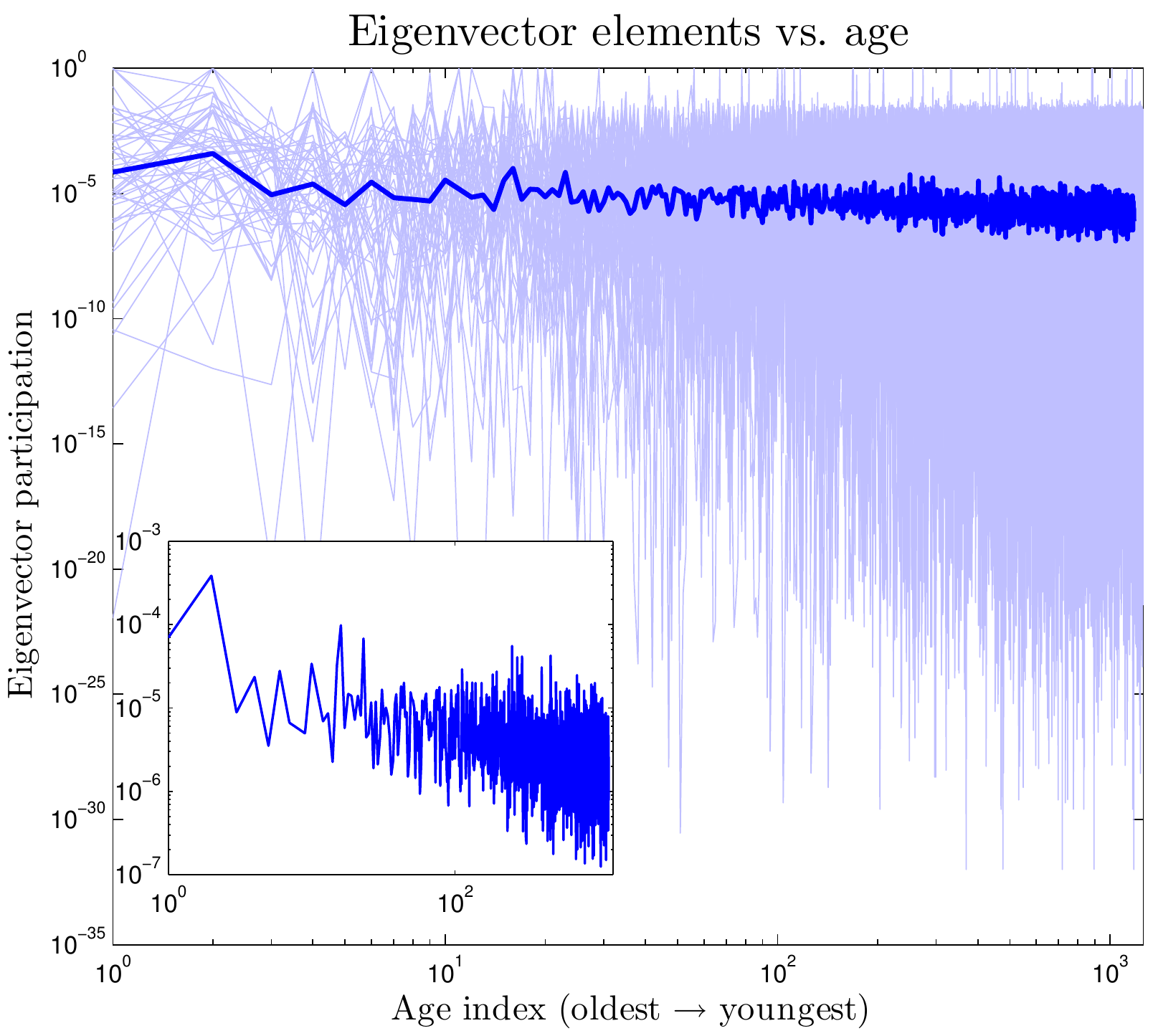}
\end{center}
\caption{{\bf Laplacian eigenvector participation.}  Elements of the eigenvector of the Laplacian matrix (defined as $\mathbf{K} - \mathbf{A}$, where $\mathbf{K}$ is a diagonal matrix with the degree of node $i$ as element $K_{ii}$) associated with the largest eigenvalue vs.~protein age index (time of introduction) in the yeast simulation.  Details of this method are discussed in \cite{Zhu_2011}.  Heavy lines are the median values from 50 simulations, and light lines are results of individual simulations.  The inset plot shows the trend line with a rescaled $y$-axis.}
\label{fig:lap_age}
\end{figure}

\begin{figure*}
\begin{center}
\includegraphics[width=0.32\textwidth]{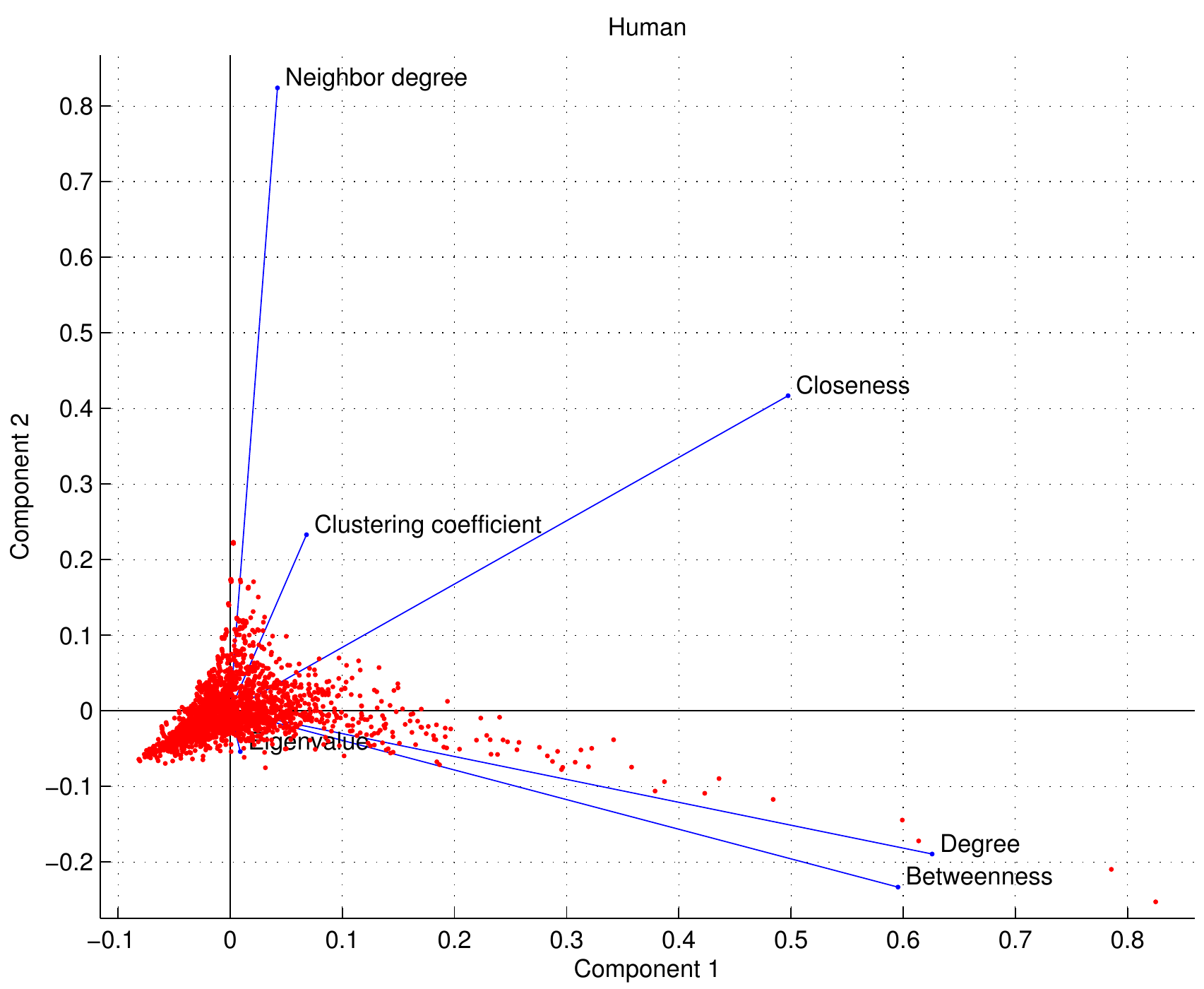}
\includegraphics[width=0.32\textwidth]{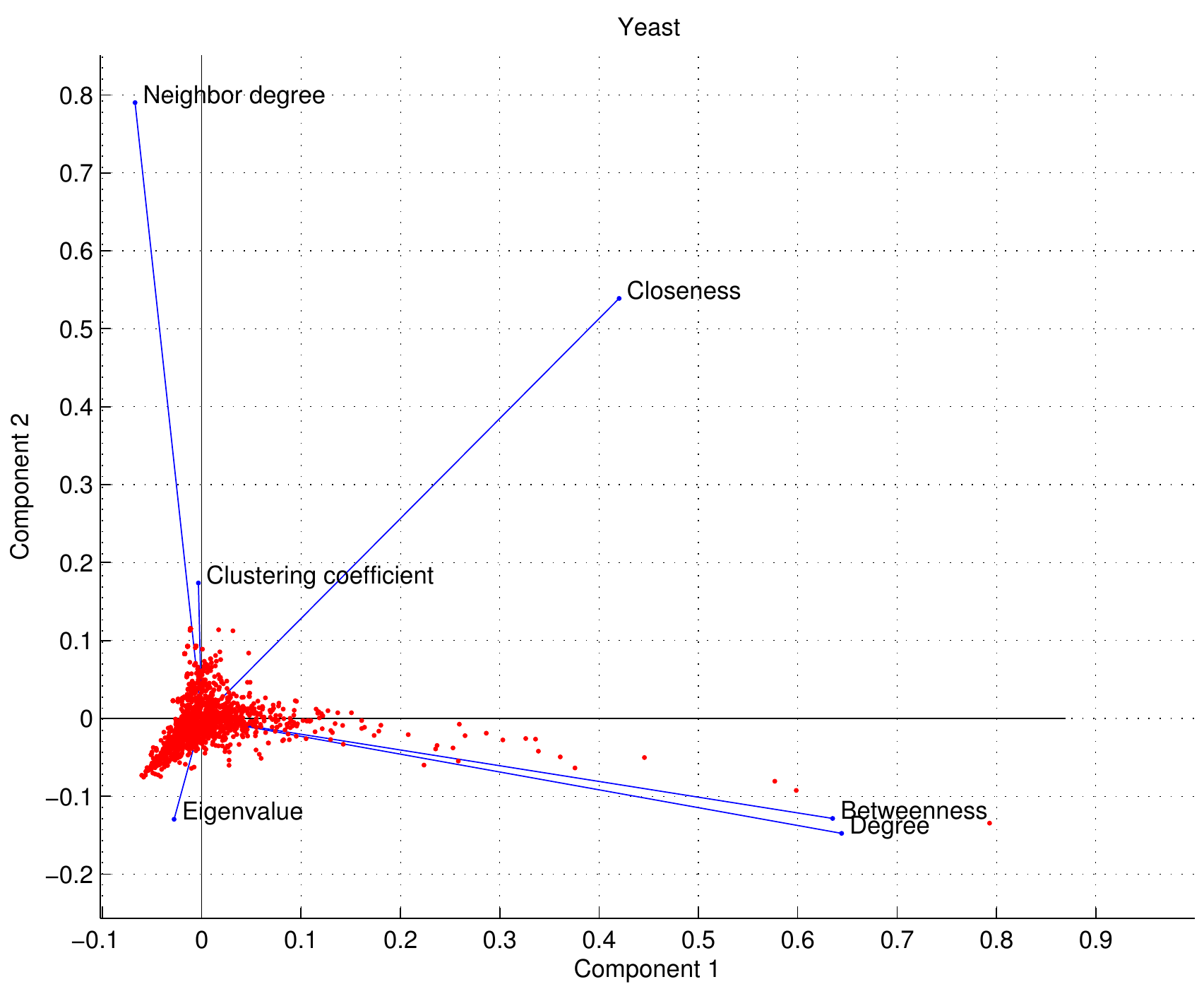}
\includegraphics[width=0.32\textwidth]{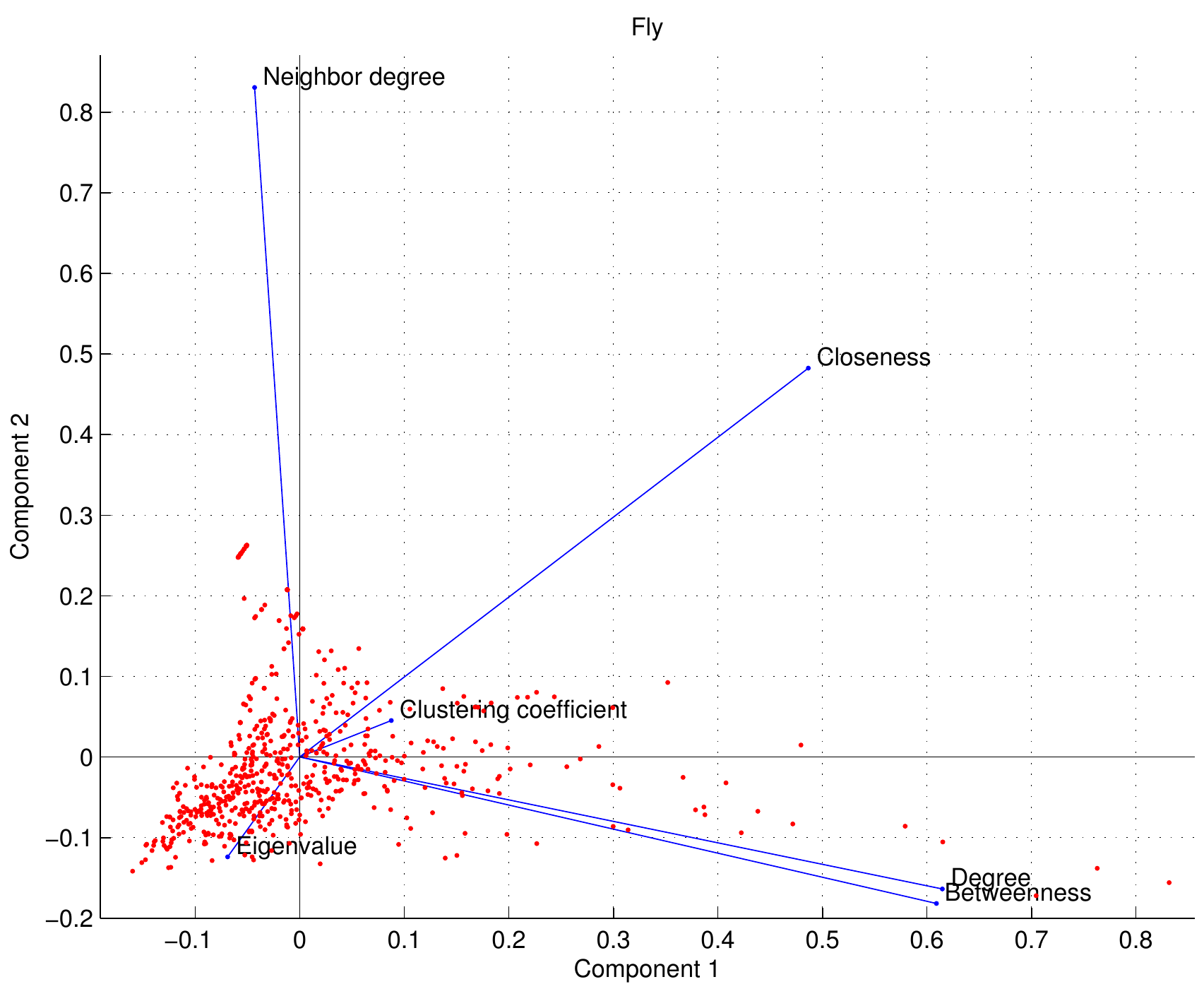}
\end{center}
\caption{{\bf Principal component analysis.}  Shown are the factor loadings and scores on the first two principal components.  Data scores are shown in red, and blue lines represent feature loadings.}
\label{fig:pca}
\end{figure*}

To perform PCA, we diagonalized each correlation matrix,
\begin{equation}
\mathbf{C} = \mathbf{S \Lambda S}^\mathrm{T},
\end{equation}
where $\mathbf{\Lambda}$ is a diagonal matrix of eigenvalues and $\mathbf{S}$ has the eigenvectors of $\mathbf{C}$ as its columns.  As shown in Figure~\ref{fig:pca}, the degree and betweenness show similar loadings on the first two principal components, reflecting the nearly linear relation between these centrality scores (Figure~\ref{fig:bvk}).

The eigenvalue matrices $\mathbf{\Lambda}$ are given by:
\begin{equation}
\mathbf{\Lambda}_\text{h} =
\begin{bmatrix}
& 2.27 & & & & & & \\
& & 1.23 & & & & & \\
& & & 1.02 & & & & \\
& & & & 0.98 & & & \\
& & & & & 0.40 & & \\
& & & & & & 0.11 &
\end{bmatrix},
\end{equation}
\begin{equation}
\mathbf{\Lambda}_\text{y} =
\begin{bmatrix}
& 2.20 & & & & & & \\
& & 1.30 & & & & & \\
& & & 1.01 & & & & \\
& & & & 0.98 & & & \\
& & & & & 0.43 & & \\
& & & & & & 0.08 &
\end{bmatrix},
\end{equation}
\begin{equation}
\mathbf{\Lambda}_\text{f} =
\begin{bmatrix}
& 1.95 & & & & & & \\
& & 1.24 & & & & & \\
& & & 1.13 & & & & \\
& & & & 0.91 & & & \\
& & & & & 0.44 & & \\
& & & & & & 0.32 &
\end{bmatrix}.
\end{equation}
(Zeros have been suppressed for clarity.)  The fraction of variance explained by the $i$th principal component is given by $\Lambda_{ii}/\sum_j \Lambda_{jj}$.  The closer the number of components required to explain most of the variance is to the total number of input signals, the more independent the signals are.  In yeast and humans, 4 components are required to explain 90\% of the variance; in fruit flies, it requires 5 components.  Linear transformations are able to only modestly reduce the dimensionality of the problem, suggesting that each feature contributes unique information about the network's structure.  This does not, of course, rule out the possibility of the existence of other independent, informative features, a far more complicated question which is outside the scope of this current work.

\begin{figure*}
\begin{center}
\includegraphics[width=0.32\textwidth]{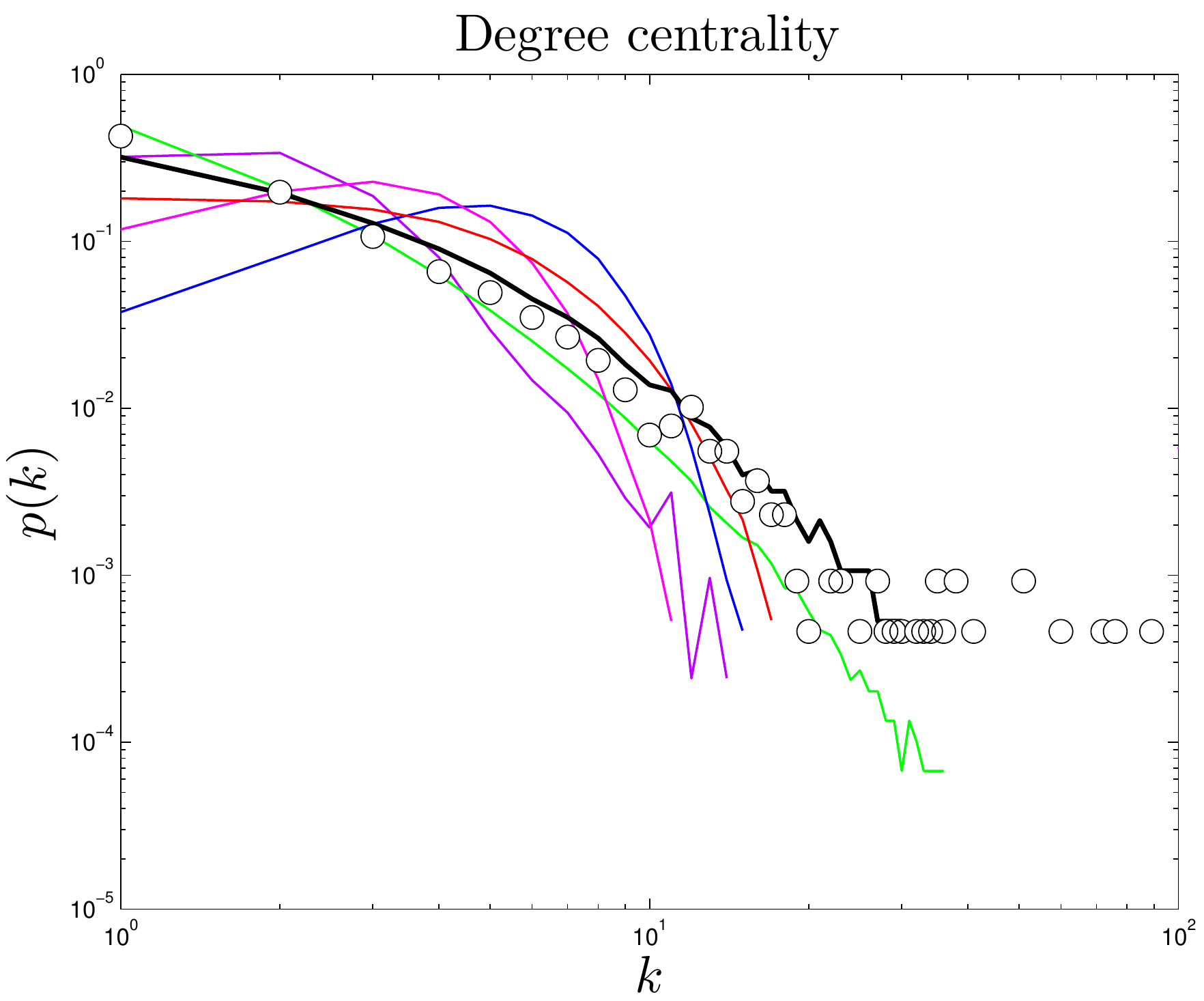}
\includegraphics[width=0.32\textwidth]{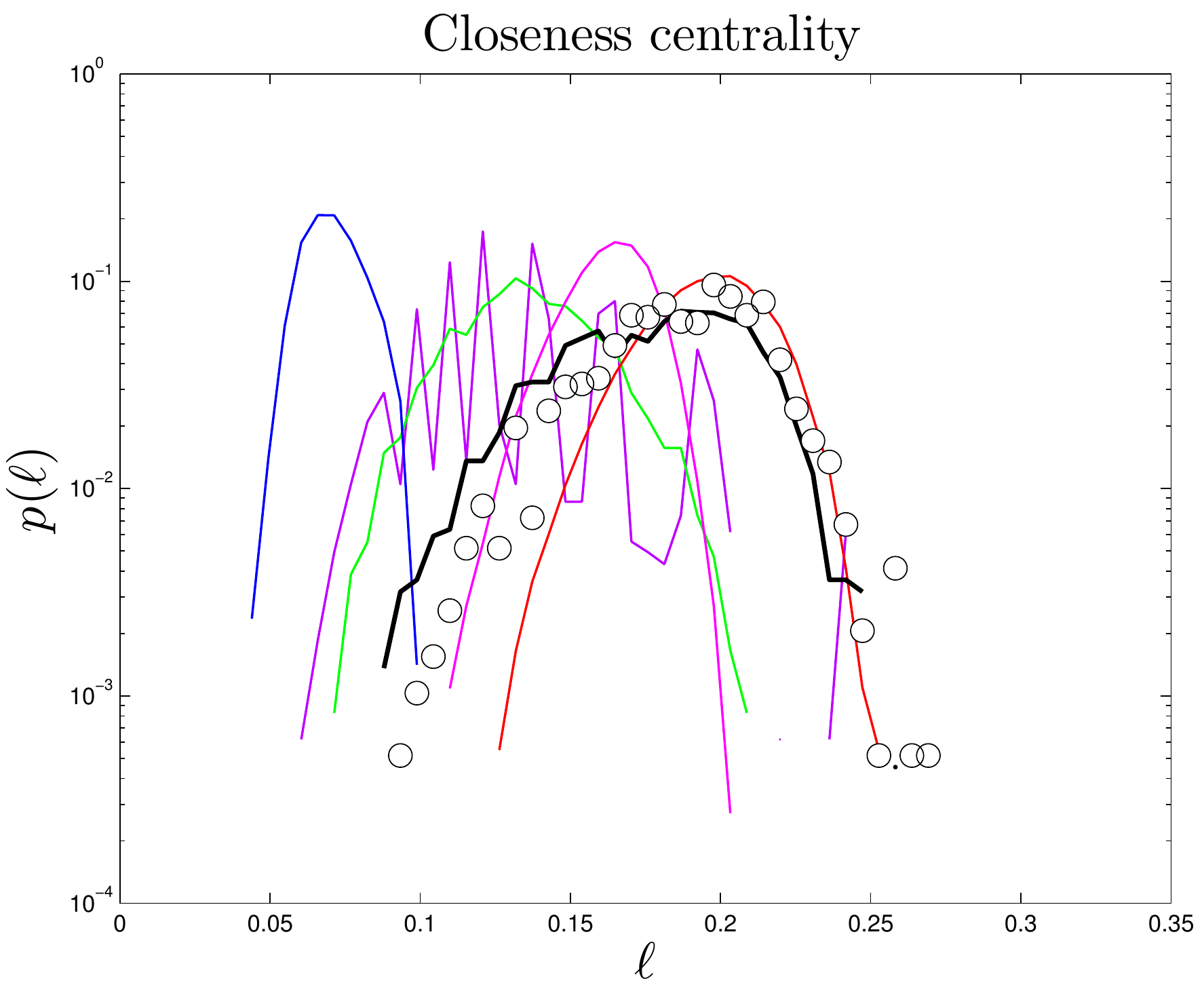}
\includegraphics[width=0.32\textwidth]{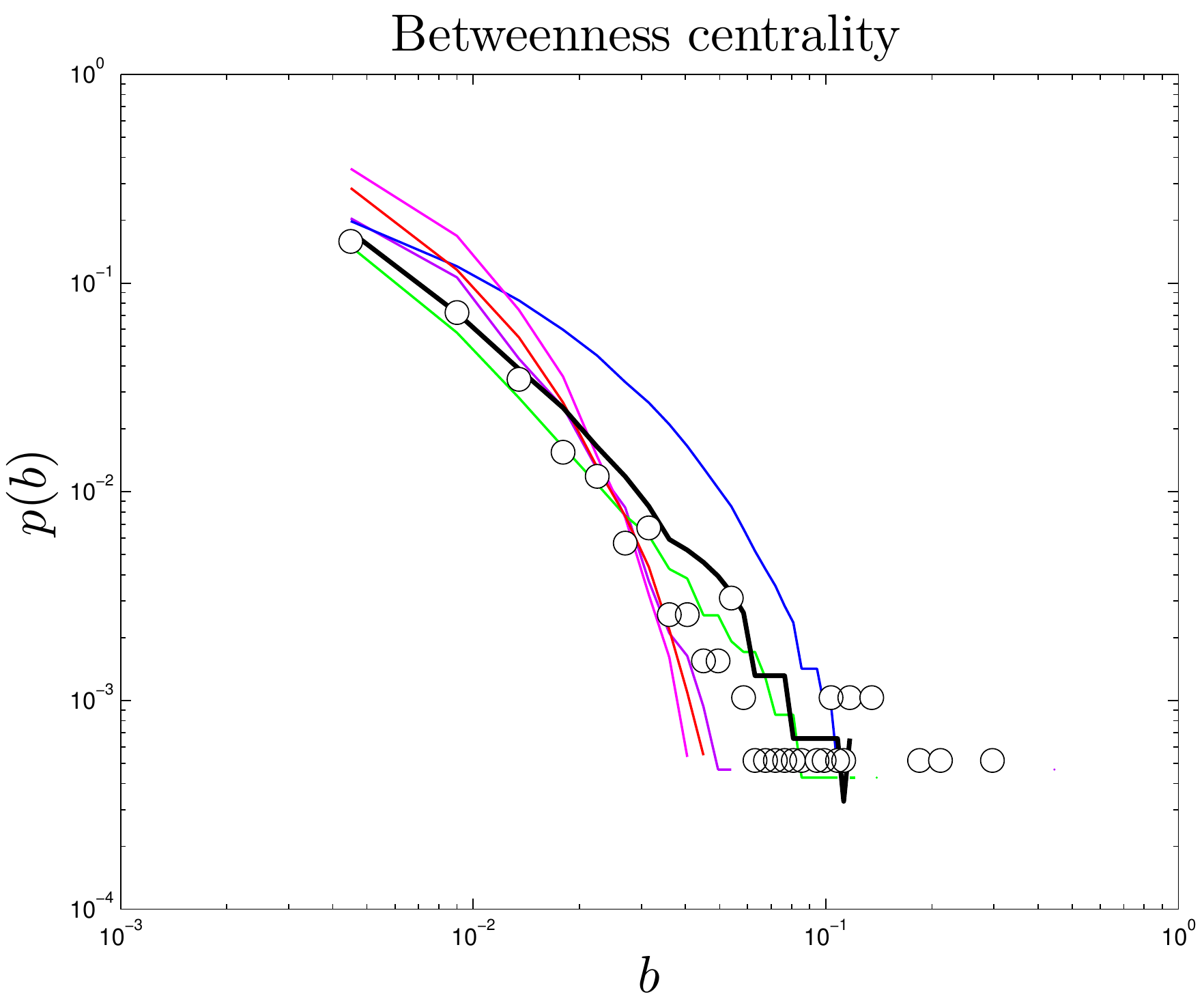}
\includegraphics[width=0.32\textwidth]{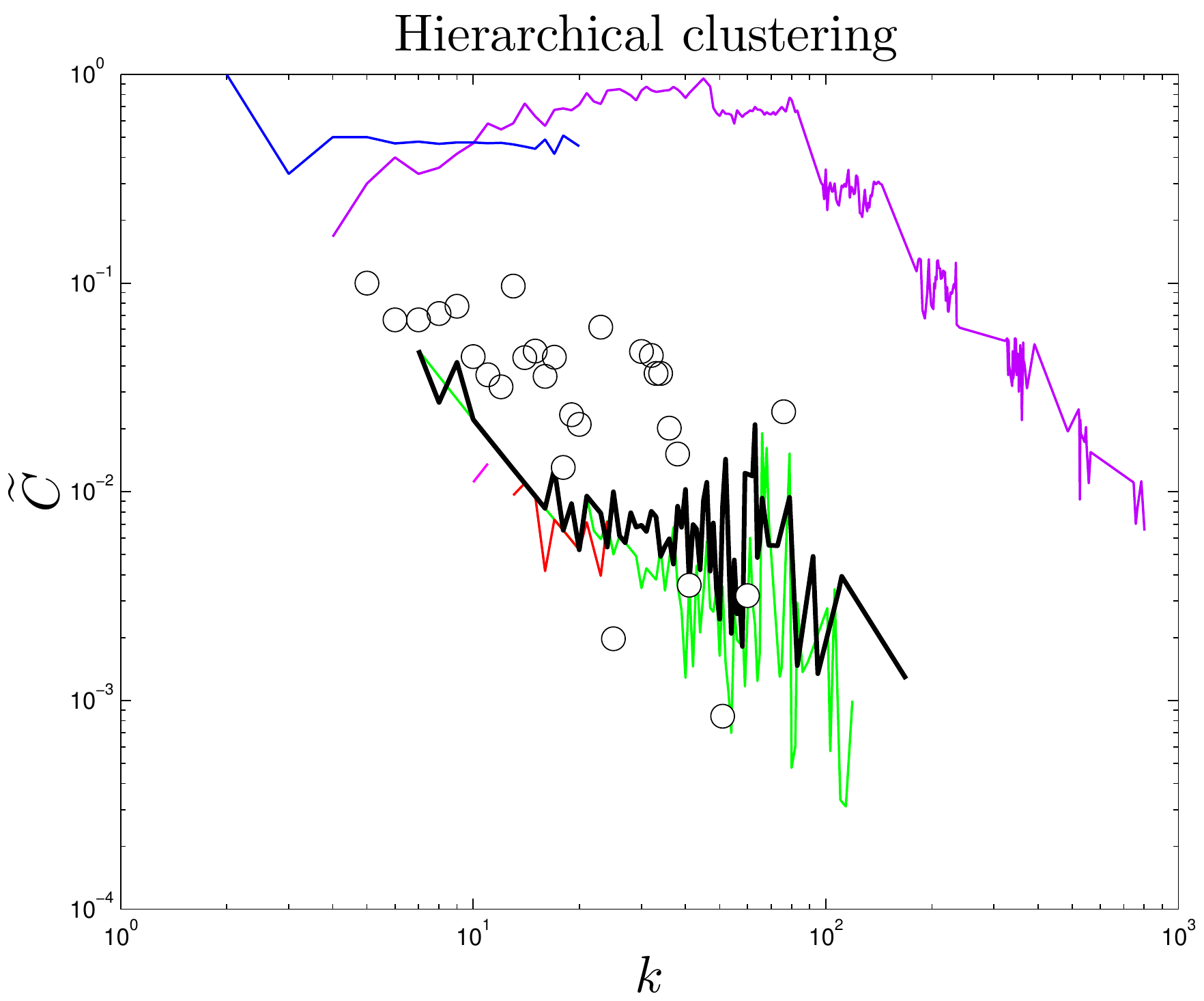}
\includegraphics[width=0.32\textwidth]{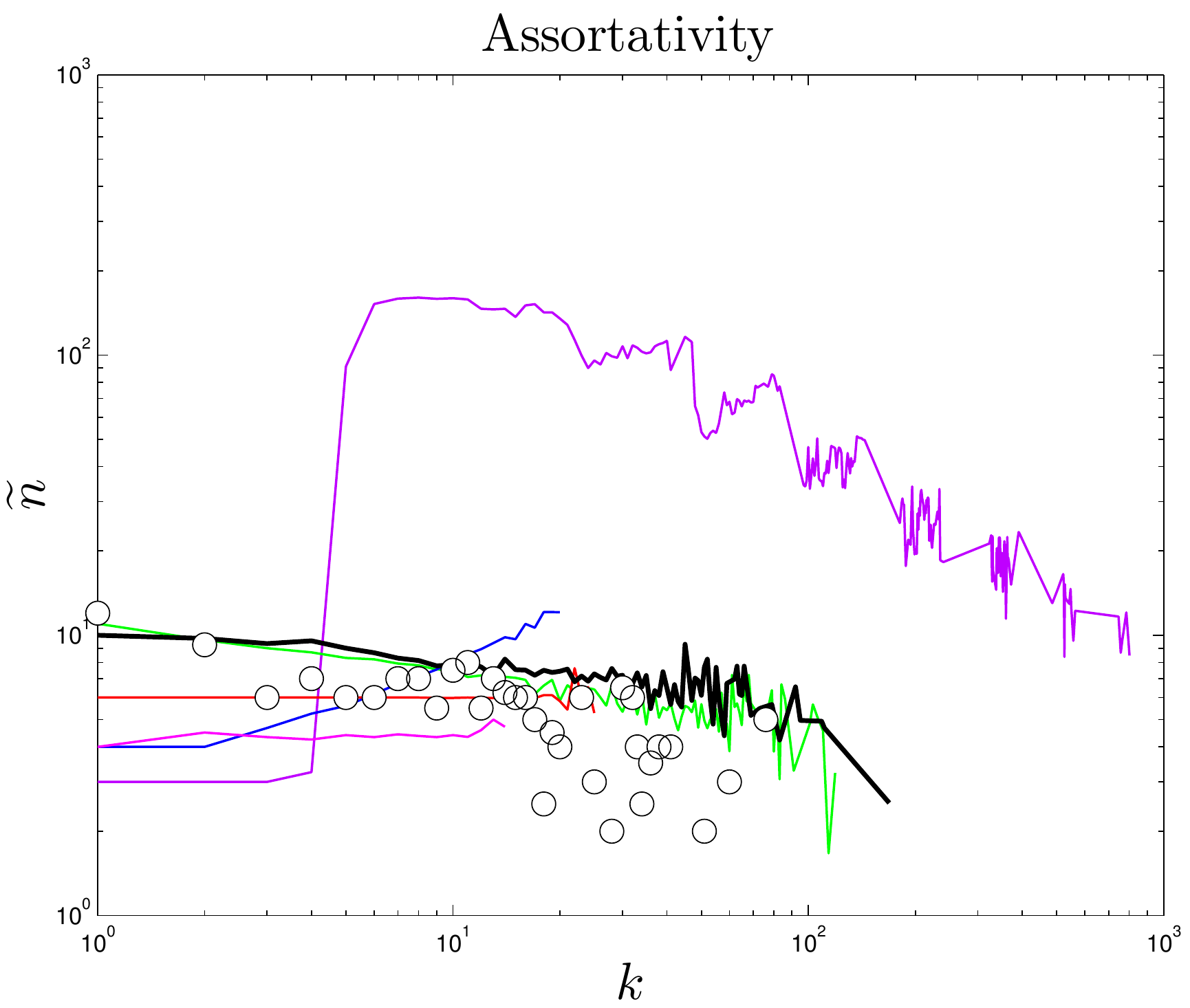}
\includegraphics[width=0.32\textwidth]{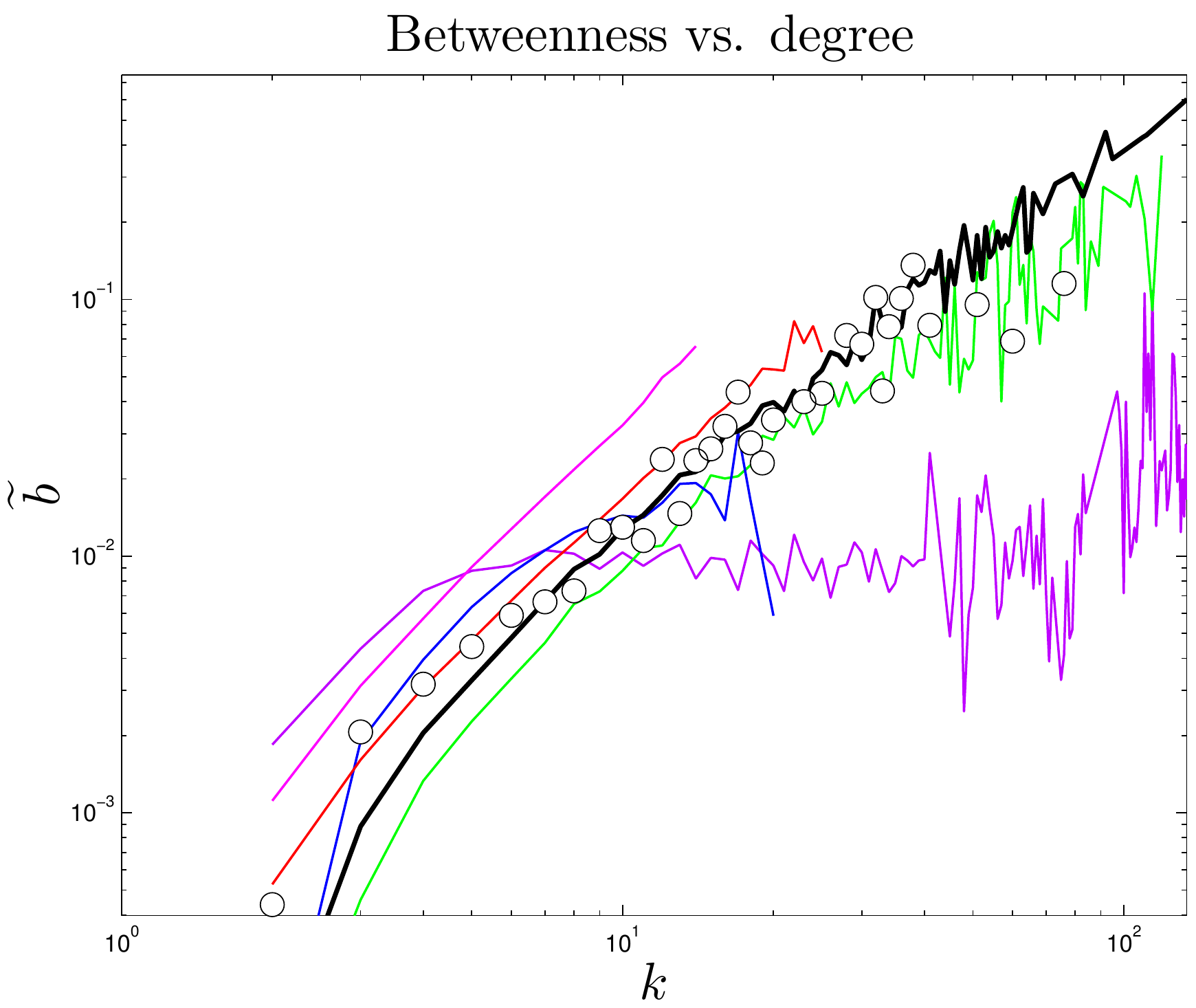}
\includegraphics[width=0.32\textwidth]{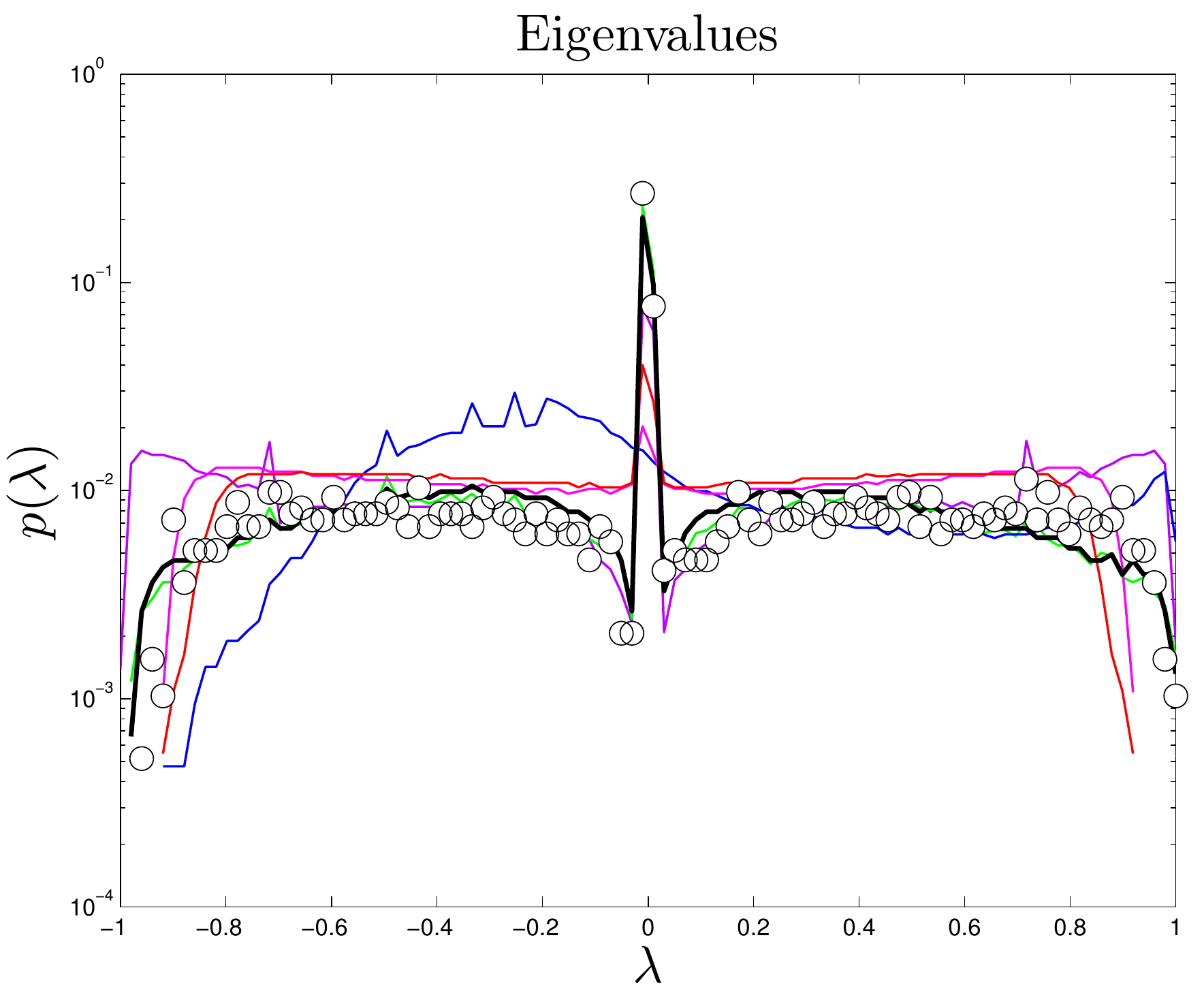}
\includegraphics[width=0.32\textwidth]{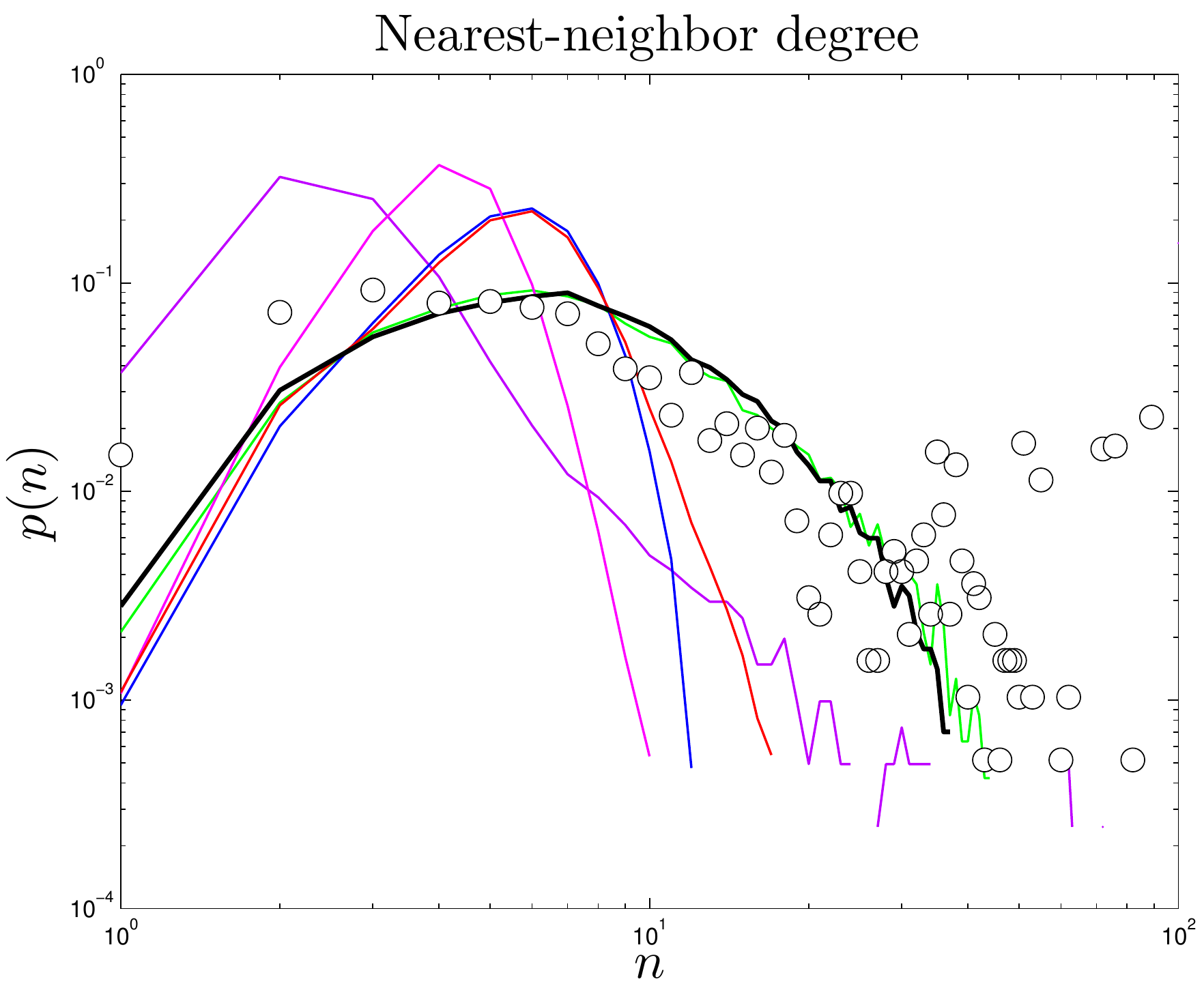}
\includegraphics[width=0.32\textwidth]{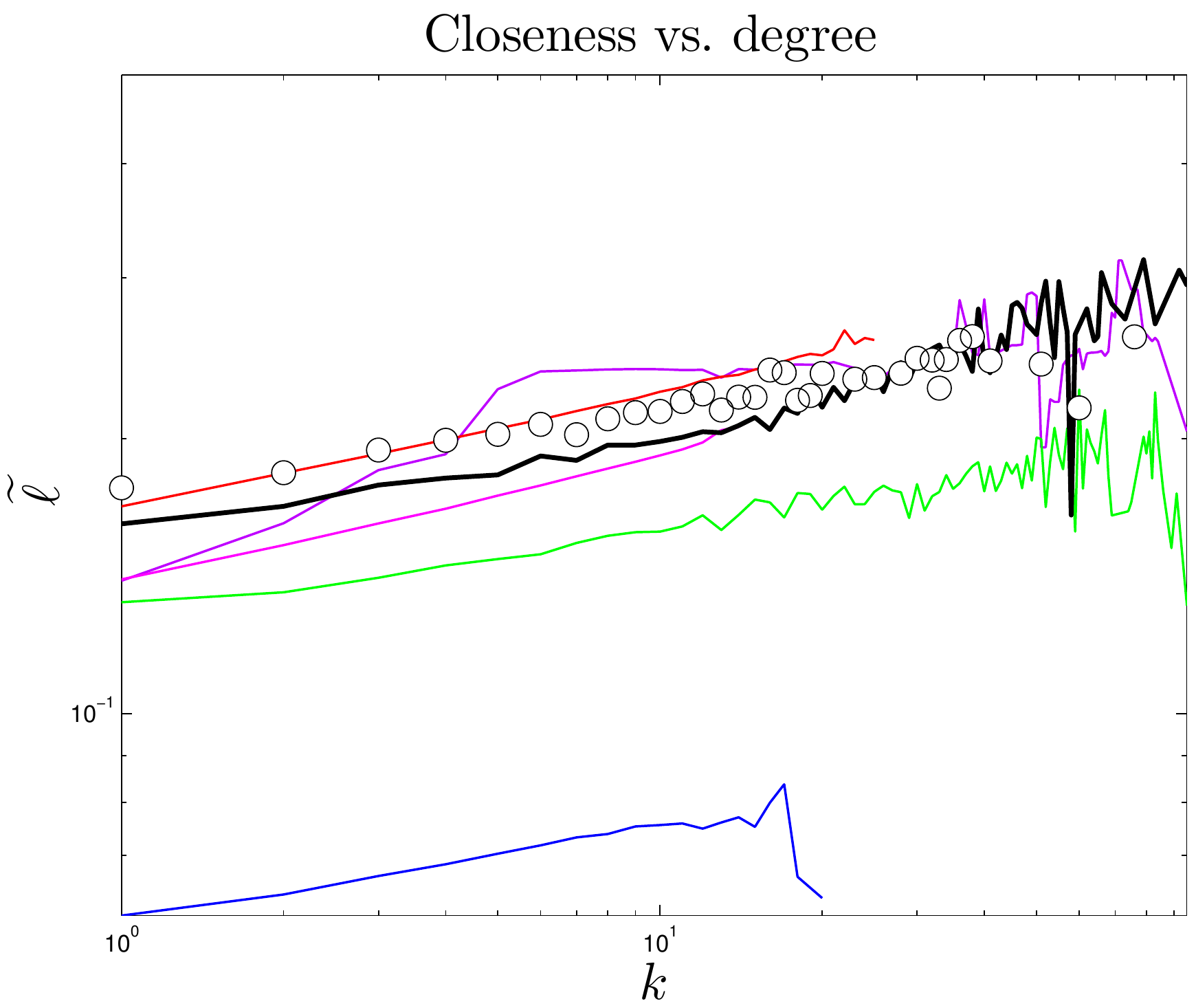}
\end{center}
\caption{{\bf Model comparison.}  Comparison of five other models to the yeast PPI network: V\'azquez \cite{Vazquez_2003} (green), Berg \cite{Berg_2004} (red), random geometric \cite{Przulj_2004} (dark blue), MpK desolvation \cite{Deeds_2006} (purple), and ER random graph \cite{Erdos_1960} (brown).  For reference, DUNE model results are shown as a black line.  Dots represent high-confidence experimental yeast data, and solid lines are median values over 50 simulations.}
\label{fig:othermodels}
\end{figure*}

\begin{table*}[t]
\begin{center}
\begin{tabular*}{\hsize}{@{\extracolsep{\fill}}| l | l l l l l l l l l l l | l |}
\hline
 		& $p(k)$ 	& $p(\ell)$	& $p(b)$	& ${C}(k)$ 	& ${n}(k)$	& ${b}(k)$	& $p(\lambda)$	& E.T.		& E.T. ($k$)	& $p(n)$	& ${\ell}(k)$ & Total \\
\hline
DUNE		& {\bf0.47}	& {\bf0.37}	& {\bf0.36}	& 0.60 	& 0.29	& 0.24	& 0.14		& 0.10	& 0.20	& 0.56	& 0.14	& {\bf3.48} \\
V\'azquez	& 0.52 	&  0.81	& 0.39	& 0.62	& 0.27	& 0.34	& {\bf0.12}		& 0.10	&  {\bf0.15}	& {\bf 0.50}	& 0.26	& 4.11 \\
Berg		& 0.70 	& 0.45	& 0.72	& 0.67	& {\bf0.15}	& {\bf0.19}	& 0.29		& 0.12	& 0.30	& 0.86	& {\bf0.06}	& 4.51 \\
RG		& 0.81 	& 0.98	& 0.61	& 0.83	& 0.22	& 0.23	& 0.33		& 0.31	& 0.45	& 0.89	& 0.49	& 6.15 \\
MpK		& 0.82 	& 0.79	& 0.68	& 0.89	& 0.88	& 0.58	& 0.18		& 0.18	& 0.18	& 0.82	& 0.16	& 6.18 \\
ER		& 0.83 	& 0.70	& 0.78	& {\bf0.53}	& 0.22	& 0.41	& 0.28		& {\bf0.07}	& 0.32	& 0.93	& 0.07	& 5.14\\
\hline
\end{tabular*}
\end{center}
\caption{Symmetric mean absolute percentage error (SMAPE) of simulation versus experiment in yeast (Eq.~\ref{eq:SMAPE}).  `E.T.' is the error tolerance curve with random protein removal, and `E.T. ($k$)' is the error tolerance curve with highest-degree proteins removed first.  `DUNE' is the model described here, `V\'azquez' is the DU-only model of \cite{Vazquez_2003}, `Berg' is the link dynamics model \cite{Berg_2004}, `RG' is random geometric \cite{Przulj_2004}, `MpK' is the physical desolvation model presented in \cite{Deeds_2006}, and `ER' is an Erd\H{o}s-R\'{e}nyi random graph \cite{Erdos_1960}.  For each comparison, the lowest value is shown in bold.}
\label{tab:SMAPE}
\end{table*}

\subsection*{Comparison to other models}

Our model is rooted in previous modeling efforts.  The basic framework for our model combines the gene duplication mechanism described in \cite{Vazquez_2003} with a link creation mechanism inspired by \cite{Berg_2004}.  The principal difference between our model and previous models is that our model considers duplication and mutation simultaneously.  The previous models we examined attempted to construct the PPI network from a single mechanism.  Another significant difference is our assimilation mechanism.  To the best of our knowledge, previous work has not explicitly modeled proteins integrating into biological pathways.

We compare the DUNE model to four models previously proposed for PPI networks.  Two were evolutionary models: (1) the V\'azquez model of DU followed by rapid loss-of-function mutations \cite{Vazquez_2003} and (2) the Berg `link dynamics' model of point mutations coupled with a PA-like `rich-get-richer' rule for assigning new interactions \cite{Berg_2004}.  (A slightly different DU model is presented by Pastor-Satorras \cite{Pastor-Satorras_2003}.  However, because the V\'azquez model has been shown to be a better fit to experimental data \cite{Middendorf_2005}, we have limited our DU-only comparison here to the V\'azquez model.)  Two others were static models (models of present-day networks that do not simulate the network's evolutionary path) that consider the primary organizing principle to be nonspecific interactions between proteins: (1) random geometric (RG), a mathematical model where proteins are randomly scattered in a 2 to 4 dimensional box, and any proteins close enough to one another form an interaction \cite{Przulj_2004} and (2) the `MpK' desolvation model, which assigns interactions based on proteins' exposed hydrophobic surface areas \cite{Deeds_2006}.  For reference, we also calculated results for an Erd\H{o}s-R\'{e}nyi (ER) random graph with $N$ and $\langle k \rangle$ set by the data \cite{Erdos_1960}.

These models were originally validated against different features of the empirical network, making it difficult to directly compare them.  To characterize these models in greater detail, we coded each of these models, and ran 50 simulations of each model with identical parameters and starting conditions.  Using Matlab, we coded the V\'azquez \cite{Vazquez_2003}, Berg \cite{Berg_2004}, RG \cite{Przulj_2004}, and MpK \cite{Deeds_2006} models as described in the original papers.  Since each model was originally parametrized for older yeast PPI data sets, we re-optimized the parameters for our yeast data as follows.  We used a Monte Carlo simulation to adjust each model's parameters to minimize the total symmetric mean absolute percentage error values (SMAPE; see below) for the yeast HitPredict data set.

For the V\'azquez model, we used a value of 0.582 for the post-duplication divergence probability, and a value of 0.083 for the dimerization probability.  As noted by previous authors, duplication-only simulations produce networks which are extremely fragmented~\cite{Hallinan_2004}.  We observed that the Vazquez simulations typically had around 20\% of their nodes in the largest connected component (Table~\ref{tab:SVF}).  Since most of the network features we examined are limited to the largest component, in order to make a reasonable comparison of the Vazquez simulation results to the data, we allowed the simulated network to grow until its number of links met or exceeded 5 times the number of links in the data, $K \ge 5K_\text{data}$.  Since the largest component is not always exactly 20\% of the total nodes, this stopping condition is somewhat arbitrary; however, results for this model seem robust to small changes in the stopping condition.  For the Berg model, we used the empirically estimated duplication rate of 0.01/gene/Myr, and found best-fit values of 24.5/gene/Myr for the mutation rate, and $N_\text{data}-98$ proteins for the initial network size.  For the RG model, we used a $45.5 \times 45.5 \times 45.5$ `box' with a maximum interaction radius of 3.92.  For the MpK model, the number of exposed surface residues was 19, the fraction of exposed hydrophobic residues was $M = 0.230 \pm 0.110$ (mean $\pm$ standard deviation), and the best-fit linear equation relating $M$ to the binding threshold was $1.09 M + 1.04$.

The V\'azquez simulations were initialized with 2 connected nodes, and the simulation was allowed to run until $K\ge 5 K_\text{data}$.  The Berg simulations were initialized with $N_\text{data}-98$ randomly connected nodes, then run until $N = N_\text{data}$.  The RG and MpK models (which are not evolutionary models and therefore create the network all at once) were set up as described in the original papers.

To characterize the networks, we computed several network properties:
\begin{description}
\item[Single-value:] modularity $Q$, diameter $D$, fraction of nodes in the largest component $f_1$, global clustering coefficient $\langle C \rangle$, and average protein degree in the largest component $\langle k \rangle$ (Table~\ref{tab:SVF})
\item[Distributional:] degree $p(k)$, betweenness $p(b)$, closeness $p(\ell)$, eigenvalue $p(\lambda)$, and nearest-neighbor degree $p(n)$ distributions
\item[Scatter plot:] closeness vs.~degree ${\ell}(k)$, clustering coefficient vs.~degree ${C}(k)$, betweenness vs.~degree ${b}(k)$, median nearest-neighbor degree vs.~degree ${n}(k)$, and error tolerance curves
\end{description}
and compared these features to those of empirical data from yeast.  As shown in Figure~\ref{fig:othermodels}, we found that none of the previous models capture the full set of network properties.

To quantify agreement with the data for non-single-value features, we calculated the symmetric mean absolute percentage error (SMAPE) between simulation and experiment \cite{OConnor_1998,Hibon_2000}:
\begin{equation}\label{eq:SMAPE}
\text{SMAPE} = \frac{1}{Y} \sum_i^{Y} \frac{\left| y_i - y^\text{data}_i \right|}{y_i + y^\text{data}_i},
\end{equation}
where $Y$ is the number of data points, and $y_i$ and $y^\text{data}_i$ denote the $i$th point of the response variable (Table~\ref{tab:SMAPE}) in the simulated and experimental data, respectively.  For the distributional features, $Y$ is the number of bins (arbitrarily chosen to be 100) minus the number of bins in which $y_i + y_i^\text{data} = 0$.  For non-distributional (scatter plot) features, $Y$ is the number of $k$ values with values for both simulation and experiment.  There are many possible measures of accuracy (such as the widely-used root mean squared error); we used SMAPE for two reasons.  First, because it relies on absolute value, SMAPE does not over-emphasize the impact of outliers.  Second, dividing by $y_i + y_i^\text{data}$ ensures that the magnitude of the response variable does not overwhelm the sum.  This is significant for the non-distrbutional features.  For example, in a plot of betweenness vs.~degree (Figure~\ref{fig:bvk}), we are just as interested in the overlap of the low-betweenness, low-degree region of the curve as we are with the high-betweenness, high-degree region.  SMAPE values are collected in Table~\ref{tab:SMAPE}.  As shown in Tables \ref{tab:SVF} and \ref{tab:SMAPE}, while previous models accurately reproduce certain features of the PPI network, only the DUNE model provides a reasonable across-the-board fit.

\end{appendix}

\end{document}